\PassOptionsToPackage{table}{xcolor}

\documentclass[11pt]{article}
\usepackage{jheppub}
\usepackage{graphicx} 
\usepackage{pgfplots}
\pgfplotsset{compat=1.18}
\usepackage{amsmath}
\usepackage{amsfonts}
\usepackage{hyperref}
\usepackage{braket}
\usepackage{physics}
\usepackage{comment}
\usepackage{dsfont}
\usepackage{hhline}
\usepackage{float}
\usepackage{bm}
\usepackage{tikz}
\usepackage{enumitem}
\usetikzlibrary{shapes,arrows,cd,chains,decorations.markings,decorations.pathmorphing,calc,positioning,patterns}
\usepackage{tikz-3dplot}

\tdplotsetmaincoords{80}{120}
\usepackage{mathrsfs}
\usepackage{adjustbox}

\usepackage{pgfplots}

\usepackage{xcolor}
\usepackage{booktabs, array, makecell, multirow}

\tikzset{
->-/.style args={#1rotate#2}{decoration={markings, mark=at position #1 with {\arrow[scale=1.5,rotate = #2 ]{stealth}}}, postaction={decorate}}
}
\tikzset{
-r-/.style args={#1rotate#2}{decoration={markings, mark=at position #1 with {\arrow[scale=1,rotate = #2 ]{>}}}, postaction={decorate}}
}

\DeclareMathOperator{\bbZ}{\mathbb{Z}}
\DeclareMathOperator{\bbR}{\mathbb{R}}
\DeclareMathOperator{\bbC}{\mathbb{C}}

\DeclareMathOperator{\calA}{\mathcal{A}}

\DeclareMathOperator{\calD}{\mathcal{D}}

\DeclareMathOperator{\calH}{\mathcal{H}}

\DeclareMathOperator{\calN}{\mathcal{N}}

\DeclareMathOperator{\calO}{\mathcal{O}}

\DeclareMathOperator{\calL}{\mathcal{L}}

\DeclareMathOperator{\calV}{\mathcal{V}}

\DeclareMathOperator{\calT}{\mathcal{T}}

\DeclareMathOperator{\HT}{{\mathrm{HT}}}

\title{Revisiting Schr\"odinger CFTs: Factorization, Massless Particles, and a Path to the Bootstrap}

\author[1]{Mathieu Boisvert,}
\author[1]{Shehab Hossam Fadda,}
\author[1,2]{Justin Kulp,} 
\author[2,3]{and Ramtin M.Yazdi}

\affiliation[1]{Yang Institute for Theoretical Physics, Stony Brook University, Stony Brook, NY 11794, USA}
\affiliation[2]{Simons Center for Geometry and Physics, Stony Brook University, Stony Brook, NY 11794, USA}
\affiliation[3]{Department of Physics and Astronomy, Stony Brook University, Stony Brook, New York 11794-3800, USA}
\emailAdd{mathieu.boisvert@stonybrook.edu}
\emailAdd{shihab.fadda@stonybrook.edu}
\emailAdd{jkulp@scgp.stonybrook.edu}
\emailAdd{ramtin.mohasselyazdi@stonybrook.edu}

\abstract{We revisit Schr\"odinger CFTs from a modern point of view. We introduce the ``harmonic trap geometry,'' analogous to the cylinder picture in relativistic CFTs, and demonstrate a state-operator correspondence that applies to all operators, including descendant, massless, and ``normal-ordered operators.'' A thermofield double construction plays an extremely important role. We systematically classify all physical spectra in the harmonic trap and their unitarity bounds, extending earlier results to include both massless and massive states of all spins, providing a new analytic treatment of unitarity bounds, and establishing foundations for a bootstrap. In our reformulation, previously known perturbative non-renormalization theorems follow immediately from non-perturbative factorization at fixed points and along RG flows. Massless states are described by an effective 1d CFT, as predicted by DLCQ, and violate the non-renormalization theorems. We include a self-consistent review of Schr\"odinger CFTs in our framework, making the paper accessible to anyone with a field theory background.}

\begin{document}

\maketitle

\section{Introduction, Background, and Summary}
QFT provides a general framework for describing physical systems at low energies or mesoscopic length scales relative to some microscopic scale. Additional assumptions like Lorentz invariance, unitarity, and locality further constrain IR physics in remarkably rigid ways, often making QFT a generic (if not unique) description of low-energy phenomena in many situations. A distinguished role is played by QFTs which are invariant under scale transformations, linking microscopic and macroscopic physics, and describing critical phenomena. When full conformal symmetry is present, kinematics become so constraining that it is possible to ``bootstrap'' the space of consistent theories, matching experimental results with stunning accuracy and providing one of few scenarios to understand strongly coupled systems (see \cite{Poland:2018epd} and references within).

However, none of the preceding ingredients -- Lorentz invariance, unitarity, locality/cluster decomposition, and especially conformal symmetry -- are strictly necessary for QFT. This is especially clear in condensed matter systems and/or lattice models; there is no reason for Lorentz symmetry to emerge in an effective field theory description of systems at mesoscopic length scales. On the contrary, a natural guess is that low energy physics will be \textit{anisotropic} in space and time, carrying an emergent ``non-relativistic'' (aka ``Galilean'') symmetry: invariant under space $P_i$ and time $P_0$ translations, Galilean boosts $K_i$, and spatial rotations $M_{ij}$; with a central element called mass $M$. At criticality, we expect such systems to have emergent anisotropic (aka ``Lifshitz'') scale symmetries $D$:
\begin{equation}
    e^{i\lambda D}\!: (t,x) \mapsto (\lambda^z t, \lambda x)\,,
\end{equation}
where $z$ is known as the ``dynamical critical exponent,'' and masses have length dimension $[m]_L = z-2$. See \cite{hornreich1975critical, grinstein1981anisotropic, Gegenwart:2008ttt, Hoyos:2013qna, Chapman:2015wha, Arav:2016akx, Chen:2017tij, Arav:2019tqm} for a mixture of experimental and theoretical results or \cite{Baiguera:2023fus} for a review.

Schr\"odinger field theories are a special class of non-relativistic conformal field theories, with $z=2$ scaling and special conformal invariance $C_0$ in time (for commutation relations, see Section \ref{sec:KinematicsAndReview}). Such theories emerge in a diverse range of scenarios in experiment, simulations, and theory, from cold atoms and nuclear physics to string theory. Moreover, their special $z=2$ scaling and conformal symmetry enable good theoretical control, making them an excellent experimentally motivated target for investigations by the conformal bootstrap. 

On the experimental side, Schr\"odinger field theories have seen tremendous success in describing cold atoms, phonons, and vortices in a ``harmonic trap'' tuned to a ``Feshbach resonance'' (effectively, tuned to criticality). In these cases, experiment \cite{roberts1998resonant, chin2001high, loftus2002resonant, o2002observation, Zwierlein:2004zz, regal2004observation, nascimbene2010exploring} and numerics \cite{chang2007unitary, von2007bec} neatly match theoretical predictions for anomalous dimensions \cite{Son:2005rv, Mehen:2007dn, Nishida:2007pj, nikolic2007renormalization, Nishida:2010tm}. Important applications also arise in describing deuterons and heavy ion EFTs \cite{Kaplan:1998tg, Kaplan:1998we, Mehen:1999nd, Kaplan:2009kr, Kobach:2018nmt}.

The ubiquity of Schr\"odinger field theories follows from their emergence in non-relativistic systems ``at unitarity'' \cite{regal2004observation, Nishida:2007pj}. In general, any non-relativistic Hamiltonian with two-body interactions has an asymptotic wavefunction of the form (see \cite{Collins:2019ozc} for a review):
\begin{equation}
    \psi_0(r) \stackrel{r\to\infty}{\sim} e^{i\vec{k}\cdot \vec{r}} + f(k) \frac{e^{i\vec{k}\cdot \vec{r}}}{r}\,,
\end{equation}
where $f(k)$ is the scattering amplitude. At energies much lower than the effective range of the two body potential $k \ll R_{\rm eff}^{-1}$, the system is fully described by the $s$-wave ($\ell = 0$) scattering amplitude
\begin{equation}
    f_0(k) = (k \cot \delta_0(k) - i k)^{-1}\,.
\end{equation}
The scattering length $a$ is defined by $f_0(k\to 0) =: -a$, and the ``unitarity limit'' is when $a\to \infty$.\footnote{This can be seen neatly in the BCS-BEC crossover (see e.g. \cite{LeggettOriginal, roberts1998resonant, chin2001high, loftus2002resonant, von2007bec}), describing effective spin-$\tfrac{1}{2}$ atoms in a magnetic field. The Zeeman effect splits the two energy levels and one computes the scattering length to be
\begin{equation}
    a = a_{\rm bg}\left(1-\frac{\Delta}{B - B_0}\right)\,,
\end{equation}
where $a_{\rm bg}$ is the scattering length in the absence of the magnetic field. For small $a < 0$ the system is BCS, and when $a > 0$ the system is BEC. When $B$ is tuned to the Feshbach resonance $B_0$, then $a \to \infty$ and we are in the unitarity limit. This critical/unitarity point is described by a Schr\"odinger CFT.
}
As $a\to\infty$, the $s$-wave cross section becomes $\int_{S^2}|f_0|^2 = 4\pi/k^2$ and completely saturates the optical theorem bound for the total cross section, i.e. $\sigma_{s} = \sigma_{\mathrm{tot}}$, thus the name ``unitarity limit.'' Relatedly, EFTs organize corrections to $f_0(k)$ as a power series in $k/R_{\rm eff}$. We give a lightning review of these experimental connections in Section \ref{sec:FermionsAtUnitarity} because they play such a major motivational role in our paper and the theory of Schr\"odinger systems more broadly. 

On the theoretical side, Schr\"odinger field theories have been the subject of intense scrutiny. Being analogues of conformal field theories, they have a classification of operators into primaries and descendants with similarly strongly constrained correlation functions. We review these theoretically important kinematic facts in Section \ref{sec:KinematicsAndReview}. They have also been studied as non-relativistic limits \cite{Jackiw:1991je, Henkel:1993sg, Henkel:2003pu, Henkel:2005dj}, in large $N$ \cite{nikolic2007renormalization, Bekaert:2011qd} and large charge regimes \cite{Kravec:2018qnu, Favrod:2018xov, Kravec:2019djc}, and with defects \cite{Raviv-Moshe:2024yzt}. They also connect to quantum hall physics, supersymmetry, and stochastic/out-of-equilibrium dynamics \cite{Ardonne:2003wa, Son:2013rqa, Chapman:2015wha}. Schr\"odinger CFTs are also expected to arise in the lightlike/null reduction of Lorentzian CFTs in one higher dimension \cite{Duval:1994qye, Duval:2008jg, Maldacena:2008wh, Bekaert:2011qd, NullDefects} and in non-relativistic string theory \cite{Gomis:2020fui, Gomis:2020izd}. This is the conformal analogue of the well-known fact that lightcone quantization and/or null reduction leads to Galilean symmetries in the reduced system \cite{Weinberg:1966jm, Susskind:1967rg, Kogut:1969xa, Kogut:1972di, Banks:1996vh, Venugopalan:1998zd}.

Given their experimental and theoretical importance, Schr\"odinger field theories should make an excellent subject for conformal bootstrap techniques. However, despite the experimental and theoretical successes, a number of critical foundational issues are unresolved. For example:
\begin{itemize} \setlength\itemsep{0em}
    \item While a classification of local operators into primaries and descendants exists, the construction of local primary operators is claimed to only work for non-zero masses $M\neq 0$. Moreover, primaries with $M = 0$ are only presently understood as composite operators in Lagrangian theories, and satisfy rich interlocking conservation laws \cite{Nishida:2007pj, Bekaert:2011qd, Golkar:2014mwa}.
    \item Many experimental successes of non-relativistic CFTs have come from using a ``state-operator correspondence'' to match the scaling dimensions of operators to the energy of states in a ``harmonic trap geometry'' \cite{Nishida:2007pj, Mehen:2007dn}. Relatedly, Schr\"odinger CFTs are treated as if they have a convergent operator product expansion \cite{Golkar:2014mwa, Goldberger:2014hca}. A simple consideration of non-relativistic geometry indicates the existence of a state-operator correspondence (and OPE convergence) is more non-trivial than previously believed (see Section \ref{sec:HTGeometry}).
    \item Relatedly, any current understanding of the state-operator correspondence does not actually apply to all primaries \cite{Bekaert:2011qd, Golkar:2014mwa, Goldberger:2014hca}. Simple universal objects, like the number density $n$, probability current $J_i$, stress tensor $T_{ij}$, or any other ``normal ordered'' composite primary, have no dual state.
    \item It is not known if there are RG monotonicity theorems for NR CFTs.\footnote{On one hand, the same argument that RG ``zooms out'' and loses degrees of freedom should imply non-relativistic RG monotonicity theorems. On the other hand, monotonicity theorems forbid limit cycles in standard 4d CFTs \cite{Luty:2012ww}, while NR CFTs are believed to have limit cycles \cite{Nishida:2007de}.}
    \item Since $M$ is central in Galilean field theories and Schr\"odinger CFTs then, for $M\neq 0$, conservation of mass is equivalent to conservation of particle number. This leads to perturbative non-renormalization theorems because virtual particles are forbidden from being created in loop diagrams \cite{Bergman:1991hf, Klose:2006dd, Auzzi:2019kdd, Chapman:2020vtn}. The validity of such non-renormalization theorems is more nebulous non-perturbatively.
    \item Many interesting and important NR CFTs are obtained by null reduction of conformal systems. Famously, non-perturbative information about the Lorentzian CFT is hidden in the $P_- = M = 0$ sector in the null reduction (see e.g. \cite{Hellerman:1997yu}), which we have no control over. While much has been understood from null reductions, some additional care is required in matching the causal structure of non-relativistic theories with null-reduction (see e.g. \cite{Herzog:2008wg} and references within for a nice discussion).
    \item Putative holographic duals to non-relativistic CFTs have been proposed and thoroughly studied in a number of references \cite{Son:2008ye, Goldberger:2008vg, Balasubramanian:2008dm, Barbon:2008bg, Maldacena:2008wh, Kachru:2008yh, Herzog:2008wg}. However, almost all studies are purely kinematical (but see \cite{Dorey:2022cfn, Dorey:2023jfw, Mouland:2023gcp}), mostly matching symmetries and not recovering dynamical information (like three-point functions) from explicit bulk dynamics. Current proposals also do not give satisfying explanations for the emergence of ``creation'' and ``annihilation'' operators in the CFT or non-renormalization theorems.
\end{itemize}

The origin of many of these foundational issues stems from a poor understanding of the massless $M=0$ (or ``neutral'') sector \textit{and} the splitting of CFT operators into creation and annihilation operators which annihilate the harmonic trap vacuum state (on at least one side). Indeed, these two problems are completely independent, but often conflated and blamed for each other's issues. Thus a better understanding of the massless sector and a canonical definition of the operator algebra of Schr\"odinger CFTs should shed some light on the important conceptual issues which currently preclude a formal bootstrap approach. The importance of understanding $M=0$ primaries is underscored further when we note that $M=0$ operators precisely constitute the good deformations of the theory. i.e. they are the hermitian observables.

The main objective of this paper is to introduce a systematic framework for discussing Schr\"odinger field theories. In doing so, we resolve some longstanding definitional issues surrounding the $M=0$ sector and the polarization of the theory into creation/annihilation operators, and thus provide resolutions to many of the problems above. Our approach follows by making analogies between Lorentzian CFT and the (experimentally successful) harmonic trap geometry. Consequently, we are able to extend the construction of local primaries and the state-operator correspondence to $M=0$ operators, and capture the previously known $M=0$ ``composite operators'' non-perturbatively. We give non-perturbative arguments for non-renormalization theorems and the existence of canonical ``normal ordered'' composite operators.  We also give evidence for the existence of these new $M=0$ operators, and show how they support famous claims of emergent ``conformal quantum mechanics'' in null reductions, and spoil perturbative non-renormalization theorems.

In the remainder of this introduction we provide a review of the literature and setup the framework for our formalism. In Section \ref{sec:KinematicsAndReview}, we review the Schr\"odinger algebra, definitions of primaries, and correlation functions in real-time Schr\"odinger CFTs. In Section \ref{sec:FermionsAtUnitarity}, we review how this framework can be used to compute the scaling dimensions of ``fermions at unitarity,'' in the $4-\epsilon$ and $2+\bar{\epsilon}$ expansions and review the perturbative non-renormalization theorem. While nothing in Sections \ref{sec:KinematicsAndReview} and \ref{sec:FermionsAtUnitarity} is fundamentally new, our review provides a new useful abstract reframing of the relevant ingredients and arguments, enabling anyone with a field theory background to understand the subject. In Section \ref{sec:Outline}, we give a more detailed discussion of the problems mentioned above, and an outline and summary of the remainder of the paper.

\subsection{Non-Relativistic Schr\"odinger CFTs}\label{sec:KinematicsAndReview}
We are interested in $(d+1)$-dimensional field theories whose spacetime symmetries include the Schr\"odinger algebra $\mathfrak{sch}_d$. As mentioned above, theories with Schr\"odinger symmetry describe ``$z=2$'' non-relativistic CFTs. We will briefly review the most famous interacting example, called fermions at unitarity, in Section \ref{sec:FermionsAtUnitarity}. Here we set kinematic conventions and comment on differences between ``particle number'' and ``mass'' charges, which are often omitted because massless states are typically ignored in Schr\"odinger CFTs.

\paragraph{Schr\"odinger Symmetry.} The Galilean algebra in $(d+1)$-dimensions consists of translations in time $P_0$ and space $P_i$, Galilean boosts $K_i$, and rotations $M_{ij}$. The Galilean algebra admits a unique central extension by an element $M$, called the ``mass,'' satisfying
\begin{equation}
    [K_i,P_j] = i \delta_{ij} M\,.
\end{equation}
The centrally extended Galilean algebra describes the spacetime symmetries of non-relativistic quantum systems, with the mass $M$ encoding a familiar Heisenberg uncertainty relation for position and velocity.

The Schr\"odinger algebra $\mathfrak{sch}_d$ can be viewed as a conformal extension of the non-relativistic Galilean algebra, where a 1d conformal algebra $\mathfrak{sl}(2,\bbR)$ is adjoined to the time direction.\footnote{An enhanced $SL(2,\bbR)$ symmetry in time is not completely exotic, as with the dynamical symmetries of magnetic monopoles in (3+1)d \cite{Jackiw:1980mm} or vortices in (2+1)d \cite{Jackiw:1989qp}; however, neither example possesses full Schr\"odinger symmetry. On the other hand, full Schr\"odinger symmetry does emerge upon making the gauge field dynamical, leading to non-relativistic Chern-Simons theories \cite{Jackiw:1990mb}.} Specifically, we add a special conformal generator $C_0$ and a $z=2$ dilatation operator $D$, which scales space and time anisotropically $e^{i\lambda D}\!:(t,x^i) \mapsto (\lambda^2 t, \lambda x^i)$. Altogether, the commutation relations are:
{\begin{alignat}{3}
    [D,P_0] 
        &= 2i P_0\,,\quad
    &[C_0,P_0] 
        &= i D\,,\quad
    &[D,C_0] 
        &= -2i C_0\,,\nonumber\\
    [D,P_i] 
        &= i P_i\,,\quad
    &[K_i,P_j] 
        &= i \delta_{ij} M\,,\quad
    &[D,K_i] 
        &= -i K_i\,,\nonumber\\
    [C_0,P_i] 
        &= i K_i\,,\quad
    &&&[P_0,K_i] 
        &= -i P_i\,,\label{eq:SchrodingerAlgebra}\\
    [M_{ij},P_k] 
        &= i(\delta_{ik} P_j-\delta_{jk}P_i)\,,\mkern-36mu
    &&&[M_{ij},K_k]
        &= i(\delta_{ik} K_j-\delta_{jk}K_i)\,,\nonumber
\end{alignat}
\vskip -1.8em
\begin{equation}
    \quad[ M_{ij}, M_{kl}]
        = -2i(\delta_{j[l} M_{k]i}-\delta_{i[l} M_{k]j})\nonumber\,,
\end{equation}}
forming an algebra
\begin{equation}
    \mathfrak{sch}_d = (\mathfrak{sl}(2,\bbR) \times \mathfrak{so}(d)) \ltimes \mathfrak{h}_{d}\,,
\end{equation}
where the $\mathfrak{sl}(2,\bbR)$ is spanned by $\{P_0, D, C_0\}$, the $\mathfrak{so}(d)$ are the usual spatial rotations, and the $\mathfrak{h}_d$ is a $d$-dimensional Heisenberg algebra spanned by boosts and translations with mass as the central element $\{K_i, P_i, M\}_{i=1,\dots, d}$. The non-trivial action of $\mathfrak{so}(d)$ is the obvious rotation action on the $d$-component vectors in $\mathfrak{h}_d$, and the $\mathfrak{sl}(2,\bbR) \cong \mathfrak{sp}(2,\bbR)$ acts  on any fixed triple $\{K_i, P_i, M\} \cong \mathfrak{h}_1$ by canonical transformations of position and momentum.

While generic non-relativistic systems only have Galilean symmetry, Schr\"odinger symmetry emerges in systems with (a form of) conformal symmetry. We caution that the $z=2$ non-relativistic Schr\"odinger CFTs are \textit{not} the CFTs that emerge from taking the non-relativistic $c\to\infty$ limit of the conformal algebra in $(d+1)$-dimensions -- they do not even have the same number of generators. The In\"on\"u-Wigner contraction of the usual conformal algebra gives a $z=1$ ``Galilean conformal algebra'' instead (see e.g. \cite{Bagchi:2009my}).\footnote{In this Galilean conformal algebra, the translations and special conformal transformations (not boosts) commute to a central element $M_{\mathrm{GCA}}$, which is not physically related to the Schr\"odinger $M$. We expect some of our algebraic/kinematic results on the $M=0$ sector to port over to Galilean conformal theories with relatively little difficulty (see also \cite{Chen:2020vvn, Chen:2022jhx}), but with different physical interpretations.\label{footnote:MGCA}}

One can consider generalizations of the Schr\"odinger symmetry mentioned above to theories with arbitrary dynamical exponent $z$. In this case, we have the same generators as the Schr\"odinger algebra, without the special conformal generator $C_0$, and with modified commutation relations:
\begin{equation}
    [D,K_i] = i(1-z)K_i\,,\quad
    [D,M] = i(2-z)M\,.
\end{equation}
We note that $z=2$ scaling is distinguished by admitting a full $SL(2,\bbR)$ symmetry, as opposed to just $z$-scaling, as well as a central element $M$. We will focus on the case with $z=2$ for the bulk of this document.

The usual conformal algebra $\mathfrak{so}(2,d+1)$ can be understood as the algebra of conformal isometries of $\bbR^{1,d}$ or the standard isometries of $\mathrm{AdS}_{1,d+1}$, where scale transformations are geometrized by a radial ``bulk'' coordinate, as in the usual AdS/CFT correspondence. Likewise, the Schr\"odinger algebra $\mathfrak{sch}_d$ can be understood as the conformal isometries of certain non-relativistic spacetimes (discussed in Section \ref{sec:HTGeometry}), or as the isometries of AdS spacetimes with a particular gravitational pp-wave wave profile \cite{Duval:1984cj, Duval:1985cd, Duval:1990hj, Duval:2008jg}. However, a genuine dynamical holographic correspondence is not as clear in these pictures.

\paragraph{Mass and Particle Number.} As mentioned above, the Schr\"odinger algebra is particularly distinguished amongst Lifshitz scaling systems by the central element $M$. This is sometimes also denoted ``$N$'' and interchangeably called ``particle number.'' However, it is more correct to think of it as a mass.

For example, consider the Schr\"odinger field theory describing a free boson of mass $m$ in ($d+1$)-dimensions, with Lagrangian
\begin{equation}\label{eq:NRLag}
    \calL_{0} = \phi^\dagger\left(i\partial_t + \frac{\nabla^2}{2m}\right) \phi\,.
\end{equation}
This theory can also be obtained as a non-relativistic limit of the relativistic free boson. Under a finite Galilean transformation, with rotation $R$, boost by $\vec{v}$, and translation by $\vec{x}$, the free field transforms as \cite{Hagen:1972pd, Bergman:1991hf, Baiguera:2023fus}
\begin{equation}
    \phi(x) \mapsto \phi'(x') = e^{i (\frac{1}{2}m\vec{v}^2 t + m\vec{v}\cdot R\cdot \vec{x})} \phi(x)\,.
\end{equation}
Working out the infinitesimal forms of the generators, one easily shows that
\begin{equation}
    [K_i, P_j] = i \delta_{ij} M \,,
\end{equation}
where the generator $M$ is
\begin{equation}
    M = m \int d^{d}x \, \phi^\dagger \phi = mN\,,
\end{equation}
and $N$ is the particle number. $M$ generates the phase $\phi \mapsto e^{i m \xi} \phi$.

Of course, we are free to rescale our expressions so that everything is in units of particle number. But, when there are many massive fields $\phi_i$, it is important to remember that it is the total mass,
\begin{equation}
    M_{\mathrm{tot}} := \sum_i M_i = \sum_i m_i N_i\,,
\end{equation}
and not total particle number,
\begin{equation}
    N_{\mathrm{tot}} := \sum_{i} N_i\,,
\end{equation}
which is central in the algebra. Thus, $M=0$ states in the harmonic trap are massless states, not $0$-particle states. 

\paragraph{Primaries and Correlation Functions.} Mimicking usual relativistic CFTs, one can introduce the concept of primary operators and study their behaviour inside correlation functions. It is helpful to briefly recall these definitions (in this section, we largely follow \cite{Nishida:2007pj, Goldberger:2008vg}).

As with relativistic CFTs, Schr\"odinger CFTs start with local operators on\footnote{In all our discussion of non-relativistic physics, it is important to remember that $\bbR^{d+1}$ is a non-relativistic spacetime, and does not have e.g. the usual lightcone/causal structure of a Minkowski metric. See Section \ref{sec:HTGeometry}.} $\bbR^{d+1}$. We define a Schr\"odinger primary local operator at the origin $\calO(t=0,\vec{x}=0)$ to satisfy:
\begin{equation}\label{eq:PriConds}
\begin{gathered}
\relax    [D, \calO(0)] = i\Delta \calO(0)\,,\quad
    [M, \calO(0)] = m \calO(0)\,, \\
    [C_0,\calO(0)] = [K_i,\calO(0)] = 0\,.
\end{gathered}
\end{equation}
More generally, at any point, a primary operator transforms as:
\begin{equation}\label{eq:ActionOnOps}
\begin{alignedat}{2}
    [D,\calO(x)]      &= i(2t\partial_t+x^i\partial_i+\Delta)\calO(x) \,,\quad&
    [M,\calO(x)]      &= m \calO(x)\,, \\[6pt]
    [C_0,\calO(x)]    &= -i(t^2\partial_t+tx^i\partial_i+t\Delta+i\tfrac{m}{2}\vec{x}^2)\calO(x)\,,\quad&
    [P_0,\calO(x)]    &= -i\partial_t \calO(x)\,, \\[6pt]
    [K_i,\calO(x)]    &= -i(t\partial_i +i m x_i)\calO(x)\,,\quad&
    [P_i,\calO(x)]    &= i\partial_i \calO(x)\,, \\[6pt]
    \multicolumn{4}{c}{$[M_{ij},\calO(x)] = i(x_i\partial_j-x_j\partial_i)\calO(x) + S_{ij}^R\cdot\calO(x)\,.$}
\end{alignedat}
\end{equation}

As a result of the $z=2$ scaling, spatial translations $P_i$ and boosts $K_i$ respectively change the conformal weight of an operator $\calO$ by $1$, while the timelike translation $P_0$ and SCT $C_0$ changes the conformal weight by $2$, i.e.
\begin{alignat}{3}
    [D,[P_{i},\calO(0)]] 
        &= i (\Delta + 1) [P_i,\calO(0)]\,,\quad
    [D,[K_{i},\calO(0)]] 
        &&= i (\Delta - 1) [K_i,\calO(0)]\,,\\
    [D,[P_{0},\calO(0)]] 
        &= i (\Delta + 2) [P_0,\calO(0)]\,,\quad
    [D,[C_{0},\calO(0)]] 
        &&= i (\Delta - 2) [C_0,\calO(0)]\,.
\end{alignat}
Applying $P_\mu$ to local operators builds up modules of descendants around our lowest weight operator $\calO(0)$, just as in relativistic CFTs. The fact that they are lowest weight modules graded by $D$ implies non-negativity of the spectrum (as we will see in Section \ref{sec:HarmonicTrap}).

However, a key difference with relativistic theories comes from the central element $M$: in the $M=0$ sector, $[K_i, P_j] = 0$. Consequently, any $M=0$ primary operator $\calO$ generates an infinite sea of primary-descendants:
\begin{equation}\label{eq:descendantSea}
    P_0^{k_0}P_{1}^{k_1} \cdots P_d^{k_d} \calO(0)\,.
\end{equation}
Not least for this reason, the literature has largely ignored the $M=0$ sector. We will return to this immediately in Section \ref{sec:PositiveUIRs} when we construct all non-negative energy unitary irreducible representations of the Schr\"odinger algebra in the harmonic trap.

Finally, we wrap up with some facts about general matrix elements. These can be computed in the standard way: by solving the associated Schr\"odinger Ward identities. Assuming $-m_1, m_2 >0$, so that correlation functions do not immediately vanish, the non-trivial Wightman functions take the form \cite{Nishida:2007pj, Goldberger:2014hca}:
\begin{align}
    \mel*{\Omega}{\calO_1(x_1)\!\calO_2^\dagger(x_2)}{\Omega} 
        &= \delta_{\calO_1\!\calO_2} \frac{c}{(t_{12}-i\epsilon_{12})^{\Delta_1}} e^{i\frac{m_2\vec{x}_{12}^2}{2(t_{12}-i\epsilon_{12})}}\,,\label{eq:2ptcorrelator}\\
    \mel*{\Omega}{\calO_1(x_1)\calO_2(x_2)\!\calO_3^\dagger(x_3)}{\Omega} 
        &= \frac{F(v_{123}) \,\,e^{i\frac{m_1 \vec{x}_{13}^2}{2t_{13}}+i\frac{m_2 \vec{x}_{23}^2}{2t_{23}}}}{t_{12}^{(\Delta_{1} + \Delta_{2}-\Delta_{3})/2}t_{23}^{(\Delta_{2} + \Delta_{3}-\Delta_{1})/2}t_{13}^{(\Delta_{1} + \Delta_{3}-\Delta_{2})/2}}\label{eq:3ptcorrelator}\,.
\end{align}
Here, $\calO_2$ can be any operator with $m_1 + m_2 + m_3 = 0$ and we have suppressed $i\epsilon$'s in the three point function. As we will show later, the naive $m \to 0$ limits of these correlation functions do describe correlation functions of $m=0$ operators.

A few points are in order. First, as with usual QFT, $i\epsilon$ prescriptions for the operators should be taken so that operators are ordered correctly in imaginary/Euclidean time; we revisit what is meant by Euclidean time in Section \ref{sec:HTGeometry} and discuss the analyticity of \eqref{eq:2ptcorrelator} briefly in Section \ref{sec:OPE}. Second, given the different scaling between space and time, we find it useful to define the dimensionless ratios $z := \vec{x}^2/t$, $z_{ij} := \vec{x}_{ij}^2/t_{ij}$, etc. Relatedly, we already find an unknown functional dependence on the new Schr\"odinger conformal cross-ratio:
\begin{equation}
    v_{ijk} := \frac{1}{2}(z_{jk}+z_{ij}-z_{ik})\,,
\end{equation}
in the three-point function \eqref{eq:3ptcorrelator}. Thus, the three-point functions are not determined by structure constants for the Schr\"odinger CFT but by ``structure functions'' for generic primaries, and the OPE changes accordingly. For $M=0$ operators simplifications occur, as we discuss in Section \ref{sec:M0WardIdentities}.

\subsection{Example: Non-Renormalization and Fermions at Unitarity}\label{sec:FermionsAtUnitarity}
We end our review with the prototypical example of a Schr\"odinger CFT: fermions at unitarity \cite{Nishida:2007pj} (see also \cite{LeggettOriginal, Nishida:2006br, Nishida:2006eu,  Nishida:2010tm}). This success of the example supports the theoretical claim that the harmonic trap spectrum should \textit{define} Schr\"odinger CFTs. We also use the example to demonstrate the perturbative non-renormalization theorem for non-relativistic field theories \cite{Bergman:1991hf, Klose:2006dd, Auzzi:2019kdd, Chapman:2020vtn}.

Our goal is to understand an interacting fixed point of non-relativistic fermions in the unitarity limit, given by the four-fermi Lagrangian in ($d+1$)-dimensions
\begin{equation}\label{eq:fourfermi}
    \calL = i \psi^\dagger_{\sigma} \partial_t \psi_{\sigma}
    - \frac{1}{2} |\nabla \psi_{\sigma}|^2 + \mu_{\sigma} |\psi_{\sigma}|^2 - \frac{c_0}{2}\psi^\dagger_{\uparrow} \psi^\dagger_{\downarrow}\psi_{\downarrow} \psi_{\uparrow}\,,
\end{equation}
where $\psi_\sigma$ are spin-1/2 fermions of dimension $\Delta_\psi = d/2$. We will assume $\mu_{\sigma} = 0$, i.e. there is no chemical potential, but we will revisit this again in Section \ref{sec:CouplingTo1d}. In the unitarity limit, one tunes $c_0$ so that the $s$-wave scattering amplitude saturates the optical theorem unitarity bound by itself and the $s$-wave scattering length diverges $a_s \to \infty$, erasing microscopic length scales. The resulting system is described by a Schr\"odinger CFT. This strongly interacting universality class governs ultracold atomic clouds, dilute neutron matter, and other scale‑invariant Fermi systems held in harmonic traps.

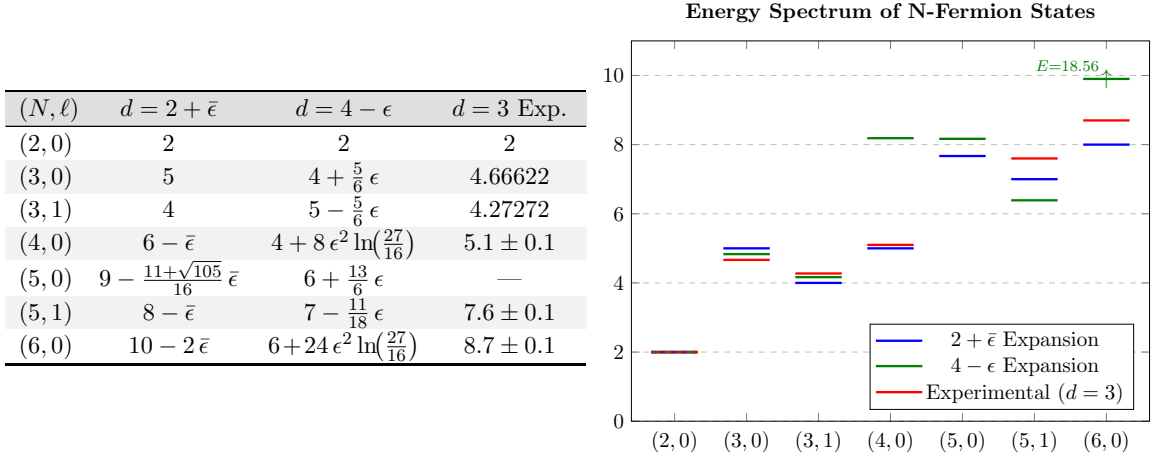
\begin{figure}[t]
\centering
\begin{minipage}{0.45\textwidth}
\centering
\resizebox{1.15\textwidth}{!}{%
{
\rowcolors{2}{gray!10}{white}
\begin{tabular}{
    >{\centering\arraybackslash}p{0.8cm} 
    >{\centering\arraybackslash}p{2.3cm} 
    >{\centering\arraybackslash}p{2.4cm} 
    >{\centering\arraybackslash}p{2.0cm}
}
\specialrule{1.2pt}{0pt}{0pt}
\rowcolor{gray!25}
\((N,\ell)\) & \(d=2+\bar{\epsilon}\) & \(d=4-\epsilon\) & \(d=3\) Exp. \\
\specialrule{0.8pt}{0pt}{0pt}
\((2,0)\) & \(2\) & \(2\) & \(2\) \\
\((3,0)\) & \(5\) & \(4 + \tfrac{5}{6}\,\epsilon\) & \(4.66622\) \\
\((3,1)\) & \(4\) & \(5 - \tfrac{5}{6}\,\epsilon\) & \(4.27272\) \\
\((4,0)\) & \(6 - \bar{\epsilon}\) & \(4 + 8\,\epsilon^2 \ln\!\bigl(\tfrac{27}{16}\bigr)\) & \(5.1\pm0.1\) \\
\((5,0)\) & \(9 - \tfrac{11 + \sqrt{105}}{16}\,\bar{\epsilon}\) & \(6 + \tfrac{13}{6}\,\epsilon\) & --- \\
\((5,1)\) & \(8 - \bar{\epsilon}\) & \(7 - \tfrac{11}{18}\,\epsilon\) & \(7.6\pm0.1\) \\
\((6,0)\) & \(10 - 2\,\bar{\epsilon}\) & \(6 + 24\,\epsilon^2 \ln\!\bigl(\tfrac{27}{16}\bigr)\) & \(8.7\pm0.1\) \\
\specialrule{1.2pt}{0pt}{0pt}
\end{tabular}
}
}
\end{minipage}
\hfill
\begin{minipage}{0.48\textwidth}
\centering
\resizebox{\textwidth}{!}{%
\begin{tikzpicture}
\begin{axis}[
    width=11cm, height=8.5cm,
    title={\textbf{Energy Spectrum of N-Fermion States}},
    symbolic x coords={N2L0, N3L0, N3L1, N4L0, N5L0, N5L1, N6L0},
    xticklabels={{$(2,0)$},{$(3,0)$},{$(3,1)$},{$(4,0)$},{$(5,0)$},{$(5,1)$},{$(6,0)$}},
    xtick=data,
    xticklabel style={anchor=south, yshift=-7mm},
    ymin=0, ymax=11,
    ymajorgrids=true,
    grid style=dashed,
    legend pos=south east,
    ytick={0,2,4,6,8,10},
    yticklabels={0,2,4,6,8,10}
]
\addplot[blue, thick, only marks, mark=|, mark options={rotate=90, scale=6, line width=1.2pt}] 
coordinates {(N2L0,2.0) (N3L0,5.0) (N3L1,4.0) (N4L0,5.0) (N5L0,7.67) (N5L1,7.0) (N6L0,8.0)};
\addplot[green!50!black, thick, only marks, mark=|, mark options={rotate=90, scale=6, line width=1.2pt}] 
coordinates {(N2L0,2.0) (N3L0,4.833) (N3L1,4.167) (N4L0,8.186) (N5L0,8.167) (N5L1,6.389) (N6L0,9.9)};
\node[above left, green!50!black] at (axis cs:N6L0,9.9) {$\scriptstyle E=18.56$};
\node[green!50!black] at (axis cs:N6L0,9.9) {$\uparrow$};
\addplot[red, thick, only marks, mark=|, mark options={rotate=90, scale=6, line width=1.2pt}] 
coordinates {(N2L0,2.0) (N3L0,4.66622) (N3L1,4.27272) (N4L0,5.1) (N5L1,7.6) (N6L0,8.7)};
\addplot[blue, very thick, only marks, mark=|, mark options={rotate=90, scale=6, line width=1.2pt,dash pattern=on 3pt off 2pt}] coordinates {(N2L0,2.0)};
\addplot[green!50!black, very thick, only marks, mark=|, mark options={rotate=90, scale=6, line width=1.2pt,dash pattern=on 2pt off 1pt}] coordinates {(N2L0,2.0)};
\addplot[red, very thick, only marks, mark=|, mark options={rotate=90, scale=6, line width=1.2pt,dash pattern=on 1pt off 1pt}] coordinates {(N2L0,2.0)};
\legend{$2+\bar{\epsilon}$ Expansion,$4-\epsilon$ Expansion,Experimental ($d=3$)}
\end{axis}
\end{tikzpicture}
}
\end{minipage}
\caption{%
Left, the scaling dimensions of $N$-fermion states with orbital angular momentum $\ell$ at one-loop order in the $\bar{\epsilon}$ and $\epsilon$ expansions. Table reproduced from \cite{Nishida:2007pj}; we have provided an independent check of the composite entries up to $N=3$. Right, the energy spectrum of the operators in the harmonic trap, with $\omega = 1$ and $\epsilon = \bar{\epsilon} = 1$. We see the Schr\"odinger CFT matches the harmonic trap with great success at leading order.
}
\label{fig:EnergyTable}
\end{figure}

We would like to understand this system in $d=3$ spatial dimensions. To do this, we study the weakly coupled fixed points in $d = 4 - \epsilon$ and $d = 2 + \bar{\epsilon}$ dimensions and then extrapolate to $d=3$. The importance of these limits was also explained in \cite{nussinov2004bcs, nussinov2006triviality, Nishida:2010tm}. The scaling dimensions of operators $\calO$ in this flat space theory can then be matched to the energy spectrum of the theory in a harmonic potential, using the relation
\begin{equation}\label{eq:HTEnergy}
    H_{\HT}\ket{\psi} = \omega \Delta_{\psi}\ket{\psi}\quad \text{for any}\quad \ket{\psi} = \psi^\dagger(i/\omega) \ket{\Omega}\,,
\end{equation}
where $\omega$ is some tunable trap frequency and $\ket{\Omega}$ is the harmonic trap ground state $H_{\HT}\ket{\Omega} = 0$. This is sometimes called a state-operator correspondence for non-relativistic CFTs because it relates the energy of a state in the harmonic trap to the scaling dimensions of a local operator in the plane. We will investigate it more carefully in Sections \ref{sec:HTGeometry} and \ref{sec:HarmonicTrap}; for now, we just take it as a fact. Using this correspondence, this two‑sided approximation of $d=3$ energy levels already reproduces the first few multi‑particle energy levels to within a few percent error, see Figure \ref{fig:EnergyTable}.

An important ingredient in these calculations follows from the explicit splitting of fields in non-relativistic systems into creation/daggered and annihilation/undaggered operators. With Lagrangian descriptions like \eqref{eq:fourfermi}, this follows immediately from the classical definition of the theory. \textit{In perturbation theory, this splitting implies that all anomalous dimensions are acquired independently in the daggered and undaggered sectors}, e.g.
\begin{equation}\label{eq:ExampleNonRenorm}
    \Delta_{(\psi^{\dagger})^{10}\psi^6} = \Delta_{(\psi^\dagger)^{10}} + \Delta_{\psi^6}\,.
\end{equation}
We call this the non-relativistic non-renormalization theorem. The argument is simple: since particle number $N$ is conserved,\footnote{As previously mentioned, it is actually mass $M$ that is conserved, so we can already anticipate the failure of the non-renormalization theorems.} time-ordered two-point functions carry Heaviside theta functions in time, and/or momentum space propagators have only a single pole in momentum space, i.e.,
\begin{equation}
    D_\psi(k) \propto \frac{1}{\omega - \vec{k}^2 + i \epsilon}\,.
\end{equation}
Thus, directionally closed loop diagrams vanish because residues from opposite-moving lines cancel.\footnote{In null reductions, this is identical to lightcone or ultraboosted non-renormalization theorems \cite{Weinberg:1966jm}.} Pictorially we have
\begin{equation}\label{eq:pertNRdiag}
\begin{tikzpicture}[baseline=-0.5ex, scale=0.6]
  \draw[->-=0.54rotate0, thick] (0,0) .. controls (1,1) and (3,1) .. (4,0);
  \draw[->-=0.46rotate180, thick] (4,0) .. controls (3,-1) and (1,-1) .. (0,0);
\end{tikzpicture}
\;\neq\; 0\,,
\qquad
\begin{tikzpicture}[baseline=-0.5ex, scale=0.6]
  \draw[->-=0.54rotate0, thick] (0,0) .. controls (1,1) and (3,1) .. (4,0);
  \draw[->-=0.54rotate0, thick] (4,0) .. controls (3,-1) and (1,-1) .. (0,0);
\end{tikzpicture}
\;=\; 0\,,
\qquad
\begin{tikzpicture}[baseline=-0.5ex, scale=0.6]
  \draw[->-=0.46rotate180, thick] (0,0) .. controls (1,1) and (3,1) .. (4,0);
  \draw[->-=0.54rotate0, thick] (4,0) .. controls (3,-1) and (1,-1) .. (0,0);
\end{tikzpicture}
\;\neq\; 0\,.
\end{equation}
These arguments can be modified in non-$N$-invariant states. Now let us see explicitly how the energy-dimension correspondence in \eqref{eq:HTEnergy} and this non-renormalization theorem apply in perturbation theory.

\paragraph{$\bm{d = 4 - \epsilon}$:}
In $d=4$ spatial dimensions, the four-fermi interaction is marginal, and it is helpful to make a Hubbard-Stratanovich transform of \eqref{eq:fourfermi} to:
\begin{equation}
\mathcal{L} =
    i \psi^\dagger_{\sigma} \partial_t \psi_{\sigma}
    - \frac{1}{2} |\nabla \psi_{\sigma}|^2
    + i \phi^* \partial_t \phi
    - \frac{1}{4} |\nabla \phi|^2 + g (\psi^\dagger_{\uparrow} \psi^\dagger_{\downarrow} \phi
    + h.c.)
    - \frac{g^2}{c_0} |\phi|^2\,.
\end{equation}
The bosonic fields are interpreted as dimer or molecular fields, with only relevant Yukawa-type interaction vertices. We note a simple but important point: the Yukawa interactions are $\psi^\dagger\psi^\dagger \phi$ and $\psi\psi\phi^*$, i.e. $\phi$ is annihilated for two $\psi$'s or vice versa, and we must be careful about the orientation of lines and loops.

Now we wish to understand effects of renormalization in $4-\epsilon$ dimensions. Working in perturbation theory, we would expect divergences to be absorbed into the renormalization of $Z_\psi$, $Z_\phi$, $Z_g g$, and $c$. However, at one loop order, we are not able to draw any diagrams correcting $Z_\psi$ and $Z_g$ leaving only $Z_\phi$ and $c$. 

The unitarity limit/criticality is obtained by demanding $\phi$ to be gapless, so we take the renormalization condition
\begin{equation}
    0 \stackrel{!}{=} -\frac{1}{c} 
        = -\frac{1}{c_0}
         +\int \frac{d^4 k}{(2\pi)^4}\,\frac{1}{k^{2}}\,.
\end{equation}
From here, a straightforward 1-loop calculation of $\langle\phi \phi \rangle(p)$ over the momentum shell $e^{-s} \Lambda < k < \Lambda$ leads to $Z_{\phi} = 1 - \frac{g^2}{8\pi^2} s$. The resulting beta function $\beta(g)$ and anomalous dimension $\gamma_\phi$ are
\begin{equation}
    \beta(g) := \frac{\partial g}{\partial s} = \left(2 - \tfrac{d}{2} - \gamma_\phi\right)g = \frac{\epsilon}{2} g - \frac{g^3}{16 \pi^2} + O(g^4)\,,\quad
    \gamma_\phi(g) = - \frac{1}{2} \frac{\partial \ln{Z_{\phi}}}{\partial s} = \frac{g^2}{16\pi^2}\,,
\end{equation}
and, at the fixed point,
\begin{equation}
    g_{*}^2 = 8\pi^2 \epsilon\,,\quad
    \gamma_{\phi} = \frac{\epsilon}{2}\,.
\end{equation}

At the fixed point, we can use the energy-dimension relation \eqref{eq:HTEnergy} to find $N$-fermion states in the harmonic trap. For example, a 1-fermion state \(\psi\) is not renormalized and \( \Delta_{\psi} = \frac{d}{2} \ \Rightarrow \ E_{\psi} = (2 - \frac{\epsilon}{2}) \omega \). The composite dimer field \(\phi\) gives us a 2-fermion bound state with spin \(\ell=0\) and dimension $\Delta_{\phi} = \frac{d}{2} + \gamma_{\phi} = 2$, exact to all orders in \(\epsilon\). 

From 3‑fermion states onward, the associated operator is a composite of the Lagrangian field and ordinary action counterterms cannot cancel short-distance divergences on their own, requiring additional work whose anomalous dimensions we now quote. The simplest 3-fermion operator is \(\phi\psi_{\uparrow}\) with $\Delta_{\phi \psi_{\uparrow}} = \Delta_{\psi} + \Delta_{\phi} + \frac{5}{6} \epsilon$. However, the 3-fermion ground state in $d=3$ is experimentally known to have spin-1. If we write a general linear combination of 3-fermion spin-1 operators, \(\alpha \ \psi\nabla\phi + \beta \ \phi\nabla\psi\), and demand that it is a NR CFT primary using \eqref{eq:PriConds}, this leads to:
\begin{equation}
    \calO_{(3,1)} := 2 \phi\nabla\psi_{\uparrow} - \psi_{\uparrow}\nabla \phi\,,
    \quad \Delta_{(3,1)} = \Delta_{\psi} + \Delta_{\phi} + 1 - \frac{\epsilon}{3}\,,
\end{equation}
which is indeed a ground state! We summarize the results of the calculations and depict the HT spectrum in Figure \ref{fig:EnergyTable}.

\paragraph{$\bm{d = 2+ \bar{\epsilon}}$:} In $d=2+\bar{\epsilon}$ dimensions, the four-fermi interaction is weakly relevant and we do not need to introduce the dimer field $\phi$. Thus, the Lagrangian is just \eqref{eq:fourfermi} with zero chemical potential again $\mu_\sigma = 0$. Repeating the previous perturbative analysis, one finds a fixed point at $g_{*}^2 = 2\pi\bar{\epsilon}$ and 1 and 2 fermion states are identical to the $4-\epsilon$ case. 

The results differ for the simplest 3 fermion operator. Since we only have fermions, the Pauli exclusion principle suggests that the simplest spin-0 operator is \(\psi_\uparrow \psi_\downarrow \partial_t \psi_\uparrow\) of dimension $3\Delta_{\psi} + 2$, while the simplest spin-1 operator is $\psi_{\uparrow}\psi_{\downarrow}\partial_i\psi_{\uparrow}$ of dimension $3\Delta_{\psi} + 1$. Note that $\partial_t$ derivatives do not increase spin since we do not have boosts mixing space and time.

\paragraph{Normal Ordered Composites.} In both $d=4-\epsilon$ and $d=2+\bar{\epsilon}$, we see from \eqref{eq:pertNRdiag} that a normal ordered composite operator, like the number density $n_{\psi} = :\!\psi_{\sigma}^\dagger \psi_{\sigma}\!:$, will have dimension $d$ to all orders in perturbation theory, but will be invisible to the harmonic trap spectrum by \eqref{eq:HTEnergy}. Despite this, it still corresponds to a good observable in non-trivial states, reflecting the ability for these bound states to have interesting dynamics with other particles in the harmonic trap. Moreover, it also corresponds to an important physical deformation: turning back on the chemical potential $\mu_\sigma$ in \eqref{eq:fourfermi}!

\subsection{Technical Goals, Summary, and Future Directions}\label{sec:Outline}
The previous discussions reveal a number of obvious holes in our understanding of non-relativistic CFTs:
\begin{itemize}\setlength\itemsep{0em}
    \item In \eqref{eq:PriConds} we give an algebraic definition of local primary operators. Trying to construct conformal families as lowest-weight modules in \eqref{eq:descendantSea}, we found an infinite family of primary descendants when $M=0$ because $[K_i,P_j] =i \delta_{ij}M$, and we were forced to stop. Can we complete the construction of $M=0$ operators?
    \item In \eqref{eq:HTEnergy} we claim that creation operators $\calO^\dagger(0)$ can act on the HT vacuum state $\ket{\Omega}$ to produce an eigenstate of the HT Hamiltonian with energy $\omega \Delta_{\calO}$. This is sometimes called the $M\neq 0$ state-operator correspondence for Schr\"odinger CFTs. To what extent is there actually a one-to-one correspondence between states and operators? And is the assumption $M \neq 0$ important?
    \item We use the term ``creation operator'' as if it is canonical. Can we define this in a general theory without a Lagrangian or non-relativistic limit? What properties do such objects have?
    \item Even with an alleged state-operator correspondence for creation operators (and annihilation operators by Hermitian conjugation)
    \begin{equation}
        \ket{\psi} \stackrel{?}{\longleftrightarrow} \psi^\dagger(0)\,,
    \end{equation}
    important operators like the number density $n(x) = :\!\!\phi^\dagger\phi\!\!:\!\!(x)$ clearly annihilate $\bra{\Omega}$ and $\ket{\Omega}$ and so have no dual bra or ket. Note: this issue is completely unrelated to the fact that $M=0$, and happens even for e.g. $:\!\!(\phi^\dagger)^3\phi^7\!\!:$ as well.
    \item Important operators like $n(x)$ above are defined as a ``normal-ordered'' composite. How do we define this in a general theory? It is presumably very scheme dependent. 
    \item In a Lagrangian theory, anomalous dimensions of composite operators built from both daggered and undaggered operators were argued to renormalize and add separately in \eqref{eq:ExampleNonRenorm}. This follows from diagrammatics in \eqref{eq:pertNRdiag}. Is this true in general?
    \item In deforming theories, the interaction terms we add to a Lagrangian are generally going to be $M=0$. How do we make sense of perturbation theory if we cannot understand such operators even at fixed points?
\end{itemize}

\paragraph{Outline and Summary of Paper.}
In the remainder of the paper, we address the above problems in a systematic way, resolving many of the conceptual problems in the introduction. Along the way, we are led to a number of neat mathematical structures reminiscent of Lorentzian CFTs (and logarithmic CFTs, see \cite{Pyramids}), and shedding light on null reductions and holographic interpretations for future investigations. The outline and summary of the subsequent sections are as follows:

\begin{enumerate}\setlength\itemsep{0em}
    \item[{\hyperref[sec:HTGeometry]{$\S$.2.}}] In Section \ref{sec:HTGeometry} we describe the geometry of non-relativistic CFTs and the existence of a state-operator correspondence. In Section \ref{sec:NRGeometry} we review non-relativistic geometry and the conformal isometries of the non-relativistic plane. In Lifshitz and Schr\"odinger systems, Weyl transforms of spacetime are completely controlled by their action on time, with spatial directions largely behaving as spectators, leading to analogies with 1d and defect CFTs. In Section \ref{sec:ImportantFacts} we briefly review the relation between Wick rotation and different quantization schemes in relativistic CFTs, and the emergence of thermofield doubles in 1d CFTs.
    \vskip -0.0cm
    Using this, we develop the ``state picture'' for Schr\"odinger CFTs in Section \ref{sec:HTG}. We define the \textit{harmonic trap spacetime}
    \begin{equation}
    M_{\HT} := (\bbR \,\cup\, \{\infty\}) \times S^0 \times \bbR^{d}\,,
    \end{equation}
    whose time translations are naturally generated by the harmonic trap Hamiltonian
    \begin{equation}
        H_{\HT} := P_0 + \omega^2 C_0\,.
    \end{equation}
    The $S^0$ factor means $M_{\HT}$ has two disconnected ``branches,'' interpreted as a thermofield double for flat spacetime, and the Hilbert space naturally factorizes as
    \begin{equation}
        \calH_{\mathrm{TFD}} = \mathcal{H}^* \otimes \mathcal{H}\,.
    \end{equation}
    This gives a geometric origin for creation and annihilation operators in Schr\"odinger CFTs. The conventional harmonic trap vacuum state $\ket{\Omega}$ is obtained by tracing over the thermofield double state in the $\beta \to \infty$ limit. In Section \ref{sec:NRSOC} we use these results to derive a correspondence between operators and states on glued squashed-hemispheres (or ``lemons'') so that: \textit{local operators are dual to states in a thermofield double spacetime}.
    \vskip 0.25cm
    \item[{\hyperref[sec:HarmonicTrap]{$\S$.3.}}]  In Section \ref{sec:HarmonicTrap} we consider the operator algebra and representation theory of Schr\"odinger CFTs. In Section \ref{eq:RaisingLowering} we discuss the superselection structure of $\calH$ and prove the spectrum $H_{\HT}$ organizes into lowest weight representations labelled by dilatation eigenvalues. Then, in Section \ref{sec:GenuinePrimaries}, we describe the natural polarization of the HT spectrum and use it to identify creation operators $\calO^{\dagger}$ and annihilation operators $\calO$ dual to states, called ``genuine'' operators. Operators which annihilate the harmonic trap vacuum are called ``non-genuine.''
    \vskip -0.0cm
    In Section \ref{sec:PositiveUIRs}, we use techniques from representation theory to classify non-negative energy representations of the Schr\"odinger group and, thus, potential multiplet structures in a Schr\"odinger CFT, as well as their associated unitarity bounds. Given a state labelled by scaling $\Delta$, spin $\rho$, and mass $m$, unitarity demands that
    \begin{equation}
        \Delta_{\text{Massive}} \geq \frac{d}{2}\,,
        \quad
        \Delta_{\text{Massless}} \geq 0\,,
        \quad
        m \geq 0\,.
    \end{equation}
    We confirm these results by method of induction in Section \ref{sec:GalileanParticles}. In Section \ref{sec:M0WardIdentities} we discuss the Ward identities of operators with zero mass and find that they are indeed restricted to behave like 1d CFT correlation functions. Crucially, we find that: \textit{genuine $M=0$ operators can exist, are completely spatially topological, and behave like a 1d CFT in time}. 
    \vskip 0.25cm
    \item[{\hyperref[sec:NonRenorm]{$\S$.4.}}] In Section \ref{sec:NonRenorm} we discuss the consequences of our new TFD perspective on creation and annihilation operators and massless states. In Section \ref{sec:OPE}, we briefly discuss the OPE and the analytic structure of correlation functions. Then, using our new technology, in Section \ref{sec:NRTheorem} we give a non-perturbative argument for non-renormalization between creation and annihilation operators at fixed points, when no non-trivial massless states are present in the theory. Specifically, we argue that if $\calO_1^\dagger$ and $\calO_2$ are operators in the theory, there exists a canonically defined ``normal ordered composite operator'' with additive scaling dimension
    \begin{equation}
        (\calO_1^{\dagger}\!\calO_2)(x)\,,\quad 
        \Delta = \Delta_1 + \Delta_2\,.
    \end{equation}
    We then show how any $M$-preserving deformation/RG flow necessarily also has this non-renormalization property.
    \vskip -0.0cm
    In Section \ref{sec:GenuineMassless} we argue that theories with non-trivial massless states should exist, may generally be non-Lagrangian, and give some examples. In particular, we claim that the null reduction of the (3+1)d free scalar contains genuine massless states in its spectrum. Then, in Section \ref{sec:CouplingTo1d}, we show that abstract massless theories can -- in principle -- be consistently coupled to a non-relativistic CFT in conformal perturbation theory while maintaining conformality.
\end{enumerate}

\paragraph{Future Directions.} In this paper we revisit the fundamentals of non-relativistic Schr\"odinger CFTs, giving formal arguments for some lore in the non-relativistic cannon, and re-contextualizing and reinterpreting a number of earlier results. There are many important and interesting conceptual issues that we do not address, which we mention throughout the paper. Here we underscore four directions which we anticipate are more tractable in light of the results of our paper:
\begin{itemize}\setlength\itemsep{0em}
    \item \textbf{Higher Conservation Laws.} Since $[K_i, P_j] = \delta_{ij} M$, composite massless operators in Schr\"odinger CFTs possess an interesting ``pyramidal'' module structure as $K_i$ and $P_j$ commute. For example, consider the number density operator $n := (\psi^\dagger\psi)$ in the free fermion theory. Algebraically, $n$ is a primary operator and generates a tower of descendants by applying derivatives $\partial_\mu$. However, there are also a number of ``alien operators,'' like the probability current $J_i$ and stress tensor $T_{ij}$, which are neither primary nor descendant but intertwine with the $n$-multiplet because they descend to $n$ under the lowering operation
    \begin{equation}
        [K_i,J_i] = n\,, \quad
        [K_i, T_{ij}] = J_j\,,\quad \text{etc.}
    \end{equation}
    This leads to a number of rich algebraic properties of operators in the neutral sector, as well as a number of powerful interlocking conservation laws \cite{Bekaert:2011qd, Golkar:2014mwa}. It is also worth noting that these ``alien operators,'' like $J_i$ and $T_{ij}$ have some formal similarity to double trace/twist operators in relativistic CFTs. We comment on this further in \cite{Pyramids}.
    \item \textbf{RG Monotonicity Theorems.} Some of the most important quantities for organizing the space of QFTs are monotonic functions which decrease along RG flows. Physically, RG theorems should still exist, and a proof of some monotonicity theorems would be extremely important for the development of non-relativistic CFTs. We note that trace anomalies have been classified \cite{Auzzi:2015fgg, Auzzi:2016lxb, Pal:2016rpz, Auzzi:2016lrq, Auzzi:2017jry} (see also \cite{Baiguera:2023fus} for more details), but the connection to monotonicity is not clear. We also note that in any dimension the Schr\"odinger algebra admits an extension to a Schr\"odinger-Virasoro algebra $\mathfrak{sv}_c$, however, the unitary representations appear to be so strongly constrained that it effectively forces $M=0$ and reduces to Virasoro representation theory \cite{Roger:2006rz, li2008representations, unterberger2011schrodinger, zhang2013unitary}. Finally, there are also tensions with the existence of limit cycles \cite{Nishida:2007de} and entanglement entropy arguments \cite{Mintchev:2022xqh}.
    \item \textbf{Holographic Interpretations.} Holographic duals of Schr\"odinger CFTs and more general Lifshitz systems have been studied in great depth. However, many studies have been at the level of kinematics, e.g. matching symmetries to constrain correlation functions \cite{Son:2008ye, Goldberger:2008vg, Balasubramanian:2008dm, Barbon:2008bg, Maldacena:2008wh, Kachru:2008yh, Herzog:2008wg, Adams:2009dm}. An important step would be to explicitly compute bulk/boundary dynamics in a particular null reduction of a CFT e.g. $\calN=4$ SYM. Moreover, in this work, we propose a strong analogy with 1d CFT -- and thus $\mathrm{AdS}_2$ -- in any dimension, via our thermofield double. It would be interesting to revisit Schr\"odinger holography in light of these claims.
    \item \textbf{The Conformal Bootstrap.} As mentioned many times in the introduction, given the physical and theoretical importance of Schr\"odinger CFTs, they make an interesting target for the conformal bootstrap. On the numerical side, it would be interesting to bootstrap the structure functions $C_{ijk}(z)$. Despite generic issues with positivity for three operators, the extra kinematic constraints on the $M=0$ sector could make some particular OPEs more tractable. On the analytic side, there also exists a number of analogies to light ray operators and double twist trajectories which warrants further attention.
\end{itemize}

\section{The Harmonic Trap and State-Operator Correspondence}\label{sec:HTGeometry}
We gave the conditions for a local operator in a Schr\"odinger CFT to be a primary or descendant in \eqref{eq:PriConds}. We also claimed that the scaling dimension of Schr\"odinger primary local operators matched the energy level of states in a harmonic trap, as in \eqref{eq:HTEnergy}. This matching of energy levels to scaling dimensions of primaries is sometimes called a state-operator correspondence, whence questions of OPE convergence can also be subsequently considered. In this section we will discuss the geometric validity of a full state-operator correspondence for non-relativistic CFTs.

The relation between harmonic trap energies and scaling dimensions in the plane suggests an analogue of the conformal cylinder in relativistic CFTs. That is, we expect there to be some harmonic trap geometry $M_{\HT}$ related to flat space by a Schr\"odinger-Weyl transform, and for states on $M_{\HT}$ to be related to local operators in non-relativistic $\bbR^{d+1}$. Moreover, in this geometry, harmonic trap time translations $\tau$ should be generated by $H_{\HT}$ and be related to the flat space generators by
\begin{equation}\label{eq:HarmonicTrap}
    H_{\HT} := P_0 + \omega^2 C_0\,.
\end{equation}
Indeed, the Hamiltonian \eqref{eq:HarmonicTrap} is exactly the analogue of the Luscher-Mack conformal Hamiltonian in Lorentzian CFT \cite{Luscher:1974ez, Mack:1975je, Mack:1976pa}. In the same way that placing a nice (compact) relativistic CFT on the cylinder $\bbR \times S^d$ gives a gapped spectrum, we have a similar expectation for $M_{\HT}$. In other words, compact Schr\"odinger CFTs are, definitionally, those with a discrete spectrum on $M_{\HT}$.\footnote{The name ``harmonic trap'' refers to the fact that $C_0$ effectively adds $\vec{x}^2 \calO^\dagger\!\calO$ terms to the Hamiltonian density in Lagrangian theories, describing bound states in a quadratic potential. This can be confirmed explicitly by writing $C_0$ in Section \ref{sec:FermionsAtUnitarity}. This sometimes also goes by the name ``oscillator frame.''}\footnote{The harmonic trap can also be understood, in part, from certain non-relativistic limits of AdS spacetime, see e.g. \cite{Craps:2017eyp, Bizon:2018frv, Evnin:2021buq, Maxfield:2022hkd}.}

The fact that a harmonic trap geometry is related in some way to the cylinder is not new \cite{Nishida:2007pj, Goldberger:2014hca, Baiguera:2024vlj}, but we will see that some extra care in its treatment reveals a number of extremely useful features and important questions (and thus why we delay our presentation of the definition of $M_{\HT}$). In particular, in Section \ref{sec:NRGeometry} we give a lightning discussion of non-relativistic geometry and use it to argue for the ``uniqueness'' of the HT geometry. We then recount some important facts from the standard relativistic CFTs in Section \ref{sec:ImportantFacts}, where the relationships between states, local-operators, and analytic continuation are very clear. We also recall the famous fact that 1d CFTs naturally have a ``doubled'' state-operator correspondence (highlighted especially in \cite{Sen:2011cn}). In Section \ref{sec:HTG} we mimic the standard discussion (with more detail) for non-relativistic CFTs, and apply it to a state-operator correspondence in Section \ref{sec:NRSOC}, demonstrating that:
\begin{itemize}\setlength\itemsep{0em}
    \item The energy of states in the harmonic trap are related to scaling dimensions of local operators -- even for non-primaries and all masses.
    \item There is a natural emergent thermofield double geometry, implying a generic splitting of \textit{any} Schr\"odinger CFT into ``daggered'' and ``undaggered'' operators (with exceptions explained in the next section).
    \item There exists a state-operator correspondence for non-relativistic CFTs, that applies to all operators of any mass and charge.
\end{itemize}
Armed with a state-operator correspondence for any mass and charge, this essentially implies the existence of a convergent OPE expansion for all operators.

\subsection{Non-Relativistic Geometry and Conformal Transformations}\label{sec:NRGeometry}
If a state-operator correspondence exists for Schr\"odinger CFTs, we would like to know: To which operators does it apply? What do we mean by Lorentzian and Euclidean signature? And how does it relate to the natural polarization (dagger/undagger splitting) of observables in Lagrangian examples? To understand some of these questions, it's worth a minor detour into non-relativistic (aka ``Newton-Cartan'') geometry, see e.g. \cite{Hartong:2022lsy, Baiguera:2023fus}.

Our non-relativistic spacetime is specified by three pieces of data: a smooth manifold $M$; a two-component contravariant tensor $h$; and a closed ``clock'' 1-form $c$, generating the kernel of $h$. We are primarily interested in flat non-relativistic spacetime $M = \bbR^{d+1}$, with local coordinates $\{t,x^i\}$, equipped with clock 1-form $c_{\mu}$ and inverse spatial metric $h^{\mu\nu}$, where
\begin{equation}
    c = dt\,,\quad h = \delta^{ij} \partial_i \partial_j\,.
\end{equation}

Oftentimes, one will supplement these intrinsic data $(M, c, h)$ with additional information, such as an inverse ``velocity'' vector $v^\mu$, satisfying $c_\mu v^\mu = \pm 1$, or an actual spatial metric $h_{\mu\nu}$. One advantage of this is that it enables the construction of a standard invertible metric $g_{\mu\nu} = c_{\mu}c_{\nu} + h_{\mu\nu}$ on the spacetime \cite{Baiguera:2023fus}. However, even with a seemingly natural choice of spatial metric, like $h_{ij} = \delta_{ij} dx^i dx^j$ in flat space, the combination of a clock 1-form and spatial metric into an invertible metric is still arbitrary. For example, in flat space
\begin{equation}
    g_{\mu\nu} = \pm c_{\mu}c_{\nu} + h_{\mu\nu} = \mathrm{diag}(\pm1,1,1,\dots,1)\,,
\end{equation}
and so, the traditional concept of a ``Lorentzian'' or ``Euclidean'' structure on non-relativistic spacetime is completely arbitrary. We can, however, still make sense of analytic continuation of time and/or the clock one-form, e.g. making it a complex 1-form $c\mapsto c_{\bbC}$. Thus we can still relate real-time/oscillatory/unitary evolution to ``Euclidean time''/exponentially damped/heat kernel evolution by analytic continuation -- which we do take advantage of in the remainder of the text.

In order to understand non-relativistic conformal field theory, we should understand conformal transformations of our non-relativistic spacetime. Such transformations should only involve the intrinsic data $(M, c, h)$, and should a priori not relate $c$ or $h$. Thus, a non-relativistic conformal transformation is a diffeomorphism of $\varphi: M \to M$ that preserves $c$ and $h$ up to a scale
\begin{equation}
    \varphi^* c = e^{\Omega_{c}} c\,,\quad
    \varphi^* h = e^{-2\Omega_{h}} h\,,
\end{equation}
for some independent conformal factors $\Omega_c$ and $\Omega_h$ \cite{Duval:2009vt, Duval:2016tzi}. Non-relativistic Weyl transformations are defined similarly, with physical rescalings of the clock and inverse metric.

At this point, we can compute the conformal transformations of the flat non-relativistic spacetime $\bbR^{d+1}$ in the standard way. The conformal Newton-Cartan algebra is isomorphic to
\begin{equation}
    \mathfrak{chr}(\bbR^{d+1}) \cong (\mathfrak{gl}(2,\bbR) \times \mathfrak{so}(d)) \ltimes \bbR^{2d}\,.
\end{equation}
This is essentially the Schr\"odinger algebra $\mathfrak{sch}_d$, with two differences: first, the mass element is missing because it is a central element without a geometric interpretation (unless we use a larger embedding space), turning $\mathfrak{h}_d \to \bbR^{2d}$, generated by boosts $K_i$ and translations $P_i$; and second, the timelike $\mathfrak{sl}(2,\bbR)$ conformal isometries are enhanced to a a full $\mathfrak{gl}(2,\bbR)$. The reason for this enhancement is clear: a general non-relativistic conformal transformation is allowed to scale space and time independently, i.e. arbitrary non-relativistic scalings are locally generated by both
\begin{equation}
    v_{c} = t\partial_t \quad \text{and}\quad v_{h} = x^i \partial_i\,,
\end{equation}
whereas Lifshitz scalings with dynamical critical exponent $z$ lock time and space scaling to
\begin{equation}
    v_z = z t \partial_t + x^i \partial_i\,.
\end{equation}
This specialization reduces the $\mathfrak{gl}(2,\bbR)$ to the usual Schr\"odinger $\mathfrak{sl}(2,\bbR)$ symmetry when $z=2$ (or $ax+b$ symmetry when $z \neq 2$).

In a Schr\"odinger field theory (resp. Lifshitz), we only expect correlation functions to be related on Schr\"odinger-Weyl equivalent spacetimes (resp. Lifshitz-Weyl), not under general non-relativistic-Weyl transformations. As above, this time-space scale locking leads to a very strong constraint on what kind of spacetimes can emerge, because conformal factors must satisfy 
\begin{equation}
    \Omega_h = z \Omega_c\,.
\end{equation}

To see the importance of this, let us first consider some very general transformation of our flat non-relativistic spacetime $\bbR^{d+1}$, schematically we write:
\begin{equation}
    t = f^0(\tau, y^i)\,,\quad x^i = f^i(\tau, y^i)\,.
\end{equation}
In this case $c = dt = \partial_\tau f^0 d\tau + \partial_i f^0 dy^i$. To locally keep the direction of time, we should have $\partial_i f^0 = 0$ and thus we are forced to consider only ``usual'' 1d conformal transformations in time $t = f^0(\tau)$. Now we turn to the spatial transformations $f^i$. As mentioned, the time-space locking now essentially constrains the form of the $f^i(\tau, y^i)$ to match the timelike rescaling, thus -- modulo some otherwise uninteresting global spatial Galilean isometries -- Lifshitz-Weyl equivalent spacetimes will be obtained by coordinate transforms of the form:
\begin{equation}\label{eq:coordinatePrinciple}
    t = f(\tau)\,,\quad  x^i = (\partial_\tau f(\tau))^{1/z} y^i\,.
\end{equation}
This severely limits the transformations we should consider. In particular, we are essentially forced to consider 1d conformal transformations of time, and thus: \textit{Schr\"odinger CFTs behave like 1d CFTs in time, with spectator spatial directions}. The spectator spatial directions are extremely important because they can carry non-trivial pressures, and allow non-vanishing stress tensors for our as-if 1d CFT. But all ``conformal'' aspects of the theory are related to what we do in time.

\subsection{Some Important Facts from Relativistic CFT}\label{sec:ImportantFacts}
Now we briefly recall the different geometries involved in standard relativistic CFT. Since this is review, we will be terse.

\paragraph{In General Dimensions.} Let's start with a standard real-time, i.e. Lorentzian, CFT on Minkowski space $M=\bbR^{1,d}$. Conformal transformations do not map Minkowski space $M$ to itself, so we must consider the conformal compactification $M_c := S^1\times S^{d}/\bbZ_2$, with $\bbZ_2$ acting by antipodal identification on both spheres. This space has closed timelike curves and so is not suitable for physics, hence we consider the universal cover (assuming $d \neq 2$) \cite{Mack:1975je, Mack:1976pa, Luscher:1974ez, Kravchuk:2018htv}
\begin{equation}
    \widetilde{M} := \bbR \times S^d\,,
\end{equation}
which is the Lorentzian cylinder. The Lorentzian cylinder carries a natural transitive action of the universal cover of the Lorentzian conformal group $\widetilde{SO(2,d)}$ and Wightman functions on $M$ can be analytically continued to all of $\widetilde{M}$. 

A natural set of global coordinates on $\widetilde{M} = \bbR \times S^d$ are given by $(\tau, \hat{e})$, where $\tau \in \bbR$ and $\hat{e}$ is a unit vector in $\bbR^{d+1}$. Usual Minkowski space $M$ embeds as a Poincar\'e patch on $\widetilde{M}$, with 
\begin{equation}\label{eq:PoinPatch}
    x^0 = \frac{\sin\tau}{\cos\tau + e^{d+1}} \,,\quad
    x^i = \frac{e^{i}}{\cos\tau + e^{d+1}} \,,
\end{equation}
see \cite{Luscher:1974ez, Kravchuk:2018htv} for more details. Real time evolution on the cylinder $\partial_\tau$ can be pulled back to the plane and is generated by the Luscher-Mack conformal Hamiltonian:
\begin{equation}
    H_{\mathrm{LM}} = \frac{1}{2}(P_0 + K_0)\,. 
\end{equation}

Now we can Wick rotate real cylinder time to Euclidean cylinder time $\tau_E = i\tau$, with metric
\begin{equation}
    ds_{E,\mathrm{cyl}.}^2 = d\tau_{E}^2 + d\Omega_{d}^2\,,
\end{equation}
where $\Omega_d$ are the usual angular variables on $S^d$ obtained from $\hat{e}$. The Euclidean cylinder can be Weyl transformed to the Euclidean plane (with point at infinity) by the radial map
\begin{equation}
    \tau_E = \log r\,.
\end{equation}
Famously, time evolution on the Euclidean cylinder $\partial_{\tau_E}$ becomes
\begin{equation}
    H_{E,\mathrm{cyl}.} = D\,,
\end{equation}
where $D$ generates dilatations in this Euclidean plane. In this sense, $H_{\mathrm{LM}} = i D$, which can then be used to relate the spectrum of the Luscher-Mack Hamiltonian to scaling dimensions, as in \cite{Mack:1975je, Mack:1976pa, Luscher:1974ez}. The fact that the infinite past/future on the Euclidean cylinder becomes a single point at the origin/infinity in $\bbR^{d+1} \,\cup\, \{\infty\}$ leads to a state operator correspondence, as all information in a state in radial quantization can be propagated backwards/forwards to a single local modification of spacetime -- a local operator.

\begin{figure}
    \centering
    \begin{minipage}{0.48\textwidth}
        \centering
        \begin{adjustbox}{width=\linewidth,center}{
        \begin{tikzcd}[cramped, column sep=small, row sep=normal]
          {\begin{matrix}
             \text{Lorentzian $\mathbb{R}\times S^d$}\\
             H_{\mathrm{LM}}=\tfrac12(P_0+K_0)
           \end{matrix}}
          &&
          {\begin{matrix}
             \text{Minkowski $\mathbb{R}^{1,d}$}\\
             (t,\vec x)
           \end{matrix}}\\[2ex]
          {\begin{matrix}
             \text{Euclidean $\mathbb{R}\times S^d$}\\
             H_{E,\mathrm{cyl}.}=D
           \end{matrix}}
          &&
          {\begin{matrix}
             \text{Euclidean $\bbR^{d+1}$}\\
             H_{\mathrm{NS}}=\tfrac12(P_{E,0}+K_{E,0})
           \end{matrix}}\\[2ex]
          &
          {\begin{matrix}
             \text{Euclidean $\bbR^{d+1}$}\\
             H_E=D
           \end{matrix}}
          \arrow["{\tau_E = i \tau}"', from=1-1, to=2-1]
          \arrow[hook',"{\begin{matrix}\text{Embeds as}\\ \text{Poincar\'e patch}\end{matrix}}", from=1-3, to=1-1]
          \arrow["{t'_E=it}", from=1-3, to=2-3]
          \arrow["{\begin{matrix}\tau_E=\log r\end{matrix}}"', from=2-1, to=3-2]
\arrow[leftrightarrow, "{\text{SCT}}"'{pos=0.5}, from=3-2, to=2-3]
        \end{tikzcd}
        }
        \end{adjustbox}
    \end{minipage}%
    \hfill
    \begin{minipage}{0.48\textwidth}
        \centering
        \begin{adjustbox}{width=1.08\linewidth,center}{
        \begin{tikzcd}[cramped, column sep=small, row sep=normal]
          {\begin{matrix}
             \text{Real Time Trap $S^0 \times \bbR^{d+1}$}\\
             H_{\HT}=P_0+\omega^2 C_0
           \end{matrix}}
          &&
          {\begin{matrix}
             \text{Real Time $\mathbb{R}^{d+1}$}\\
             (t,\vec x)
           \end{matrix}}\\[2ex]
          {\begin{matrix}
             \text{Euclidean Trap $S^0 \times R^{d+1}$}\\
             H_{E}=\omega D
           \end{matrix}}
          &&
          {\begin{matrix}
             \text{Euclidean $\bbR^{d+1}$}\\
             H_{\mathrm{NS}}=P_{E,0}+\omega^2 C_{E,0}
           \end{matrix}}\\[2ex]
          &
          {\begin{matrix}
             \text{Euclidean $\bbR^{d+1}$}\\
             H_E=\omega D
           \end{matrix}}
          \arrow["{\tau_E = i \tau}"', from=1-1, to=2-1]
          \arrow[hook',"{\begin{matrix}\text{Embeds as}\\ \text{patch}\end{matrix}}", from=1-3, to=1-1]
          \arrow["{t'_E=it}", from=1-3, to=2-3]
          \arrow["{\begin{matrix}\tau_E=\log\abs{t_E}\end{matrix}}"', from=2-1, to=3-2]
\arrow[leftrightarrow, "{\begin{matrix}\text{M\"obius}\\ \text{transform}\end{matrix}}"'{pos=0.5}, from=3-2, to=2-3]
        \end{tikzcd}
        }
        \end{adjustbox}
    \end{minipage}%
    \caption{Left, a commuting diagram explaining the relation between different geometries in relativistic CFT. Right, a commuting diagram explaining the analogous geometries in Schr\"odinger CFT. Schr\"odinger-Weyl transformations effectively only act on time, deforming space in a way completely determined from the transform in time.}
    \label{fig:CommutingQuantization}
\end{figure}
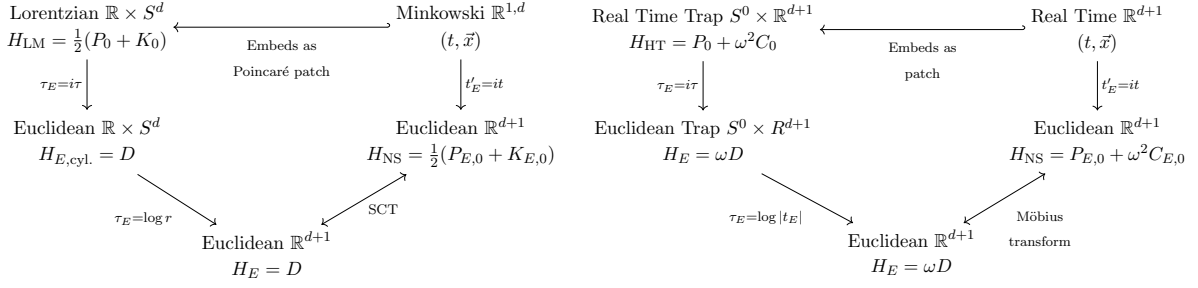

At risk of belabouring the point, let us note that Wick rotating on the cylinder then using the radial map is not the same as Wick rotating on the plane. If we were to Wick rotate directly in the plane, then the Luscher-Mack Hamiltonian of course becomes
\begin{equation}
    H_{\mathrm{LM}} = i (P_{E,0} + K_{E,0}) = i H_{\mathrm{NS}}\,,
\end{equation}
which is the Hamiltonian of NS quantization and can be related to the radial quantization scheme by a special conformal transformation \cite{Rychkov:2016iqz}. We summarize this discussion in Figure \ref{fig:CommutingQuantization}.

\paragraph{Specialization to (0+1)d.} In $d=0$ the story simplifies greatly, but let us spell out a few important points related to the doubling of the geometry. Lorentzian $M = \bbR$ is not closed under conformal transformations and the ``Lorentzian cylinder'' is now
\begin{equation}
    \widetilde{M} = \bbR \times S^0\,,
\end{equation}
which is instead two disconnected copies of the real line. This is the familiar doubling that also happens for line defects and $\mathrm{AdS}_2$ holography \cite{Sen:2011cn, NullDefects}. We use coordinates $\tau \in (-\infty,\infty)$ on $\bbR$ and $s=\pm 1$ to distinguish the branches. The original $M=\bbR$ covers a patch of one of the branches $\widetilde{M}$ by (compare to \eqref{eq:PoinPatch}):
\begin{equation}\label{eq:1dCyl}
    t 
        = \frac{\sin \tau}{\cos\tau + s} 
        = \begin{cases}
            \tan(\tau/2) & \text{if $s = +1$}\\
             -\cot(\tau/2) & \text{if $s = -1$}
        \end{cases}\,.
\end{equation}
Note that planar time evolution $\partial_t \sim P_0$ is orientation reversed along the two branches.

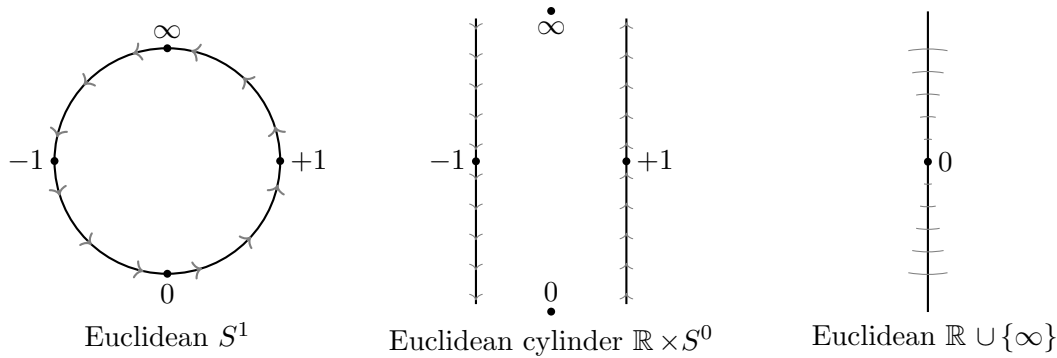
\begin{figure}
  \centering
  \begin{minipage}{.32\textwidth}
        \centering
\begin{tikzpicture}[scale=1.0,baseline={(current bounding box.center)}]
  \def\r{1.5}
  \draw[thick] (0,0) circle (\r);

  \fill (0,-\r) circle (1.5pt) node[below] {$0$};       
  \fill (-\r,0) circle (1.5pt) node[left] {$-1$};
  \fill (\r,0) circle (1.5pt) node[right] {$+1$};
  \fill (0,\r) circle (1.5pt) node[above] {$\infty$};

\foreach \angle in {15,45,75,105,135,165,195,225,255,285,315,345} {
    \pgfmathsetmacro\x{\r*cos(\angle)}
    \pgfmathsetmacro\y{\r*sin(\angle)}
    \pgfmathsetmacro\dx{-\r*sin(\angle)*0.05} 
    \pgfmathsetmacro\dy{\r*cos(\angle)*0.05}
    
    \draw[gray,->,thick,
        postaction={decorate}] (\x,\y) -- ({\x+\dx},{\y+\dy});
}

\end{tikzpicture}       
        \vskip 0.1cm Euclidean $S^1$
    \end{minipage}
    \begin{minipage}{.32\textwidth}
        \centering
\begin{tikzpicture}[scale=1.0,baseline={(0,0)}]
  \draw[thick] (-1,-1.9) -- (-1,1.9);
  \draw[thick] (1,-1.9) -- (1,1.9);

  \foreach \y in {1.8,1.4,1.0,0.6,0.2,-0.2,-0.6,-1.0,-1.4,-1.8} {
      \draw[gray,->,thin] (-1,\y) -- (-1,{ \y - 0.05});
      \draw[gray,->,thin] (1,\y) -- (1,{ \y + 0.05});
  }

  \fill (-1,0) circle (1.5pt) node[left] {$-1$};
  \fill (1,0) circle (1.5pt) node[right] {$+1$};
    \fill (0,-2.0) circle (1.5pt) node[above] {$0$};
    \fill (0,+2.0) circle (1.5pt) node[below] {$\infty$};
\end{tikzpicture}
\vskip 0.1cm Euclidean cylinder $\bbR \times S^0$
    \end{minipage}
    \begin{minipage}{.32\textwidth}
        \centering
        \begin{tikzpicture}[baseline={(current bounding box.center)}, scale=1.0]
      \draw[thick] (0,-2) -- (0,2);
    
      \fill (0,0) circle (1.5pt) node[right] {$0$};
    
\foreach \r in {0.3,0.6,0.9,1.2,1.5} {
    \draw[gray,thin] ({\r*cos(80)},{\r*sin(80)}) 
    arc[start angle=80,end angle=100,radius=\r];
    \draw[gray,thin] ({\r*cos(260)},{\r*sin(260)}) 
        arc[start angle=260,end angle=280,radius=\r];
  }
        \end{tikzpicture}
        \vskip 0.1cm Euclidean $\bbR\, \cup\, \{\infty\}$
    \end{minipage}
  \caption{Left, periodic Euclidean time on $S^1$ can be mapped to the Euclidean cylinder $\bbR \times S^0$ (center) by a Weyl transform -- splitting it into two branches. Time translations along the Euclidean cylinder coordinate are orientation reversed relative to the $-1$ branch. Right, the Euclidean ``plane'' is shown with its radial quantization surfaces, which slice the line along \textit{two} disconnected points and opposite orientations.}
  \label{fig:SenTwoBranches}
\end{figure}

The origin of this doubling is especially clear when Wick rotating the cylinder time to $\tau_E = i\tau$. There, the doubling is just the statement that
\begin{equation}\label{eq:logarithm}
    \tau_E = \log\abs{t_E}\,,\quad s_E = \mathrm{sgn}(t_E)
\end{equation}
has two branches: $t_E=0$ corresponds to $\tau_E = -\infty$, but can be approached from $t_E < 0$ or $t_E > 0$ and likewise for $t_E = \infty$, see Figure \ref{fig:SenTwoBranches}. This doubling of the geometry in the cylinder picture leads not to a state-operator correspondence in 1d CFTs, but a correspondence between local operators and states in a thermofield double \cite{Sen:2011cn}.

Finally, as before, we can consider the Wick rotation of the original Lorentzian time $t'_E = i t$ on $\bbR$. This time $t_E'$ is related to $\tau_E$ by 
\begin{equation}
    t'_E = \frac{\text{sinh}(\tau_E)}{\text{cosh}(\tau_E)\,+\,s}
\end{equation}
And this is related to the other Euclidean plane time, $t_E$, via the M\"obius transform:
\begin{equation}
    t'_E =\frac{t_E\,-\,s}{t_E\,+\,s}\,,
\end{equation}
which maps $t_E = 0$ and $t_E = \infty$ to $t'_E = -1$ and $t'_E = +1$ as we expect, see also Figure \ref{fig:RadialNS}.

\subsection{The Harmonic Trap Geometry}\label{sec:HTG}
Having understood the geometry of Schr\"odinger CFTs and the geometric state-operator story for relativistic CFT -- and especially the $d=0$ case -- we can now follow the same scheme for non-relativistic CFTs. We summarize the discussion in Figure \ref{fig:CommutingQuantization}.

\paragraph{The Real Time Harmonic Trap.} We start in flat non-relativistic spacetime $M = \bbR^{d+1}$ with real-time evolution, as described in Section \ref{sec:NRGeometry}. As we saw there, non-relativistic CFTs have no ways to mix the temporal and spatial coordinates and all Weyl transformations are effectively controlled by their effect on the timelike coordinate. To this end, we consider the action of a general finite Schr\"odinger transformation on spacetime
\begin{equation}
    t \mapsto \frac{at + b}{ct + d}\,,\quad \vec{x} \mapsto \frac{R\vec{x} + \vec{v}t + \vec{a}}{ct + d}\,.
\end{equation}
Clearly any finite time can be mapped to infinite time, and we are forced to extend $t$ to $\infty$. However, once we add the point $t=\infty$, we are also forced to add an entire spatial plane $\bbR^d$ at $t = \infty$. Just as in Lorentzian CFT, adding a point/plane at $t =\infty$ leads to closed timelike curves, and we should pass to the universal cover, the \textit{harmonic trap spacetime}:
\begin{equation}
    M_{\HT} := \bbR \times \, S^0 \times \bbR^{d}\,.
\end{equation}
Thus we have a doubling, just as in 1d Lorentzian CFT. We use coordinates $s=\pm1$ to distinguish the two branches again, and $(\tau,\vec{y})$ as the coordinates on $\bbR^{d+1}$.\footnote{Taking the universal cover removes closed timelike curves, but the non-relativistic spacetime is still ``non-distinguishing'' meaning that we cannot uniquely determine points on the manifold by knowing their causal past and causal future.}

\begin{figure}
    \centering
    \begin{minipage}{.38\textwidth}
        \centering
        \begin{tikzpicture}[baseline={(current bounding box.center)}, scale=1.0]
            \draw[->, thick] (-2.5,0) -- (2.5,0) node[right] {\(\vec{x}\)};
            \draw[->, thick] (0,-2.5) -- (0,2.5) node[above] {\(t\)};
            
            \foreach \x in {-2, -1, 1, 2} {
                \draw[gray, dashed] (\x,-2.3) -- (\x,2.3);
            }
            
            \foreach \t in {-2, -1, 1, 2} {
                \draw[gray, dashed] (-2.3,\t) -- (2.3,\t);
            }
        \end{tikzpicture}
        
        \vskip 0.1cm Flat space
    \end{minipage}
    \begin{minipage}{.22\textwidth}
        \centering
        \vspace{1em}
        \begin{align*}
            \omega t &= \tan(\omega \tau) \\
            \vec{x} &= \vec{y} \sec(\omega \tau) \\
            \\
            \omega \tau &= \arctan(\omega t) \\
            \vec{y} &= \frac{\vec{x}}{\sqrt{1+\omega^2 t^2}}
        \end{align*}
    \end{minipage}
    \begin{minipage}{.38\textwidth}
        \centering
        \begin{tikzpicture}[baseline={(current bounding box.center)}, scale=1.2]
            \draw[->, thick] (0,-2.0) -- (0,2.0) node[above] {\(\tau\)};
            \draw[->, thick] (-2.2,0) -- (2.2,0) node[right] {\(\vec{y}\)};
    
            \foreach \y in {-2,-1,1,2} {
                \draw[gray, dashed, domain=-1.57:1.57, samples=100, variable=\t] 
                    plot ({\y * cos(deg(\t))}, {\t});
            }

            \foreach \tau in {-1.11, -0.785, 0.785, 1.11} {
                \draw[gray, dashed] (-2.1,\tau) -- (2.1,\tau);
            }

            \node at (0.4,1.57) {\({\scriptstyle +\!}\frac{\pi}{2\omega}\)};
            \node at (0.4,-1.57) {\({\scriptstyle -\!}\frac{\pi}{2\omega}\)};
            \draw[thick] (-0.1,1.57) -- (0.1,1.57);
            \draw[thick] (-0.1,-1.57) -- (0.1,-1.57);
            
        \end{tikzpicture}
        \vskip 0.1cm Harmonic trap
    \end{minipage}

    \caption{Left: flat spacetime $\bbR^{d+1}$ in coordinates \((t, \vec{x})\). A non-relativistic Weyl transformation brings flat space to a patch of the harmonic trap geometry $M_{\HT}$. In this picture, $(\tau,y^i)\in (-\tfrac{\pi}{2\omega},\tfrac{\pi}{2\omega}) \times \bbR^{d}$ describes a ``Poincar\'e patch'' on one branch of the harmonic trap geometry, with future and past infinities of the patch given by the boundaries $\tau = \pm \pi/2\omega$. $M_{\HT}$ is a Schr\"odinger version of the Lorentzian cylinder.}
    \label{fig:HarmonicTrap}
\end{figure}
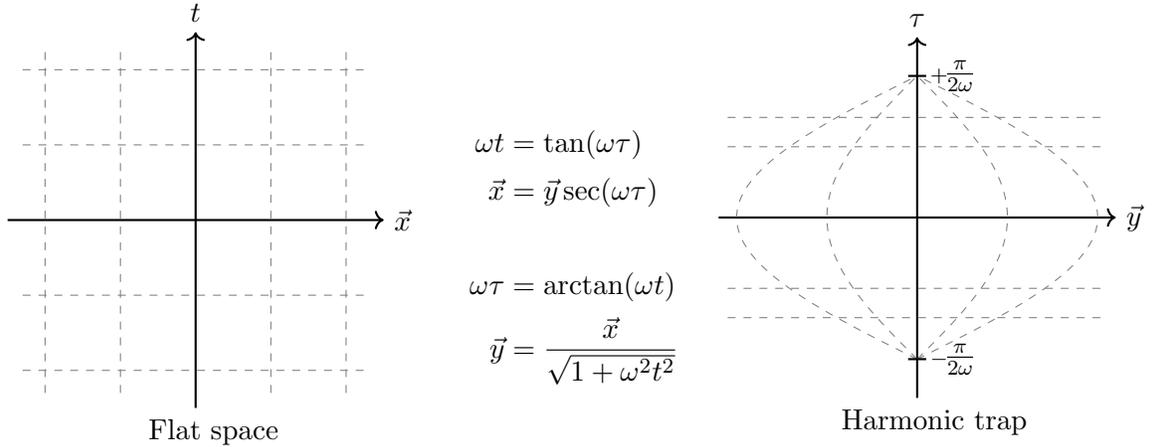

As with relativistic CFTs, real-time non-relativistic flat space embeds as a ``Poincar\'e patch'' in $M_{\HT}$. The map is is given by adapting \eqref{eq:1dCyl} to:
\begin{equation}\label{eq:HTcoords}
    \omega t 
        = \tan \omega \tau\,,\quad
    \vec{x}
        = \vec{y} \sec \omega\tau \,,
\end{equation}
where we have also added back in conventional factors of $\omega$ (analogous to changing the radius of the sphere), see Figure \ref{fig:HarmonicTrap}. This is the usual harmonic trap map described in the literature.
As with usual Minkowski space, we see that these coordinates only cover a patch $\tau \in (-\tfrac{\pi}{2\omega},\tfrac{\pi}{2\omega})$ of spacetime. The boundary points $\tau = \pm \pi/2\omega$ give analogues of future and past timelike infinity in Minkowski space.

At this point, let us recall our vector fields describing real time evolution in the flat non-relativistic geometry. They are
\begin{equation}
\begin{alignedat}{2}\label{eq:realtimeVectors}
    \mathscr{D}      &= -(2t\partial_t + x^i \partial_i) \,,\quad&
    \mathscr{M}_{ij}    &= -(x_i \partial_j - x_j \partial_i) \,, \\[6pt]
    \mathscr{C}_0    &=  t^2 \partial_t + t x^i \partial_i \,,\quad&
    \mathscr{P}_0    &=  \partial_t \,, \\[6pt]
    \mathscr{K}_i    &=  t \partial_i \,,\quad&
    \mathscr{P}_i    &= -\partial_i \,, \\[6pt]
\end{alignedat}
\end{equation}
and $M$ is non-geometric so has no vector field without an embedding space. They satisfy the same commutation relations as in \eqref{eq:SchrodingerAlgebra} after dropping the $i$. In the harmonic trap coordinates, on the $s=1$ branch, we can rewrite these vector fields as\footnote{The vector fields on the $s=-1$ branch can be computed analogously, and in practice it amounts to flipping the orientation of time and swapping sines with cosines.} 
\begin{equation}\label{eq:HTVecs}
\resizebox{0.9\textwidth}{!}{$
\begin{alignedat}{2}
    \mathscr{D}      &= -\bigl(\tfrac{1}{\omega}\sin(2\omega\tau)\partial_\tau
                          +\cos(2\omega\tau) y^i\partial_{y^i}\bigr),\quad&
    \mathscr{M}_{ij} &= -(y_i \partial_{y^j} - y_j \partial_{y^i}), \\[6pt]
    \mathscr{C}_0    &=  \tfrac{1}{\omega^2}\sin^2(\omega\tau)\partial_\tau
                          +\tfrac{1}{\omega}\sin(\omega\tau)\cos(\omega\tau)
                            y^i\partial_{y^i},\quad&
    \mathscr{P}_0    &=  \cos^2(\omega\tau)\partial_\tau
                          -\omega\sin(\omega\tau)\cos(\omega\tau)
                            y^i\partial_{y^i}, \\[6pt]
    \mathscr{K}_i    &=  \tfrac{1}{\omega}\sin(\omega\tau)\partial_{y^i},\quad&
    \mathscr{P}_i    &= -\cos(\omega\tau)\partial_{y^i}\,.
\end{alignedat}
$}
\end{equation}
All-in-all, real time translations $\partial_\tau$ in the harmonic trap are generated by
\begin{equation}\label{eq:HTVector}
    \partial_{\tau} = s (\mathscr{P}_0 +\omega^2\mathscr{C}_0)\,,
\end{equation}
note the orientation reversal on the $s=-1$ branch. As promised, Hamiltonian evolution in the real time harmonic trap $M_{\HT}$ is generated by
\begin{equation}
    H_{\HT} = P_0 + \omega^2 C_0\,,
\end{equation}
which gives the expected quadratic potential in Lagrangian theories. As claimed above, $\tau = \pm \pi/2\omega$ act like a past/future timelike infinity, and these coordinates make it clear that finite boosts act on timelike infinities by translations
\begin{equation}
    K_i: y^i \mapsto y^i \pm \frac{v^i}{\omega}\quad \text{at} \quad \tau = \pm\frac{\pi}{2\omega}\,,
\end{equation}
just as in usual flat space.

In summary, for each real trap time $\tau$, we have \textit{two} copies of $\bbR^{d}$. The two different branches are orientation reversed in time, explaining the natural polarization of the operator algebra of Schr\"odinger CFTs: the system on the left branch is the time-reversal conjugate of the system on the right branch.

\paragraph{The Euclidean Harmonic Trap.} Now we consider analytic continuation in the real-time harmonic trap coordinate $\tau$ (not the plane coordinate $t$), by $\tau_E = i\tau$. The corresponding ``Euclidean harmonic trap'' is denoted $M_{E,\HT}$
and the vector fields in \eqref{eq:HTVecs} can be realized accordingly. Now we can map to the Euclidean plane by adapting \eqref{eq:logarithm}:
\begin{equation}
\label{eq:HarmonicTrapMap}
    \omega |t_E| = \exp(2\omega \tau_E)\,,\quad \vec{x}' = \sqrt{2}\exp(\omega \tau_E) \vec{y}\,.
\end{equation}
Under this coordinate transform to the Euclidean plane, we find that Euclidean trap time evolution $\partial_{\tau_E}$ is generated by
\begin{equation}\label{eq:usefulD}
    H_{E,\HT} = -i(P_0 + \omega^2 C_0) = \omega D\,.
\end{equation}
Thus we will be able to relate the spectrum of operators in the real-time harmonic trap to scaling dimensions of operators, as with relativistic CFTs. 

\begin{figure}
    \centering
    \begin{minipage}{.45\textwidth}
        \centering
        \includegraphics[width=.95\linewidth]{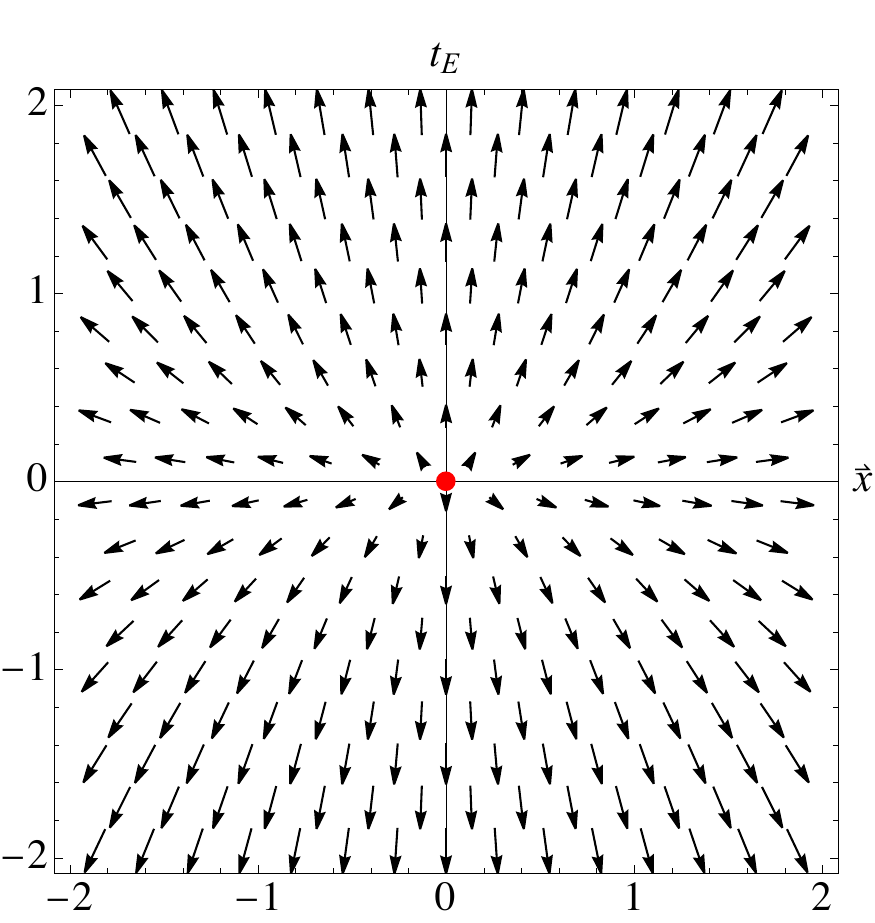}
    \end{minipage}
    \begin{minipage}{0.45\textwidth}
        \centering
        \includegraphics[width=.95\linewidth]{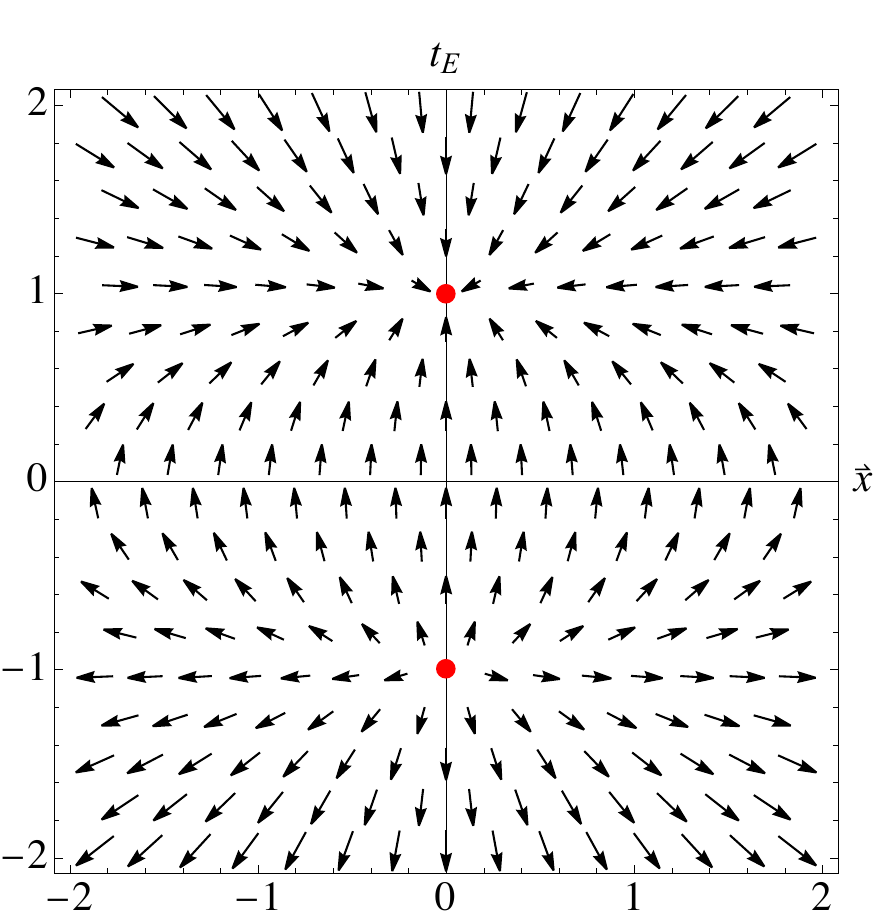}
    \end{minipage}
    \caption{Left, the dilatation vector field $\tfrac{\omega}{2}\mathscr{D}_E$ in the plane. The fixed point(s) of the vector field is at the origin (red), and infinity (not shown). Right, the NS vector field $\mathscr{P}_{E,0}+\omega^2 \mathscr{C}_{E,0}$. The fixed points (red) are now at $(t_E,\vec{x}) = (\pm 1,\vec{0})$. Plots are in units $\omega = 1$.}
    \label{fig:RadialNS}
\end{figure}

Finally, we note that the convention in the Schr\"odinger CFT literature is \textit{not} to use dilatations. Instead, one uses a non-relativistic analog of the relativistic North-South quantization scheme, which we call Nishida-Son quantization (or ``NS quantization'' for short) \cite{Nishida:2007pj}, obtained by Wick rotating directly in the non-relativistic plane. The Wick rotation of the HT Hamiltonian is then just
\begin{equation}
    H_{\HT} = i(P_{E,0}+\omega^2 C_{E,0}) = iH_{\mathrm{NS}}\,.
\end{equation}
Such a quantization scheme is obviously unitarily equivalent to the dilatation scheme described above, related by a coordinate transform mapping $(0,\vec{0})$ to $(-\omega,\vec{0})$ and $(\infty,\vec{0})$ to $(\omega,\vec{0})$. We illustrate the two vector fields in Figure \ref{fig:RadialNS}.

\subsection{State-Operator Correspondence and Thermofield Double}\label{sec:NRSOC}
Having understood the geometry of the harmonic trap spacetime and its complexification in time, we can now consider to what extent a state-operator correspondence makes sense in Schr\"odinger CFTs.

In usual relativistic CFT, a state-operator correspondence exists when we use radial quantization on the plane, with Hamiltonian flow generated by the dilatation operator $D$. All information about a (dilatation eigen)state can then be represented by a local operator at the origin in the plane, obtained by propagating the state backwards to the origin. Likewise for a point at infinity. On the cylinder, this radial foliation becomes equal time quantization, and the operator at the origin/infinity represents a state in the infinite past/future of the cylinder.

In Schr\"odinger CFTs, we have a useful analog of this fact: Euclidean trap evolution is still related to ``radial evolution'' in the plane, in the sense that 
\begin{equation}
\label{eq:EuclideanDilation}
    H_{E,\HT} = \omega D\,.
\end{equation}
As before, $D$ has two fixed points $(0,\vec{0})$ and $(\infty,\vec{0})$ in the conformally completed space, and they are related under the inversion
\begin{equation}
    t_E \mapsto \frac{1}{t_E}\,,\quad \vec{x}' \mapsto \frac{\vec{x}'}{t_E}\,.
\end{equation}
Thus, in order to proceed, we must pick a quantization scheme/foliation of spacetime.

\subsubsection{Lemon Quantization and a State-Operator Correspondence}
One candidate foliation is to consider ``radial'' leaves in the Euclidean plane, which we call lemon quantization. As we will see, these lemon leaves are useful because they describe a state-operator correspondence compatible with the polarization of observables in Schr\"odinger CFTs.

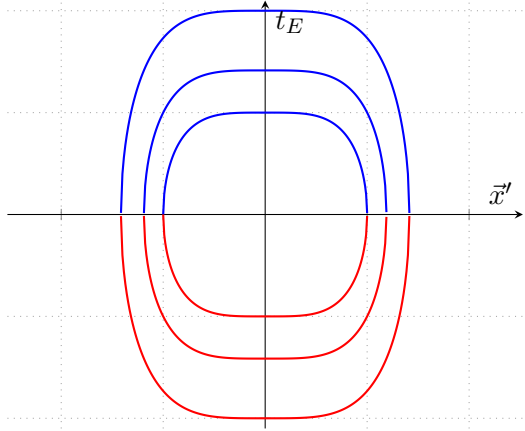
\begin{figure}
 \centering
\begin{tikzpicture}
  \begin{axis}[
    axis equal,
    axis lines=middle,
    xlabel=$\vec{x}'$,
    ylabel=$t_E$,
    samples=200,
    grid=both,
    grid style={dotted, gray!70},
    xticklabels={},
    yticklabels={},
    xmin=-2.1, xmax=2.1,
    ymin=-2.1, ymax=2.1,
  ]
    \addplot[blue, thick, domain=-1:1] {sqrt(1 - x^4)};
    \addplot[red,  thick, domain=-1:1] {-sqrt(1 - x^4)};
    \addplot[blue, thick, domain=-sqrt(sqrt(2)):sqrt(sqrt(2))] {sqrt(2 - x^4)};
    \addplot[red,  thick, domain=-sqrt(sqrt(2)):sqrt(sqrt(2))] {-sqrt(2 - x^4)};
    \addplot[blue, thick, domain=-sqrt(2):sqrt(2)] {sqrt(4 - x^4)};
    \addplot[red,  thick, domain=-sqrt(2):sqrt(2)] {-sqrt(4 - x^4)};
  \end{axis}
\end{tikzpicture}
\caption{Lemon quantization surfaces using the level sets $t^2_E+x^4 = R^4$ for $R=1,2,4$. The top half of the lemon (blue) lives on the $s=+1$ branch of $M_{E,\HT}$, while the bottom half of the lemon (red) lives on the $s=-1$ branch.} \label{fig:LemonLeaves}
 \end{figure}

Start with the non-relativistic Euclidean plane. Since our Hamiltonian is $D$ and we are interested in foliations which are radially symmetric around the origin, a general foliation should be made from level sets of the form
\begin{equation}\label{eq:fLemon}
    |t_E|^{\alpha}f\left(\frac{|\vec{x}'|^2}{|t_E|}\right) = c\,,
\end{equation}
where $\alpha \in\bbR$ and $f$ is an arbitrary function of the Schr\"odinger cross ratio.
These define Lifshitz (homogeneous) functions,
\begin{equation}
    \label{eq:Lifshitz}
    F_\Delta(\lambda^2 t_E,\lambda \vec{x})=\lambda^\Delta F_{\Delta}(t_E,\vec{x})\,.
\end{equation}
A simple choice with smooth leaves -- mimicking radial quantization -- is to consider the level sets
\begin{equation}
    L_{k}(R):\quad |t_E|^{k} + |\vec{x}'|^{2k} = R^{2k}\,.
\end{equation}
In the minimal $k=1$ case, these leaves $L_{1}(R)$ resemble lemons, with a sharp cusp at $t_E = 0$. However, this cusp is resolved by taking $k$ larger, thus foliations $L_{k\geq 2}$ provide a more suitable quantization surface, see Figure \ref{fig:LemonLeaves}.

By design, the leaves are mapped into each other under $D$ and are $M_{ij}$-invariant. In the limit as $R \to 0$, the leaves collapse on the origin $(t_E = 0, \vec{x}' = 0)$, stabilized by $D$, $M_{ij}$, $K_i$, and $C_0$. And so, we have a state operator correspondence: \textit{$D$ and $M_{ij}$ are good quantum numbers for states on the surface of the lemon, and such eigenstates are in correspondence with local operators at the origin with well-defined scaling dimension $D$ and spin $M_{ij}$.} Adding that local operators are annihilated by the stabilizers $K_i$ and $C_0$ are precisely the conditions for Schr\"odinger primaries (recall that $M$ is not generally geometric).

A priori, there is no requirement to have a Schr\"odinger CFT. Any scale invariant theory with critical exponent $z$ has such a quantization scheme, relating states on Lifshitz-homogeneous lemons to local operators at the origin. However, scale-invariant theories do not have a SCT $C_0$ (or, more generally, inversion) that maps the origin to infinity. Better yet, having the SCT $C_0$ allows us to actually relate Euclidean $D$-evolution to our real-time Hamiltonian. For example, in a Lifshitz symmetric theory, we can still map the Euclidean plane to a (doubled) harmonic trap spacetime $M_{E,\HT}$ by using a modified version of \eqref{eq:HarmonicTrap} (recall the general principle \eqref{eq:coordinatePrinciple}), thus $D$ generates Euclidean trap time translations $\partial_{\tau_E}$. Then, if we Wick rotate back to real harmonic trap time $\tau_E = i \tau$, real trap time evolution can be written as
\begin{equation}
    \partial_{\tau} \sim \partial_t + (t^2 \partial_t + \frac{2}{z}x^i\partial_i)\,,
\end{equation}
by modifying \eqref{eq:HTcoords}. If $z=2$, the second term is actually realizable by some generator $C_0$ in the plane, and so the real-time Luscher-Mack Hamiltonian is related to the $D$ spectrum as in \eqref{eq:usefulD}.

\subsubsection{Factorization and the Thermofield Double}
In the Euclidean harmonic trap $M_{E,\HT}$, non-relativistic spacetime splits into two branches and the top half of each lemon becomes a (non-compact) leaf of the $s=1$ branch of the harmonic trap, likewise for the bottom half of each lemon on the $s=-1$ branch (recall the center and right-most images of Figure \ref{fig:SenTwoBranches}). Both slices are oriented ``up'' their respective branches in time $\tau_E$, so that the time $t_E$ on the $s=-1$ branch is orientation reversed with respect to $\tau_E$. As $R$ goes from $0$ to $\infty$, these leaves map from the infinite past $\tau_E = -\infty$ to the infinite future $\tau_E = +\infty$.

If we cut $M_{E,\HT}$ along both branches, the Hilbert space factorizes into \textit{two} copies of the Hilbert space $\calH$ on $\bbR^d$. More canonically, since the $s=-1$ branch is orientation reversed, we can write the Hilbert space as
\begin{equation}\label{eq:factorizeHilb}
    \calH_{\mathrm{TFD}} = \calH^{*} \otimes \calH\,.
\end{equation}
Consequently, the state-operator correspondence implies that: \textit{local operators in Schr\"odinger CFTs are in one-to-one correspondence with states in a doubled Hilbert space $\calH^*\otimes \calH$}.\footnote{We also expect states on the different hemispheres of the lemon to require some gluing condition at the equator. In the HT spacetime, this is some condition on states as $|\vec{y}|^2 \to \infty$; presumably that they are built on top of the same vacuum, so that states on hemispheres are actually elements in conjugate Hilbert spaces.} 

As explained in Section \ref{sec:HTG}, the usual ``harmonic trap map'' \eqref{eq:HTcoords} maps to (a patch of) one branch of the harmonic trap spacetime with Hamiltonian evolution given by $H_{\HT}$. If we denote the usual HT ground state by $\ket{\Omega} \in \calH$, then the state-operator correspondence says that
\begin{equation}
    \mathds{1}\leftrightarrow \ket*{\Omega^*}\otimes\ket*{\Omega}\,.
\end{equation}
More generally, let us write a basis of states in $\calH_{\mathrm{TFD}}$ as:
\begin{equation}
 \ket{a^*}\otimes\ket{b}\in \mathcal{H}^*\otimes\mathcal{H}\,.
\end{equation}
Intuitively, local operators on $\calH$ are constructed by gluing states along North and South quantization ``hemispheres'' and propagating them towards the origin.

As alluded to above, this is just a thermofield double construction in the zero temperature limit \cite{Sen:2011cn}. More generally, instead of considering the theory on $M_{E,\HT} = (\bbR \times S^0) \times \bbR^d$, we could consider the case that the Euclidean time is a finite $S^1$ of temperature $\beta$, i.e.
\begin{equation}\label{eq:finiteTemp}
    M_{\beta,\HT} = S_{\beta}^1 \times \bbR^d\,.
\end{equation}
Then we can define the thermofield double state
\begin{equation}
    \ket{\ket{\mathrm{TFD}}} = \frac{1}{\sqrt{Z}} \sum_{\Delta,m} e^{-\beta H_{\HT} - \mu M} \ket{\Delta^*,m^*}\otimes\ket{\Delta,m}\,,
\end{equation}
where we have enriched the usual TFD state with the superselection number $M$. In this case, correlation functions on $M_{\beta,\HT}$ can either be computed by the usual path integral over the $S^1$ and written as a trace over states in $\calH$ weighted by $e^{-\beta H_{\HT} - \mu M}$, or computed as a matrix element in the TFD state.

In passing, we note that this has a possible interesting interpretation for Schr\"odinger holography. This TFD picture suggests that a generic Schr\"odinger field theory is holographically dual to a spacetime with two boundaries. A priori, we do not expect the sides to interact (this is the factorization/non-renormalization theorem) unless there are $M=0$ states in the spectrum of $\calH$, corresponding to Hermitian operators which are not ``normal-ordered.'' Interestingly, $M=0$ states should appear in null reductions (see Section \ref{sec:GenuineMassless}), and enriching the TFD state with temperature $\beta$ and chemical potential fugacities $\mu$ are like tracking null momentum, i.e. $H_{\HT} \sim P_-$ and $M \sim P_+$. We leave the pursuit of this point for future works.

Returning to our zero temperature setup, a distinguished role is played by operators dual to the ``in-states'':
\begin{equation}\label{eq:ImportantSplit}
    \mathcal{O}^\dagger_b(0) 
        \leftrightarrow \ket{\Omega^*}\otimes\ket{b}\,,\quad 
    \mathcal{O}_a (0)
        \leftrightarrow \ket{a^*}\otimes\ket{\Omega}\,.
\end{equation}
Indeed, we can use this to \textit{define} what we mean by daggered and undaggered operators in an abstract theory, as we discuss further in Section \ref{sec:HarmonicTrap}. Likewise, the ``out-states'' describe local operators at infinity:
\begin{equation}
    \mathcal{O}_b(\infty)\leftrightarrow \bra{\Omega^*}\otimes \bra{b}\,,\quad
    \mathcal{O}^\dagger_a(-\infty)\leftrightarrow \bra{a^*}\otimes \bra{\Omega}\,.
\end{equation}
Note that we have written the operators above as a function of the plane time $t_E$, and that $\pm \infty$ are really the same point on $S^1$; the sign emphasizes that the operator is defined by state propagating up the $s=\pm1$ branch. More generally, there are operators dual to the states $\ket{a^*}\otimes\ket{b}$. As we will demonstrate in Section \ref{sec:NonRenorm}, the operators dual to these general hemispherical pairings are ``normal-ordered'' products.

Usual inner products of harmonic trap states $\braket{a}{b}$ in $\calH$ correspond to placing a ``creation'' operator $\calO_b^\dagger$ at $(0,\vec{0})$ in plane coordinates, and placing an annihilation operator $\calO_a$ at $(\infty,\vec{0})$, i.e.
\begin{equation}
    \braket{a}{b} = \lim_{t_E \to \infty }t_{E}^{\Delta_a}\mel{\Omega}{\calO_a(t_E) \calO_b^\dagger(0)}{\Omega}\,.
\end{equation}
In fact, the TFD picture allows us to define a notion of inner product on all observables, even those of ``normal ordered type'' which annihilate $\bra{\Omega}$ and $\ket{\Omega}$, see also \cite{Pyramids}.

Finally, as mentioned in Section \ref{sec:HTG}, we conventionally perform a coordinate transform so that $H_{\mathrm{NS}}$ is the Hamiltonian, not $D$, then in/out-states are prepared at $t_E = \mp 1/\omega$, and the usual inner product is presented \cite{Nishida:2007pj, Goldberger:2014hca}:
\begin{equation}
    \braket{a}{b} = \mel{\Omega}{\calO_a(i/\omega) \calO^\dagger_b(-i/\omega)}{\Omega}\,.
\end{equation}

\noindent Let us summarize the upshot of all of these points:
\begin{enumerate}\setlength\itemsep{0em}
    \item A careful consideration of non-relativistic geometry and coordinate transforms indicates that $H_{\HT}$ is conjugate to $i\omega D$. Thus the spectrum of $H_{\HT}$ is related to the spectrum of $D$, as in relativistic CFTs. This is true even for non-primary operators.
    \item Schr\"odinger-Weyl transformations relate the theory on $\bbR^{d+1}$ to the Euclidean Harmonic Trap geometry $M_{E,\HT} = \bbR \times S^0 \times \bbR^d$, which has two disconnected components or ``branches.'' Thus the Hilbert space at any time $\tau_E$ in the Euclidean trap is naturally identified with two copies of the Hilbert space of flat space $\calH_{\mathrm{TFD}} = \calH^* \otimes \calH$, and local operators in Schr\"odinger CFTs are in one-to-one correspondence with states in this doubled space, or endomorphisms on $\calH$.
    \item By this correspondence, the spectrum of the theory in the harmonic trap essentially defines the space of local operators in Schr\"odinger CFTs, with operators $\calO^\dagger$ canonically creating/annihilating states on the right/left and vice-versa for $\calO$.
\end{enumerate}

At this point, a careful reader may recall that there is an ambiguity in the choice of quantization surface selected in \eqref{eq:fLemon}. In general, we could consider an $f$-lemon for some function $f$ of the Schr\"odinger cross ratio $z$. Crucially, any generic $f$ will have the doubling we describe in this section, because any foliation will cut spacetime along the two branches. For general lemons, the half-leaves are not at constant harmonic trap time $\tau_E$ (like usual radial quantization), but do become flatter as $|\vec{y}|^2 \to \infty$. This means inversion does not act nicely on a general $f$-lemon, although a general (compact) $f$-lemon does lead to a state-operator correspondence and TFD Hilbert space. One special choice is when $f \equiv 1$. On one hand, this is an equal-time plane quantization in the harmonic trap, it acts nicely with respect to time-inversions and non-relativistic geometry, and guarantees that the associated state-space is our usual plane Hilbert space. On the other hand, this choice does not obviously lead to a state-local operator correspondence. It would be nice to show that given two choices of generalized lemon, $f$ and $f'$, that the quantization schemes are effectively equivalent. In particular, we note that the confining potential of Schr\"odinger CFTs leads to an extremely sharp (exponential) localization of states in space, recall \eqref{eq:2ptcorrelator}. It is plausible that this makes different quantization schemes essentially equivalent, and we leave exploration of this technical point to future works.

\section{The Harmonic Trap Spectrum}\label{sec:HarmonicTrap}
Given the spectrum of a Schr\"odinger CFT in the harmonic trap, or space of states $\calH$, we can determine the space of local operators: local operators are in one-to-one correspondence with states in a TFD $\calH_{\mathrm{TFD}} = \calH^* \otimes \calH$. This is physically important because the standard setup for engineering Schr\"odinger CFTs is to place a system (large cold atoms, say) in a quadratic potential. Thus, in principle, we can determine the space of states and algebra of local operators of Schr\"odinger CFTs by looking at spectral lines in the harmonic trap.

In this section we discuss the ``admissible spectra'' or Hilbert spaces $\calH$ of Schr\"odinger field theories. We start in Section \ref{eq:RaisingLowering} by discussing some basic assumptions and results about $H_{\HT}$ and its spectrum on $\calH$. Then, in Section \ref{sec:GenuinePrimaries}, we introduce the terminology of ``genuine'' and ``non-genuine'' operators, which act non-trivially and annihilate the HT vacuum $\ket{\Omega}$ respectively.

In Section \ref{sec:PositiveUIRs} we use the raising and lowering algebra to classify the physical spectra that can actually arise. Mathematically, we classify the non-negative energy Unitary Irreducible Representations (UIRs) of the Schr\"odinger group, and thus the module structure of the genuine creation (and annihilation) operators.\footnote{In \cite{Pyramids} we also consider the module structure of non-genuine operators.} As we will see, representations with $M=0$ can appear in principle; these are the states that were ignored to avoid seas of descendants in \eqref{eq:descendantSea}. We also consider unitarity bounds for massive and massless states, with and without spin, showing that massless states have lower unitarity bounds. We also discuss how Lorentzian unitarity proofs can be used to prove unitarity for entire modules without studying the entire algebra of descendants. In Section \ref{sec:GalileanParticles} we give a terse complimentary perspective on the preceding discussions by uplifting constructions of ``static Galilean particles,'' i.e. using the method of induction to describe UIRs, which matches our lowest weight constructions perfectly. 

Finally, in Section \ref{sec:M0WardIdentities} we discuss the associated Ward identities to our new $M=0$ states/operators and show that massless operators behave like standard 1d CFT operators which have been delocalized in space. In particular, we discuss how they imply a weak violation of cluster decomposition, which prefaces our discussions of non-renormalization in Section \ref{sec:NonRenorm}.

\subsection{The Structure of \texorpdfstring{$\calH$}{calH}}\label{eq:RaisingLowering}
Let us consider our non-relativistic CFT in a harmonic trap. Physically, this amounts to turning on a particular quadratic potential for the quantum mechanics and probing the states of the Schr\"odinger CFT on one branch of the HT geometry $M_{\HT}$; recall Figure \ref{fig:HarmonicTrap}. Our real-time Hamiltonian is
\begin{equation}\label{eq:HarmonicTrap2}
    H_{\HT} = P_0 + \omega^2 C_0\,.
\end{equation}
Placing the theory in the harmonic trap discretizes the energy spectrum of compact Schr\"odinger CFTs by definition.

We denote the Schr\"odinger invariant vacuum state by $\ket{\Omega}$. Since we consider unitary theories, we have an entire Hilbert space $\calH$ of states built on top of $\ket{\Omega}$, and a Hermitian conjugation $\dagger$ under which all of our plane generators are unitary. The Hilbert space $\calH$ necessarily decomposes into superselection sectors labelled by the mass $M$
\begin{equation}\label{eq:SSSDecomp}
    \calH = \bigoplus_{m} \calH_{m}\,.
\end{equation}
As we will see in Section \ref{sec:Massless}, we necessarily have $m \geq 0$ in a unitary theory. Without loss of substance, we can assume that $M$ is the only interesting superselection number in our theory. For example, we will assume there is no further decomposition of the Hilbert space $\calH$ into universes or other exotic superselection quantum numbers. Combined with unitarity bounds on the $m=0$ sector, this implies that $\ket{\Omega}$ is the unique ground state of $H_{\HT}$.

By Wigner's theorem, the entire Hilbert space $\calH$ is decomposable into (projective) Unitary Irreducible Representations (UIRs) of the Schr\"odinger group. For a physical spectrum, the representations that appear should also be \textit{non-negative energy} representations for the Hamiltonian $H_{\HT}$. Generically, non-negative energy unitary representations will be lowest weight representations where the action of $H_{\HT}$ is diagonalized. Since $H_{\HT}$ is conjugate to $i\omega D$, the lowest weight representations will have a lowest eigenvalue $\omega \Delta$, and all other eigenvalues will take the form $\Delta+k$ for some $k\in\mathbb{N}$. An explicit conjugation is given by:
\begin{equation}\label{eq:ConjugateToD}
    e^{-\tfrac{\pi}{4}\left(\tfrac{1}{\omega}P_0 - \omega C_0\right)} H_{\HT}e^{\tfrac{\pi}{4}\left(\tfrac{1}{\omega}P_0 - \omega C_0\right)} = i \omega D\,.
\end{equation}

\subsection{Genuine and Non-Genuine Operators}\label{sec:GenuinePrimaries}
One consequence of the previous abstract discussion is the natural polarization of the algebra of observables, i.e. the splitting into daggered and undaggered operators (e.g. from \eqref{eq:ImportantSplit}). This is manifest in Lagrangian formulations and non-relativistic limits, but here it is completely general.

In particular, we use the observable/physical HT spectrum to define a distinguished subset of daggered operators $\calO^\dagger(0)$, dual to the states $\ket{\Omega^*}\otimes \ket{\calO}$. More precisely, $\ket{\calO}$ must have well-defined scaling dimension to be defined in the infinite past in the harmonic trap. Lowest weight states of non-negative energy UIRs of the Schr\"odinger group specifically define our local primaries $\calO^\dagger(0)$, and compactness of the spectrum prevents ambiguities in identifying and normalizing them. Annihilation operators $\calO$ are defined likewise, but with conjugation. Thus the HT vacuum state $\ket{\Omega}$ defines what is meant by daggered/undaggered operators, as we expect it to, and the splitting is canonical: there is no need to think of these as modes of a relativistic theory or fields in a non-relativistic Lagrangian.

Combined with unitarity, see Section \ref{sec:PositiveUIRs}, this recovers a number of useful expectations. For example, in a unitary theory we have:
\begin{equation}
    [M,\calO^\dagger(0)] = m \calO^\dagger(0)
    \quad \text{and} \quad 
    [M,\calO(0)] = -m \calO(0)\,,
\end{equation}
where necessarily 
\begin{equation}
    m \geq 0 \,.
\end{equation}
Moreover, we also have that:
\begin{equation}
    \bra{\Omega} \calO^\dagger(x) 
        = 0 
        = \calO(x)\ket{\Omega}\,,
\end{equation}
for all $x \in \bbR^{d+1}$ if $m > 0$. The exception to this is when $m = 0$, in which case the operators do not annihilate the vacuum on either side, leading to very strong kinematic constraints and dynamical consequences.

At risk of belabouring the point, these subclasses of operators do not exhaust all operators in the theory: there is not a state-operator correspondence with $\calH$ nor $\calH \oplus \calH^\dagger$. Intuitively, we are missing essential ``normal-ordered'' products like the number density $n(x) = \,:\!\phi^\dagger\phi\!:$ and even more complicated things like $:\!\!(\phi^\dagger)^7\phi^3\!\!:$. We expect that they are dual to general products $\ket{a^*}\otimes\ket{b}$, and will discuss this further in Section \ref{sec:NonRenorm}. 

This motivates the splitting of operators into two types:
\begin{enumerate}\setlength\itemsep{0em}
    \item \textbf{Genuine Primaries.} By definition, these primaries are operators dual to states in $\calH$ or $\calH^\dagger$.\footnote{In principle, we should define operators to be right-genuine and left-genuine, or genuine and co-genuine. However, context makes clear which objects are relevant, so there is no need for such maximally pedantic and linguistically burdened terminology -- especially given the number of other terms we introduce.} These include the operators we were calling $\calO^\dagger$ and $\calO$ above, as well as ``composites'' like $(\calO^\dagger)^k$ to the extent it is well-defined (there are generically scheme ambiguities in the definition of composites).
    \item \textbf{Non-Genuine Primaries.} These are primaries which do not create a state when acting on $\ket{\Omega}$ or $\bra{\Omega}$. These include operators like $n(x)$ and $:\!\!(\phi^\dagger)^7\phi^3\!\!:$ in the free theory. Despite not creating a state when acting on the HT vacuum $\ket{\Omega}$, these operators will still appear in OPEs of genuine primaries as ``composite operators.'' Non-renormalization theorems make clear the extent to which these composite operators are well-defined.
\end{enumerate}
We emphasize that whether an operator is genuine or non-genuine is independent of $M$.

More abstractly, assume that the observables of the theory are described by some algebra $\calA$, e.g. mathematically this could be a $*$-algebra or factorization algebra. States are normalized non-negative-definite linear functionals $\omega: \calA \to \bbC$, possibly with additional regularity constraints. When a state $\omega$ has a kernel, operators $a \in \ker(\omega)$ are the ``non-genuine operators'' for $\omega$. In our case, we privilege the ground state $\ket{\Omega}$ of $H_{\HT}$ in defining our non-genuine operators. 

It is quite unusual, from the point of view of relativistic QFT, for a local operator to annihilate the vacuum.  In usual relativistic QFT, the Reeh-Schlieder theorem states that products of local operators $\calO_{f_1} \cdots \calO_{f_n} \ket{\Omega}$ smeared on a small region (including smearing in the time direction) will generate the whole Hilbert space; in relativistic CFTs, the convergence of the OPE implies single dilatation eigenoperators are sufficient for a dense set. In other words, the vacuum vector is cyclic in the Hilbert space. The Reeh-Schlieder theorem also implies that the vacuum is separating, meaning if $\calO\ket{\Omega} = 0$, then $\calO$ is identically $0$. To prove this, consider a spacelike separated region containing local operators $\calO'$. These $\calO'$ also act on the vacuum and generate a dense set of vectors in $\calH$. Since spacelike operators commute, $\calO\calO'\ket{\Omega} = \calO'\calO\ket{\Omega} = 0$, so $\calO$ is $0$ on a dense subspace of the Hilbert space and thus is exactly the zero operator. We illustrate this in Figure \ref{fig:NullReduction}. Roughly, this means that the vacuum state of relativistic QFTs is ``maximally entangled'' \cite{Reeh:1961ujh, Witten:2018zxz, Casini:2022rlv}.

In non-relativistic CFTs $\ket{\Omega}$ is still a cyclic vector in $\calH$, but the separating property is obviously false: all non-genuine operators vanish on $\ket{\Omega}$ and $\bra{\Omega}$. So where does the preceding argument breakdown? The usual proof that $\ket{\Omega}$ is separating fails when we consider the causal structure of non-relativistic spacetime and/or analyticity of correlators. In proving the Reeh-Schlieder theorem, we must smear slightly in the timelike direction, and in non-relativistic spacetime $\bbR^{d+1}$ regions with timelike support are never spacelike separated. A similar argument, that null manifolds do not have a good analytic continuation, is claimed to evade the Reeh-Schlieder theorem in entanglement entropy proofs of the monotonicity theorems \cite{Casini:2022rlv}, and we expect similar considerations to hold for theories on non-distinguishing spacetimes like non-relativistic $\bbR^{d+1}$. On the other hand, the TFD state $\ket{\ket{\mathrm{TFD}}}$ gives a thermal state which is guaranteed to be separating for the local operators of our theory.

Curiously, light-ray operators in Lorentzian CFT also annihilate the vacuum state and appear naturally in the OPE of local operators \cite{Kravchuk:2018htv}. It is plausible that non-genuine operators in Schr\"odinger field theories could provide a simple toy-model for studies of light-ray operators, or that they could even be directly related by null-reduction (see also Section 2.3 of \cite{NullDefects}).

\subsection{Admissible Spectra and \texorpdfstring{$M=0$}{M=0} Lowest Weight States}\label{sec:PositiveUIRs}
Our goal in this section is to understand the UIRs of the Schr\"odinger group that are non-negative energy for the HT Hamiltonian $H_{\HT}$. In principle, such UIRs constitute all representations that could appear in the spectrum of a Schr\"odinger field theory. 

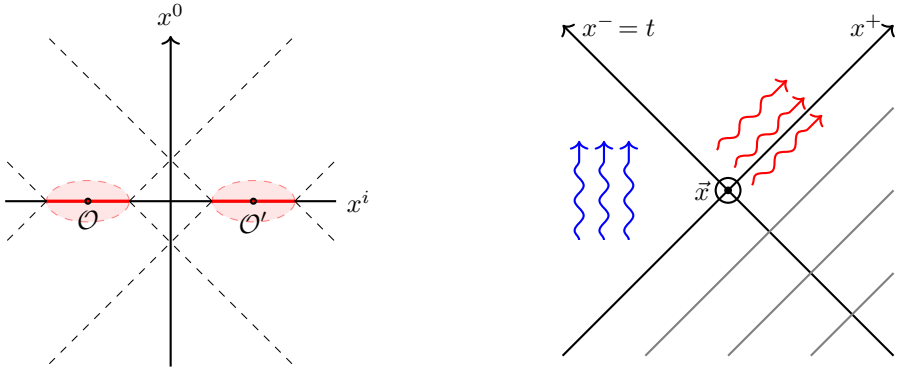
\begin{figure}
    \centering
    \begin{minipage}{0.45\textwidth}
        \centering
        \begin{tikzpicture}[baseline={(current bounding box.center)}, every node/.style={font=\small}, scale=1.1]
            \fill[fill=red!10, draw = red!50, dashed] (-1,0) ellipse (0.5 and 0.25);
            \fill[fill=red!10, draw = red!50, dashed] (1,0) ellipse (0.5 and 0.25);
            \draw[-, thick] (-2.0,0) -- (2.0,0) node[right] {\(x^i\)};
            \draw[->, thick] (0,-2.0) -- (0,2.0) node[above] {\(x^0\)};
            \draw[-, red, very thick] (-1.5,0) -- (-0.5,0);
            \draw[-, red, very thick] (+1.5,0) -- (0.5,0);
            \draw[-, dashed] (-1.5,0) -- (-2,0.5);
            \draw[-, dashed] (-1.5,0) -- (-2,-0.5);
            \draw[-, dashed] (+1.5,0) -- (+2,0.5);
            \draw[-, dashed] (+1.5,0) -- (+2,-0.5);
            \draw[-, dashed] (-0.5,0) -- (1.5,2.0);
            \draw[-, dashed] (-0.5,0) -- (1.5,-2.0);
            \draw[-, dashed] (0.5,0) -- (-1.5,2.0);
            \draw[-, dashed] (0.5,0) -- (-1.5,-2.0);
            
            \draw[thick] (-1,0) circle (0.03);
            \draw[thick] (1,0) circle (0.03);
            \node[below] at (-1,0) {\(\calO\)};
            \node[below] at (+1,0) {\(\calO'\)};
        \end{tikzpicture}
\end{minipage}
    \begin{minipage}{.45\textwidth}
        \centering
\begin{tikzpicture}[scale=1.1, every node/.style={font=\small}]
    \draw[->, thick] (-2,-2) -- (2,2) node[anchor = east] {\(x^+\)};
    \draw[->, thick] (2,-2) -- (-2,2) node[anchor = west] {\(\,\,x^-\!= t\)};
    
    \draw[thick] (0,0) circle (0.15);
    \filldraw[black] (0,0) circle (0.04);
    \node[left] at (-0.1,0) {\(\vec{x}\)};
    
    \foreach \shift in {-1,-2,-3} {
        \draw[gray, thick] 
            (-2 - \shift, -2) -- (2, 2 + \shift);
    }

    \foreach \x in {-0.212, 0.0, 0.212} {
        \draw[red, thick, decorate, ->,
              decoration={snake, amplitude=0.6mm, segment length=4mm, post=lineto, post length=1mm}]
            (-0.424 - \x+0.5, -0.424 + \x+0.7) -- (0.424 - \x+0.5, 0.424 + \x+0.7);
    }

    \foreach \x in {-0.3, 0.0, 0.3} {
        \draw[blue, thick, decorate, ->, 
              decoration={snake, amplitude=0.6mm, segment length=4mm, post=lineto, post length=1mm}]
            (\x-1.5, -0.6) -- (\x-1.5, 0.6);
    }
\end{tikzpicture}
    \end{minipage}
    \caption{\textbf{Left}, in a relativistic theory, local operators $\calO$ and $\calO'$ on spacelike separated regions (red) commute. Smearing over small pill-boxes in time (red regions) still requires finite time to communicate between excitations, as illustrated by the intersection of lightcones (dashed lines). In non-relativistic CFTs, the lightcones of such pill-boxes are flat, and so any amount of timelike smearing connects the regions instantaneously. \textbf{Right}, a ($d$+2)-dimensional Lorentzian CFT in null coordinates $(x^+, x^-, x^\perp)$. In null reduction, the null coordinate $x^-$ becomes the real Schr\"odinger time $t$. Lightlike radiation (red) parallel to the reduction direction corresponds to states with $M=0$, while timelike radiation (blue) is generally decomposed into massive states.}
    \label{fig:NullReduction}
\end{figure}

In familiar contexts, we expect such representations to be lowest weight representations of our symmetry group. This is the case e.g. for the conformal group and other semi-simple groups \cite{Mack:1975je}. While the Schr\"odinger group $S_d$ on $(d+1)$-dimensional spacetime is not semi-simple, it does embed inside the conformal group $SO(2,d+2)$ of $\bbR^{1,d+1}$. Geometrically, Schr\"odinger field theories also live on lightlike slices of relativistic spacetimes, and thus we expect compatibility with the non-negative energy representations of the lightcone conformal Hamiltonian, see Figure \ref{fig:NullReduction}. Moreover, since $H_{\HT} = P_0 + \omega^2 C_0$ depends entirely on the timelike $SL(2,\bbR)$, we expect any representations to be compatible with 1d CFT unitarity/energy constraints when restricted to this subgroup \cite{Pal:2018idc}; we will see that this is indeed the case. In Section \ref{sec:GalileanParticles} we consider a complementary perspective by induction from ``static Galilean particles.''

We start by considering lowest weight modules for the HT Hamiltonian $H_{\HT}$. It is straightforward to check that
\begin{equation}\label{eq:HRRaiseLower1}
    P_{\pm i}
        = \frac{1}{\sqrt{2\omega}}P_i \pm i \sqrt{\frac{\omega}{2}}K_i\,,\quad
    L_{\pm}
        = \frac{1}{2}\left(\frac{1}{\omega}P_0-\omega C_0 \pm i D\right)\,,
\end{equation}
act as raising and lowering operators for the HT Hamiltonian \cite{Goldberger:2014hca}. They satisfy

\begin{alignat}{3}
    [H_{\HT}, P_{\pm i}] 
        &= \pm \omega P_{\pm i}\,,\quad
    &[H_{\HT}, L_{\pm}] 
        &= \pm 2\omega L_{\pm}\,,\quad\\
    [P_{- i}, P_{+ j}] 
        &= \delta_{ij} M\,,\quad
    &[L_{-}, L_{+}] 
        &= \frac{1}{\omega}H_{\HT}\,,\\
    [P_{-i}, L_{+}] 
        &= P_{+i}\,,\quad
    &[L_{-}, P_{+i}] 
        &= P_{-i}\,,
\end{alignat}
as well as the obvious vectorial commutation relations under the action of $M_{ij}$. By definition, a lowest weight state $\ket{\calO}$ satisfies
\begin{equation}
    H_{\HT}\ket{\calO} = \omega \Delta \ket{\calO}\,,\quad
    P_{-i}\ket{\calO} = 0 = L_-\ket{\calO}\,.
\end{equation}

Using the raising operators, we can create a full Verma module of descendants
\begin{equation}
    \calV(\calO) := \{\,\cdots P_{+i} \cdots L_+ \cdots P_{+j} \cdots L_+ \cdots \ket{\calO}\,\}\,.
\end{equation}
In principle, the Verma module $\calV(\calO)$ can contain singular vectors and the module is reducible. The singular vectors correspond to null states and decouple from the theory as usual. Below, we will have to quotient out the $\calV(\calO)$ by these null states. To find null states, it is useful to note that Hermitian conjugation acts on our raising and lowering operators by:
\begin{equation}
    H_{\HT}^\dagger = H_{\HT}\,,\quad
    P_{\pm i}^\dagger = P_{\mp i}\,,\quad
    L^\dagger_{\pm} = L_{\mp}\,.
\end{equation}
Since any state corresponds to a genuine daggered (and undaggered) operator, all of the states and null conditions in the Verma module also have operator interpretations.

\subsubsection{Massive Scalars, Unitarity, and Mass Bounds}\label{sec:massiveScalar}
Consider a massive scalar lowest weight state $\ket{\calO}$. We would like to construct the full module of descendants and also understand the various null relations (if any) inside the Verma module. 

For states with $M \neq 0$, it is extremely useful to define the universal enveloping algebra elements:
\begin{equation}
    Q_{\pm} 
        := L_{\pm} - \frac{\vec{P}_{\pm}^2}{2M}\,,
\end{equation}
which commute with all of the $P_{\pm i}$ and satisfy
\begin{equation}
    [Q_-,Q_+] 
        = \frac{1}{\omega}H_{HT} - \frac{1}{2M}\{P_{-i}, P_{+i}\}
        = \frac{1}{\omega}H_{HT} - \frac{1}{M}\vec{P}_{+}\!\cdot\!\vec{P}_{-} - \frac{d}{2}\,.
\end{equation}
Then a general state in $\calV(\calO)$ is a linear combination of:
\begin{equation}
    \ket{k_0;k_1,\dots,k_d} := Q_+^{k_0} P^{k_1}_{+1}\cdots P^{k_d}_{+d} \ket{\calO}\,,
\end{equation}
and carries weight $\Delta + 2k_0 + k_1 + \dots + k_d$.\footnote{In the plane the $Q_+$ and $P_{i+}$ charges act by
\begin{equation}
    \mathcal{Q}_{+} = \frac{(i+\omega t)}{2\omega}\left(-i \omega(\Delta-\tfrac{d}{2}) + (i+\omega t)\left(i \partial_t+ \frac{\nabla^2}{2m}\right) \right)\,, \quad
    \mathcal{P}_{i+} = \frac{im\omega x_i + (i+\omega t)\partial_i}{\sqrt{2\omega}}\,.
\end{equation}
Thus a general state like $\ket{k_0;k_1,\dots,k_d}$ corresponds to a complicated mess of derivatives of an operator on the plane.} We can obtain both the scalar unitarity bound on $\Delta$ and a bound on the mass of states by using positivity of the primary $\braket{\calO}{\calO} > 0$ and its descendants. 

To obtain the scalar unitarity bound, consider the norm of the level 2 descendant:
\begin{equation}
    \braket{1_0}{1_0} = \mel*{\calO}{Q_- Q_+}{\calO} = \mel*{\calO}{[Q_-,Q_+]]}{\calO} = \left(\Delta - \tfrac{d}{2}\right) \braket{\calO}{\calO} \geq 0\,.
\end{equation}
Thus we obtain the usual non-relativistic scalar unitarity bound $\Delta \geq \tfrac{d}{2}$, as is known from \cite{Nishida:2007pj, Goldberger:2014hca}.

When a relativistic scalar primary is tuned to the unitarity bound $\Delta = \tfrac{d-2}{2}$, it becomes the free scalar via a shortening condition $P^2 \ket{\phi} = 0$ \cite{Erramilli:2019njx}. The same thing happens here. When the null state $\ket{1_0}$ saturates the unitarity bound $\Delta = \frac{d}{2}$, the explicit plane representation
\begin{equation}
    \mathcal{Q}_{+}\mathcal{O}(t,\vec{x})  = \frac{(i+\omega t)}{2\omega}\left(-i \omega(\Delta-\tfrac{d}{2}) + (i+\omega t)\left(i \partial_t+ \frac{\nabla^2}{2m}\right) \right)\mathcal{O}(t,\vec{x}) 
\end{equation}
implies the free Schr\"odinger equation
\begin{equation}
    \left( i \partial_t + \frac{\nabla^2}{2m}\right)\mathcal{O}(t,\vec{x}) = 0\,.
\end{equation}

This bound is only a necessary, but not sufficient, condition for unitarity: in principle, there could be states at higher levels in the module that spoil unitarity or demand a stronger bound. Rather than check an infinite number of positivity conditions, we can check the positivity of the entire module by smearing the corresponding operators $\calO^\dagger(x)$ with test functions $f(x)$ \cite{Mack:1975je, simmons2019tasi}.\footnote{Given that the analytic structure of non-relativistic correlation functions is not the same as relativistic ones, we do not know the appropriate space of test functions for non-relativistic CFTs. A better understanding of analytic methods in non-relativistic theories is warranted, and we leave this to future work.\label{footnote:analytic}} We define states
\begin{equation}
    \ket{\calO(f)} := \int dt \,d^d\vec{x}\, f(x) \calO^\dagger(x) \ket{\Omega}\,,
\end{equation}
then the norm is
\begin{align}
    \braket*{\calO(f)}{\calO(f)} 
        &= \int dt_1 dt_2 d^d\vec{x}_1 d^d\vec{x}_2\, f^*(x_2) f(x_1) \mel*{\Omega}{\calO(x_2)\calO^\dagger(x_1)}{\Omega}\\
        &= \int \frac{dk^0}{2\pi} \frac{d^d\vec{k}}{(2\pi)^d} |\hat{f}(k)|^2 \hat{K}(k)\,,
\end{align}
and positivity of $\braket*{\calO(f)}{\calO(f)}$ follows from positivity of the two-point function.

For the massive scalar, we can check this positivity by a direct computation
\begin{align}
    \hat{K}(k) 
        &= \int dt\, d^d\vec{x} \, e^{i k^0 t - i \vec{k} \cdot \vec{x}}\mel*{\Omega}{\calO(x)\!\calO^\dagger(0)}{\Omega} \\
        &= c \left(\frac{2\pi i}{m}\right)^{\frac{d}{2}} \, \int \frac{dt}{(t-i\epsilon)^{\Delta-d/2}} e^{i t \left(k^0 - \frac{\vec{k}^2}{2m}\right)}\\
        &= \frac{C_{d,\Delta}}{\Gamma(\Delta-d/2)}\left(k^0-\frac{k^2}{2m}\right)^{\Delta - \tfrac{d}{2} - 1} \Theta\left(k^0-\frac{k^2}{2m}\right)\,.
\end{align}
where $C_{d,\Delta}$ contains all of the global phase factors from the Gaussian integrals and time-integral. When $\Delta=d/2$, these factors cancel with all of our normalizations (see also Section \ref{sec:OPE} or compare to the usual Schr\"odinger kernel) and $C_{d,\Delta}$ is positive. For generic $\Delta > d/2$, the global phases do not cancel, it would be helpful to understand why that is the case. If we ignore this global phase factor $C_{d,\Delta}$, positivity of the kernel demands $\Gamma(\Delta-d/2) > 0$, which follows if $\Delta > d/2$. At the unitarity bound, the $\Gamma$-function has a pole which combines with the singularity of $k^0 \to k^2/2m$ to produce a $\delta(k^0 - \frac{k^2}{2m})$. Thus we see that the mass shell condition is enforced in the free theory. For some values of $\Delta < d/2$, the $\Gamma$-function becomes positive again, but the singularities from $k^0 = \frac{k^2}{2m}$ make the kernel poorly defined. Thus, up to the global phase $C_{d,\Delta}$, the unitarity bounds should follow from positivity of $\hat{K}(k)$ as in Lorentzian CFT.

We can also put bounds on the mass of states by using positivity. In particular, all massive genuine daggered operators can only creates states with $m > 0$. To see this, consider the level 1 descendant $\ket{1_i} = P_{+i} \ket{\calO}$, then:
\begin{equation}
    \braket{1_i}{1_i} 
        = \mel*{\calO}{P_{-i} P_{+i}}{\calO} 
        = \mel*{\calO}{M}{\calO} = m \braket{\calO}{\calO} \geq 0\,.
\end{equation}
Thus positivity of $\braket{1_i}{1_i}$ implies $m>0$. If $m = 0$ we would have null states, which we will return to in Section \ref{sec:Massless}. 

Having now intrinsically defined daggered operators, and showing that they necessarily have $m > 0$, we have now given an explanation of what earlier authors mean when they say they will ``assume $\calO^\dagger$ is made out of creation operators.'' Similarly, this implies that $\calO$ with $M\neq 0$ must annihilate $\ket{\Omega}$, or else violate unitarity. As a result, we can now safely call these daggered and undaggered operators as creation and annihilation operators.

\subsubsection{Massive Spinning Operators}\label{sec:massiveSpinning}
In the usual conformal algebra, the commutator of SCTs $K_\mu$ and translations $P_\nu$ is 
\begin{equation}
    [K_\mu, P_\nu] = - 2i(\eta_{\mu\nu}D + M_{\mu\nu})\,.
\end{equation} 
This fact, together with $K_\mu^{\dagger} = P_\mu$, leads to the appearance of the spin Casimir in the unitarity bounds. In our Schr\"odinger CFTs, the relevant commutator of boosts $K_i$ and spatial translations $P_i$ is central, 
\begin{equation}
    [K_i, P_j] = \delta_{ij} M\,.
\end{equation}
Moreover, the rotation generators $M_{ij}$ actually never appear in the RHS of commutators not already involving rotation generators. This implies that the spin rep of a primary operator can never appear in the norm of a descendent state, and thus doesn't feature in unitarity bounds. This also follows if we construct the massive spinning group representations explicitly (see Section \ref{sec:GalileanParticles} and \cite{Perroud:1977qh}).

As a result, the unitary bounds in Schr\"odinger field theory are independent of spin, and massive spinning operators $\calO_I$ must also have unitarity bound
\begin{equation}
    \Delta_{\calO} \geq \frac{d}{2}\,.
\end{equation}
This also means that the shortening condition at unitarity is  just the free Schr\"odinger equation again
\begin{equation}
    \left( i \partial_t + \frac{\nabla^2}{2m}\right)\mathcal{O}_{I}(t,\vec{x}) = 0\,.
\end{equation}

\subsubsection{Genuine Massless States and Unitarity}\label{sec:Massless}
We can employ the previous strategy again in the case  that $M = 0$. Take a scalar lowest weight state $\ket{\calO}$, then all states in $\calV(\calO)$ can be written as a linear combination of:
\begin{equation}
    \ket{k_0;k_1,\dots,k_d} := L_+^{k_0} P^{k_1}_{+1}\cdots P^{k_d}_{+d} \ket{\calO}\,.
\end{equation}
In the $M=0$ sector, $[P_{-i}, P_{+j}] = 0$ and thus any state with spatial descendants is null, i.e. a state is null if any $k_1, \dots, k_d \neq 0$. Quotienting out by null states leaves only states of the form:
\begin{equation}\label{eq:genuineScalar}
    \ket{k_0} := L_+^{k_0} \ket{\calO}\,.
\end{equation}

The resulting quotients are just $\mathfrak{sl}(2,\bbR)$-modules, and for $\Delta \geq 0$ these exhaust all the null states (a similar result is obtained in the ($1+1$)d case in \cite{dobrev1997lowest, Dobrev:2013kha}). By the same logic as the massive case, or by the 1d CFT unitarity bounds \cite{Pal:2018idc}, we see that: \textit{any genuine massless scalar primary must have:}
\begin{equation}
    \Delta \geq 0\,.
\end{equation} 
This provides a sufficient, but not necessary, experimental signature of genuine $M=0$ states in a physical system: its spectrum in the harmonic trap includes a line with energy $0 < \omega \Delta < \omega\tfrac{d}{2}$.

Since the resulting structure is essentially just that of an $\mathfrak{sl}(2,\bbR)$-module, we expect that genuine $M=0$ states will behave like a background 1d CFT in our system. In particular, given that $\vec{P}_{+}$-descendants are quotiented out, in operator language we expect that corresponding dual operators $\calO^\dagger(t,\vec{x}) = \calO(t,\vec{x})$ do not depend on position at all. i.e. \textit{genuine massless operators are topological in the space directions}.

The shortening condition for massless genuine operators occurs when we saturate the unitarity bound, $\Delta = 0$. In this case, $\mathcal{L}_+ \calO(t) = 0$ implies $\calO$ is independent of $t$. i.e. the analogue of the free Schr\"odinger equation is just
\begin{equation}
    \partial_{t} \calO(t) = 0\,,
\end{equation}
as we would hope for $\Delta=0$ operators. Thus, genuine $M=0$ operators saturate the unitarity bound if they are topological operators. We have already precluded the existence of other universes in Section \ref{sec:GenuinePrimaries}.

We can generalize this analysis to states with spin with no additional difficulty. The $L_{\pm}$ are scalars under the spatial rotations $M_{ij}$ and, as a result, the $\mathfrak{sl}(2,\bbR)$-modules just become $\mathfrak{sl}(2,\bbR) \times \mathfrak{so}(d)$-modules in a trivial way. Thus genuine massless states and their descendants will generally transform as in \eqref{eq:genuineScalar}, tensored with some ``internal'' $\mathrm{Spin}(d)$-degrees of freedom. We confirm this again in Section \ref{sec:GalileanParticles}.

In DLCQ, with dimensional reduction along $x^+$, we expect genuine $M=0$ states to arise from states with momentum $k_+ = 0$, see Figure \ref{fig:NullReduction}. Indeed, it has long been suspected that the DLCQ of conformal field theories gives some theories resembling a ``conformal quantum mechanics'' (see e.g. \cite{Maldacena:2008wh}), and here we see that such things essentially uniquely populate the genuine $M=0$ sector. We discuss this again in Section \ref{sec:NullReduction}.

\subsection{Uplifting Static Galilean Particles}\label{sec:GalileanParticles}
In the preceding section, we considered highest weight UIRs of the Schr\"odinger algebra, motivated by positive energy considerations. Projective UIRs of the (centrally extended) Schr\"odinger group $S_d$ were also constructed by induction in \cite{Perroud:1977qh} (with no non-negativity constraints). The method matches exactly with previous results, but provides some different physical intuition for the representations appearing.

Recall that in the method of induction for a group $G$: one first classifies UIRs of a normal subgroup $H < G$; then groups the UIRs of $H$ into orbits under $G$; and, finally, ``lifts'' the orbits of $H$ representations to $G$ e.g. in the construction of one-particle states. The strategy for constructing projective UIRs of $S_d$ (UIRs of the universal cover $\widetilde{S}_d$) is to induce twice:
\begin{equation}
    \bbR^{d+1} \to \widetilde{G}_d' \to \widetilde{S}_d\,.
\end{equation}
Here, $\bbR^{d+1}$ is the subgroup generated by spatial translations $P_i$ and the mass $M$, and $G'_d$ is the (centrally extended) Galilean group \textit{without} the time translation generator $P_0$, which we call the ``static Galilean group.'' Then we can write $\widetilde{S}_d = \widetilde{G}^{\prime}_d \rtimes \widetilde{SL}(2, \bbR) = (H_d \rtimes \mathrm{Spin}(d))\rtimes \widetilde{SL}(2, \bbR)$, where $H_d$ is the Heisenberg group generated by $\{K_i, M, P_i\}$, and perform the double induction.

We start with representations of $\bbR^{d+1} = \bbR^d \times \bbR_m$. Representations here describe plane waves in $\bbR^{d+1}$, with well-defined spatial momenta and mass $(\vec{p}, m)$. This interpretation of the mass parameter as an additional spacetime direction is quite literal when constructing Schr\"odinger field theories by null-reduction/DLCQ. Consider a plane-wave with $m\neq 0$, then a $\widetilde{G}_d'$ Galilean transformation with rotation $R$ and boost $\vec{v}$ changes the momentum to
\begin{equation}
    \vec{p} \mapsto R\cdot\vec{p} - m\vec{v}\,,
\end{equation}
but $m$ is unaffected. Thus all massive states with the same $m$ are identified under $\widetilde{G}_d'$ and form an orbit $O_m \cong \bbR^d \times \{m\} \subset \bbR^{d} \times \bbR_m$. 

Alternatively, when $m = 0$, a Galilean transform can rotate $\vec{p}$ along any direction but cannot change its magnitude. Thus all states with the same magnitude of spatial momentum $\abs{\vec{p}} > 0$ form an orbit $O_{\abs{\vec{p}}} \cong S^{d-1}_{\abs{\vec{p}}} \times \{0\} \subset \bbR^{d} \times \bbR_m$. We separate off $(\vec{p},m)=(\vec{0},0)$ as a special case, called $O_0$.

Now that we have understood how all of these $\bbR^d \times \bbR_m$ plane waves transform under $\widetilde{G}_d'$, we can induce to representations of $\widetilde{G}_d'$. This involves computing the little group of (a representative element in) each of the orbits. The results are as follows:
\begin{enumerate}\setlength\itemsep{0em}
    \item Consider the representative plane wave $(\vec{0},m) \in O_m$, the little group is $\mathrm{Spin}(d) \subset \widetilde{S}_d$. As a result, the $O_m$ orbits lift to massive spinning $\widetilde{G}_d'$-particles in a straightforward way, labelled by a mass $m$ and a spin rep $\rho$ of $\mathrm{Spin}(d)$.\label{item:list1}
    \item Next we turn our attention to $O_p$. Consider a representative momentum $\vec{p} = \abs{\vec{p}} \hat{n}$ pointing along the north pole of the celestial sphere $S^{d-1}_{\abs{\vec{p}}}$. Such a massless momentum is stabilized by all \textit{transverse} rotations $M_{ij}$ and all boosts $K_i$. Thus the little group of $O_p$ inside $\widetilde{G}_d'$ is $\mathrm{Spin}(d-1)\ltimes \bbR^{d}_K$.

    This is just like the little group stabilizing a massless momentum pointed along the North Pole in the standard construction of massless particle states in (3+1)d. As a result, we have both ``helicity'' type and ``continuous spin'' type representations (CSRs) from the $\mathrm{Spin}(d-1) \ltimes \bbR^{d-1}$. However, we notice here that there is an additional factor of $\bbR$, corresponding to the boost $K \cdot \hat{n}$ along the North pole of the celestial sphere. As a result, even the helicity-type representations possess a continuous internal degree of freedom.\label{item:list2}
    
    \item Finally, one can induce from $(\vec{0},0) \in O_0$, with little group $\mathrm{Spin}(d) \ltimes \bbR^{d}_K \ = \ 
    \widetilde{\mathrm{ISO}}(d)$ to construct both continuous spin and helicity-type $\widetilde{G}_d'$-vacua.\label{item:list3}
\end{enumerate}

The massive spinning $\widetilde{G}_d'$-particles from the $O_m$ orbit already resemble our massive spinning particles in Schr\"odinger CFTs. However, we have four different types of massless states: $O_p$ helicity and CSRs, and $O_0$ helicity and CSRs. We expect local operators with a continuum of internal degrees of freedom to be thermodynamically untenable \cite{Schuster:2024wjc} (although cannot strictly rule them out), and hence we ignore them on physical grounds.\footnote{Although we do not expect there to be local operators creating states in these representations, we do think it would be interesting to understand the physical realization of these representations. Curiously, we note that light ray operators $\mathbb{O}$ in conformal field theories are non-local continuous spin operators which annihilate the vacuum \cite{Kravchuk:2018htv}. We speculate that the $O_p$ type states could be related to the dimensional reduction of generic states carrying some momentum transverse to the light plane and/or light ray operators. The $O_0$ type corresponding to genuine $m=0$ operators are presumably a further special subset.}

The careful reader may wonder why we do not simply demand that particles transform in a trivial rep for the boost generators $K_i$. In fact, one does this if limiting to the Galilean group, $\widetilde{G}_d$. However, if we are interested in lifting the representation of $\widetilde{G}_d'$ to a representation of the Schr\"odinger group, $\widetilde{S}_d$, then assuming that the $K_i$ act trivially would result in a contradiction for the $O_p$ reps. To see this, note that $K_i \equiv 0$ implies $[P_0,K_i] = iP_i = 0$, which contradicts $|\vec{p}| \neq 0$. Thus, we can only successfully set $K_i = 0$ for $O_0$, where $\vec{p} = 0$. As a result, the only remaining representations are again the spinning massive $\widetilde{G}_d'$ particles and spinning $\widetilde{G}_d'$ vacua.

The next step is to understand how all of these representations of $\widetilde{G}_d'$ transform under $\widetilde{S}_d$. The details can be found in \cite{Perroud:1977qh}, but the final result is that the massive spinning $\widetilde{G}_d'$-particles lift to reps of $\widetilde{S}_d$, matching the massive scalar and massive spinning reps constructed in Sections \ref{sec:massiveScalar} and \ref{sec:massiveSpinning} respectively. On the other hand, the spinning $\widetilde{G}_d'$-vacua simply lift to reps of $\mathrm{Spin}(d) \times \widetilde{SL}(2,\bbR)$, which matches our findings in Section \ref{sec:Massless}.

Altogether, the only non-pathological projective unitary irreducible representations of $S_d$ are precisely those that match our lowest weight construction of non-negative energy reps in the harmonic trap.

\subsection{Ward Identities for the \texorpdfstring{$M=0$}{M=0} Sector}\label{sec:M0WardIdentities}
Let us now derive the most general form of the two and three point functions in the presence of massless operators. 

A general translation and rotation symmetric ansatz for a two-point function, compatible with $M$-symetry, is:
\begin{equation}
\mel{\Omega}{\calO_1(x_1)\!\calO^\dagger_2(x_2)}{\Omega} = \delta_{m_1,m_2}G_{12}(t_{12},r_{12}^2)\,.
\end{equation}
Now let's consider massless operators, $m_i = 0$. Boosts act on massless operators (recall \eqref{eq:ActionOnOps}) by
\begin{equation}
    [K_i,\calO(x)] = -i t \partial_i \calO(x)\,,
\end{equation}
giving the condition
\begin{align}
    0 
        &=(t_1\nabla_1 + t_2\nabla_2)G_{12}(t_{12},r_{12}^2)\\
        &= t_{12} \partial_{r_{12}} G_{12}(t_{12},r_{12}^2)\,.
\end{align}
For generic $(t,r)$, this implies that the two-point function depends only on $t_{12}$.

At this point, we can use our conformal invariance in time, generated by $D$ and $C_0$, to constrain $G_{12}$ further. The result is a function that looks identical to a CFT two-point function with the important caveat that we have $z = 2$ scaling and thus a different power of $\Delta$ in the denominator. It is also important to note that we have constrained the two-point function kinematically, but we still have to restore $i\epsilon$ prescriptions appropriately. Altogether, this gives
\begin{equation}
    \mel{\Omega}{\calO_1(t_1,\vec{x}_1)\!\calO^\dagger_2(t_2,\vec{x}_2)}{\Omega} = \frac{c}{(t_{12}-i\epsilon)^{\Delta_1}} \delta_{\Delta_1,\Delta_2}\,.
\end{equation}
This correlation function is precisely the $m \to 0$ limit of the massive two-point function \eqref{eq:2ptcorrelator}. This could have been anticipated since all the generators are analytic in $m$.

We can consider the three-point function in a similar way. Crucially, any three-point function should be the appropriate $m\to 0$ limit of \eqref{eq:3ptcorrelator}. Not only does this limit set one of the $m_i = 0$, and enforce that the other two masses satisfy $m_j = -m_k$, it also implies that dependence on position $\vec{x}_i$ disappears. For example, if we compute $\mel*{\Omega}{\calO_1(x_1)\calO_2(x_2)\!\calO_3^\dagger(x_3)}{\Omega}$ with $m_1 \to 0$ then
\begin{equation}
    \mel*{\Omega}{\calO_1(x_1)\!\calO_2(x_2)\!\calO_3^\dagger(x_3)}{\Omega} = \frac{f_{123}(z_{23}) \,\,e^{i\frac{m_2 \vec{x}_{23}^2}{2t_{23}}}}{t_{12}^{(\Delta_{1} + \Delta_{2}-\Delta_{3})/2}t_{23}^{(\Delta_{2} + \Delta_{3}-\Delta_{1})/2}t_{13}^{(\Delta_{1} + \Delta_{3}-\Delta_{2})/2}}\,.
\end{equation}
In the limit that all the operators becomes massless, then the three point function reduces to a usual three-point function in CFT (with adjusted $z=2$ scaling):
\begin{equation}
    \mel*{\Omega}{\calO_1(x_1)\!\calO_2(x_2)\!\calO_3^\dagger(x_3)}{\Omega}
        = \frac{C_{123}}{t_{12}^{(\Delta_{1} + \Delta_{2}-\Delta_{3})/2}t_{23}^{(\Delta_{2} + \Delta_{3}-\Delta_{1})/2}t_{13}^{(\Delta_{1} + \Delta_{3}-\Delta_{2})/2}}\,,
\end{equation}
with appropriate $i\epsilon$ prescriptions.

\subsubsection{Cluster Decomposition}\label{sec:Cluster}
As we have demonstrated above, $M=0$ states are constant in space but depend on time. Thus, if they correspond to local operators, the local operators should be topological in space and depend only on time. This makes sense: inserting a massless field operator at the origin in a relativistic theory creates a state which spreads out along the (future directed) null cone; in non-relativistic theories, the null cone is flat and the massless operator produces an entire isotropic and homogeneous background. We give discuss $M=0$ operators in Section \ref{sec:GenuineMassless}. Consequently, we expect some weak violation of cluster decomposition in non-relativistic theories with $M=0$ states. It is not immediately obvious to us that this is problematic, given that it is just like allowing massless particles in a standard QFT; the major difference is just that the null cone is flat.

Oftentimes, cluster decomposition is violated when there are degenerate vacua or soft modes, and the vacuum state was not chosen correctly to be a pure state. In our case, we note that $M=0$ states are not really vacua (although they are lifted from static Galilean vacua), because local operators can then map from $M=0$ state to $M=0$ state. More importantly, they are not ground states, since they definitely carry energy $E = \omega \Delta$. Thus, in practice, the $M=0$ sector behaves like a background time-dependent 1d CFT in our theory. From this point-of-view, the interesting question is then whether the $M\neq 0$ sector can couple to this 1d CFT (this is discussed in Section \ref{sec:CouplingTo1d}).

In any case, we expect non-vacuum $M=0$ states in $\calH$ to cause complications for arguments based on factorization of the Hilbert space or particle-number conservation. In most Schr\"odinger CFTs, the only $M=0$ state is $\ket{\Omega}$ itself, and since $M$ is a superselection number, we can never fluctuate from $\ket{\Omega}$ to another state in $\calH$. If we (try to) normalize $M \sim m N$ and identify it with particle number, this is the same argument that leads to the famous particle number conservation described in Section \ref{sec:FermionsAtUnitarity}: we cannot create particles in intermediate states, greatly restricting various loop diagrams. When there are non-vacuum $M=0$ states in $\calH$, this is modified. Indeed, we expect the $M=0$ sector to have all of the features of standard relativistic QFT. For example, we expect non-genuine composite operators to be renormalized with standard QFT-type divergences (see Section \ref{sec:failureFactorization}). More intrinsically, in the language of CFT, we expect that the OPE of operators $\calO$ and $\calO^\dagger$ to not be regular if there are genuine $M=0$ states, causing some form of ``factorization breaking,'' see Section \ref{sec:CouplingTo1d}.

\section{OPEs, Non-Renormalization, and Genuine Massless Theories}\label{sec:NonRenorm}
In the previous sections we introduced the idea of genuine massive and massless operators, which act on the vacuum and create a state, and constrained their quantum numbers and correlation functions. We also asserted the existence of non-genuine local operators which annihilate the vacuum. In free theories, the non-genuine operators are composites that appear as the normal ordered product of two local operators. In a general interacting CFT, the closest notion we have to a composite operator comes from operators ``on the right hand side'' of the OPE. However, this notion also carries some scheme/definitional ambiguities. 

In this section, we argue that massless and non-genuine operators actually exist and interact with a non-relativistic CFT in meaningful ways. In Section \ref{sec:OPE} we discuss the OPE of local operators in non-relativistic theories, and some analytic properties we expect it to have. Then, in Section \ref{sec:NRTheorem}, we use this OPE to argue that there exists a canonically defined non-genuine ``composite'' local operator, obtained as the leading regular term in the OPE of a creation and annihilation operator when no massless particles exist. This can be considered the non-perturbative version of the usual perturbative non-renormalization theorems. In Section \ref{sec:GenuineMassless} we argue, by way of examples, that massless particles should exist in non-relativistic CFTs. In the examples, the non-trivial $M=0$ sector is decoupled from the rest of the CFT. Thus, in Section \ref{sec:CouplingTo1d}, we show that given an $M=0$ sector, that it is possible to couple it to the massive sector of the CFT in conformal perturbation theory, while maintaining conformality.

\subsection{The Operator Product Expansion and Analytic Continuation}\label{sec:OPE}
In any local QFT, given two local operators $\calO_1(x_1)$ and $\calO_2(x_2)$ there is an asymptotic expansion of local operators
\begin{equation}\label{eq:OPEAsymptotic}
    \calO_1(x_1) \calO_2(x_2) \sim \sum_k C_{12k}(x_{12}, \partial)\calO_k(x_2)\,.
\end{equation}
The expression \eqref{eq:OPEAsymptotic} is understood to be a property of the abstract space of observables, and thus holds in any state (possibly with some additional regularity conditions on states). We stress that the expansion is only asymptotic -- valid as $x_1 \to x_2$.

With a state-operator correspondence in Schr\"odinger CFTs, standard path integral arguments can be used to justify the convergence of the OPE of (smeared) products creation operators on the HT vacuum $\ket{\Omega}$ or any other finite norm state in the HT Hilbert space $\calH$. In particular, a product $\calO^\dagger_{f_1}\! \calO^\dagger_{f_2} \!\ket{\Omega}$ defines a finite norm state in the HT Hilbert space $\calH$, and this state can be replaced by an infinite sum of energy eigenstates for $H_{\HT}$. This infinite sum over states gives the genuine operators appearing on the right hand side of the OPE on vacuum.\footnote{As mentioned in \cite{Goldberger:2014hca}, the crucial step is therefore arguing the finiteness of norm. But their proofs pass through unchanged for operators of any $M$ so long as one considers genuine operators.} Thus we can write
\begin{equation}
    \calO_1^\dagger(x_1) \calO_2^\dagger(x_2) \ket{\Omega} = \sum_{\text{{\tiny genuine}} \, g} C_{12g}(x_{12}, \partial)\calO_g^\dagger(x_2)\ket{\Omega}\,,
\end{equation}
and similarly for $\bra{\Omega}$ and undaggered operators.

By contrast, the abstract OPE \eqref{eq:OPEAsymptotic} of any two operators necessarily contains a sum over genuine and non-genuine operators for $\Omega$. A simple scenario where this becomes very relevant is in the OPE of genuine daggered and undaggered operators, then the OPEs in \eqref{eq:OPEAsymptotic} must expand as
\begin{align}
    \calO_1(x_1) \calO_2^\dagger(x_2) 
        &\sim \sum_n C_{12^\dagger n}(x_{12}, \partial)\calO_n(x_2) + \sum_g C_{12^\dagger g}(x_{12}, \partial)\calO_g(x_2)\,,\label{eq:OPE1}\\
    \calO_2^\dagger(x_1) \calO_1(x_2) 
        &\sim \sum_n C_{2^\dagger1 n}(x_{12}, \partial)\calO_n(x_2)\,,\label{eq:OPE2}
\end{align}
where $\calO_n$ are non-genuine and $\calO_g$ are genuine. Of course, it is important to remember the non-genuine terms because they still contribute to matrix elements in non-trivial states $\ket{\Psi}$. Our claim, which we justify further in the next sections, is that these non-genuine primaries are canonically defined (essentially scheme dependent) local operators, which are dual to states $\ket{a^*}\otimes \ket{b}$ under the state-operator correspondence.

Before turning to this, let us comment on some analytic properties of the OPE. This is useful for relating different OPE channels. In our non-relativistic CFTs, the analogue of a lightlike separation between two-points is a spacelike separation. Thus we expect two OPE channels, where an operator $\calO_1$ approaches $\calO_2$ from ``above'' in real time ($t_1 > t_2$) and from ``below'' in real time ($t_1 < t_2$). At least inside correlation functions, we can see that these two different channels are related by analytic continuation. 

To see this, consider two scalar operators for simplicity. The only non-zero Wightman functions are
\begin{equation}\label{eq:nonzeroWightmans}
    \mel*{\Omega}{\calO(t,\vec{x})\!\calO^\dagger(0,\vec{0})}{\Omega}
    \quad\text{and}\quad
    \mel*{\Omega}{\calO(-t,\vec{x})\!\calO^\dagger(0,\vec{0})}{\Omega}\,,
\end{equation}
where $t>0$. All other Wightman functions vanish from $\calO$ or $\calO^\dagger$ annihilating the vacuum when $M \neq 0$. When $M=0$ these Wightman functions reduce to the familiar Wightman functions of the usual 1d CFT. Both Wightman functions in \eqref{eq:nonzeroWightmans} are just generalizations of the Schr\"odinger kernel in quantum mechanics. Explicitly, we have
\begin{equation}\label{eq:Wightman1}
    \mel*{\Omega}{\calO(t,\vec{x})\!\calO^\dagger(0,0)}{\Omega} = \frac{c}{(t-i\epsilon)^{\Delta}}\exp(i\frac{M \vec{x}^2}{2(t-i\epsilon)})\,,
\end{equation}
where the $i\epsilon$-prescription moves the branch cut in $t$ and the essential singularity from the exponential to $t = +i\epsilon$. Note that one can arrive at this equation by analytic continuation of the (well-defined) Euclidean correlator
\begin{equation}\label{eq:EuclideanCorrelator}
    \expval*{\calO(t_E,\vec{x})\calO^\dagger(0,0)} = \frac{c}{t_E^{\Delta}}\exp(-\frac{M\vec{x}^2}{2t_E})\,,
\end{equation}
while keeping the Euclidean times well-ordered $t_E >0$. 

For consistency with Hermitian conjugation, complex conjugating \eqref{eq:Wightman1} implies that
\begin{equation}\label{eq:Wightman2}
    \mel*{\Omega}{\calO(-t,x)\!\calO^\dagger(0,0)}{\Omega} = \frac{c^*}{(t+i\epsilon)^{\Delta}}\exp(-i\frac{M \vec{x}^2}{2(t+i\epsilon)})\,.
\end{equation}
Alternatively, we could compute $\mel*{\Omega}{\calO(-t,x)\!\calO^\dagger(0,0)}{\Omega}$ by analytic continuation of \eqref{eq:Wightman1} to $t<0$ through the upper half complex plane (where \eqref{eq:EuclideanCorrelator} is defined), then \eqref{eq:Wightman1} becomes
\begin{align}
    \mel*{\Omega}{\calO(e^{+i\pi}t,x)\!\calO^\dagger(0,0)}{\Omega} 
        &= \frac{c}{(e^{+i\pi} t-i\epsilon)^{\Delta}}\exp(i\frac{M \vec{x}^2}{2(e^{+i\pi}t-i\epsilon)})\\
        &= \frac{c\, e^{-i \pi \Delta}}{ (t+i\epsilon)^{\Delta}}\exp(-i\frac{M \vec{x}^2}{2(t+i\epsilon)})\,,
\end{align}
so the two agree for $c^* = c\, e^{-i\pi \Delta}$. This matches the standard Schr\"odinger kernel with $\Delta = d/2$, and the familiar monodromy in relativistic CFT with a different power of $\Delta$ from the $z=2$ scaling.

Let us note a few additional points. First, when $\Delta < d/2$, the two-point function diverges like $\sim  |t|^{\frac{d}{2}-\Delta}\,\delta^{d}(\vec{x})$ as $t\to 0$ -- this is the scalar unitarity bound. On the other hand, when $\Delta = d/2$ or $\Delta > d/2$, we recover the Schr\"odinger propagator, proportional to $\delta^{(d)}(\vec{x})$, or more general ``derivatives of $\delta$'' respectively. Finally, in relativistic CFT, we expect the commutator to be proportional to $\delta$-functions on the lightcone:
\begin{equation}
    \mel*{\Omega}{[\calO(x),\calO(0)]}{\Omega}_{\rm rel.} = i (G_R(x) - G_A(x)) \propto \delta(x^2)\,.
\end{equation}
In the non-relativistic CFT case, the essential singularity represents this already in the Wightman two-point function. Of course, this is because the Wightman function is already the commutator
\begin{equation}
    \mel*{\Omega}{[\calO(t,x),\calO^\dagger(0,0)]}{\Omega} = \mel*{\Omega}{\calO(t,x)\!\calO^\dagger(0,0)}{\Omega}\,,
\end{equation}
and the lightcone has flattened to $t=0$. Similarly, the statement that particles only propagate forward in time is reflected by
\begin{equation}\label{eq:timeOrdered}
    \mel*{\Omega}{\calT\{\calO(t,x)\!\calO^\dagger(0,0)\}}{\Omega}
    = \Theta(t)\mel*{\Omega}{\calO(t,x)\!\calO^\dagger(0,0)}{\Omega}\,.
\end{equation}

\subsection{Non-Renormalization Theorems and How They Fail}\label{sec:NRTheorem}
In free theories, a canonical example of a non-genuine operator is the $M=0$ number density operator $n(x) := (\phi^\dagger\phi)(x)$, with scaling dimension $\Delta_n = 2\Delta_{\phi}$. In interacting theories, the definition of composite operators comes with a host of scheme-dependent choices.\footnote{For simplicity, consider Euclidean QFT and work in a point-splitting scheme. There are ambiguities in defining a composite operator at the origin from $\calO_1$, $\calO_2$, and $\calO_3$: do we send $\calO_1 \!\to\! \calO_2$ and then $\calO_3$ towards the pair? Or do we send all three to the origin at the same time? The space of such point-splitting schemes is in principle as large as the space of paths in the configuration space of three points (see e.g. \cite{Hollands:2009bke}). In Lorentzian signature, the configuration space should also be further divided to include potential lightcone divergences.} However, even in the interacting Schr\"odinger CFT describing bosons at unitarity, $n(x)$ can still be uniquely defined: $n(x)$ is the coefficient of the first regular term in the $\phi^\dagger(x) \phi(0)$ OPE which does not vanish in the $x \to 0$ limit. In a general CFT, such an operator almost never exists because $\Delta_n - 2\Delta_\phi$ has no reason to be integral, but, in bosons at unitarity, $\Delta_n = 2\Delta_\phi$ because of the non-renormalization theorem(s) explained in Section \ref{sec:FermionsAtUnitarity}.

Formally, the non-renormalization argument in Section \ref{sec:FermionsAtUnitarity} only works in perturbation theory around free theories; although the general physical picture of ``no-particle production'' should hold for general fixed points. We also note that it hinges on having particle number $N$ conservation, not $M$ conservation, as mentioned in Section \ref{sec:Cluster}. With our formalism we can now give a non-perturbative version of this non-renormalization theorem. We make the following claims:
\begin{enumerate} \setlength\itemsep{0em}
    \item In Schr\"odinger CFTs with no non-trivial genuine massless operators (i.e. there are no $M=0$ states in $\calH$ that are not the vacuum), local operators are renormalized entirely in the daggered and undaggered sectors, with no anomalous dimensions between the sectors. To say it differently, the OPE of a daggered and undaggered operator is regular.
    \item As a specific case, given any two operators $\calO_1^\dagger$ and $\calO_2$, there is a canonically defined composite non-genuine local primary operator $(\calO_1^\dagger \!\calO_2)$, generalizing the number density operator, with scaling dimension $\Delta = \Delta_1 + \Delta_2$. We call this primary operator the ``normal ordered composite'' or ``normal ordering'' for obvious reasons. It is dual to $\ket{\calO_1^*}\otimes\ket{\calO_2}$ under the state-operator correspondence.
    \item In perturbation theory around \textit{any} fixed point, with no non-trivial genuine massless operators, the non-renormalization theorem holds to any order in perturbation theory.
    \item In theories with genuine massless operators, the following results are all violated in a quantifiable way -- given by correlation functions of a 1d CFT in the genuine massless sector.
\end{enumerate}

\subsubsection{Non-Renormalization from Factorization at Fixed Points}
Suppose we are in a Schr\"odinger CFT with no non-trivial genuine massless operators. We further suppose, as in the preceding sections, that the spectrum is discrete in the harmonic trap, with a good filtration on scaling dimensions (e.g. no accumulation points in scaling dimension). Here we will prove that given any two genuine operators $\calO_1$ and $\calO_2$, there is a unique/canonical non-genuine local primary operator, called the normal ordered composite or normal ordering, appearing in their OPE
\begin{equation}
\label{eq:canonicalcomposite}
    (\calO_1^\dagger\! \calO_2)(x)\,,\quad \Delta := \Delta_1 + \Delta_2\,,
\end{equation}
essentially defined as the first regular term in the OPE which does not vanish in the $x \to 0$ limit. This is essentially the non-renormalization theorem of Schr\"odinger CFTs, captured in the singular behaviour of correlation functions.

To prove the argument, we use factorization and the associativity and convergence of the non-relativistic OPE. We start by considering the matrix element
\begin{equation}\label{eq:4ptfunc}
    A_{11^\dagger22^\dagger}(x_i) := \mel*{\Omega}{\calO_1(x_1)\! \calO_1^\dagger(x_2)\! \calO_2(x_3)\! \calO_2^\dagger(x_4)}{\Omega}\,.
\end{equation}
We can expand this matrix element in two different ways using the OPE, setting the positions of the spectator operators $\calO_1(x_1)$ and $\calO_2^\dagger(x_4)$ so that the OPEs between (12), (23), and (34) are guaranteed to converge (without analytic continuation) -- in principle a time ordering would suffice $t_1 \!> \!t_2 \!>\! t_3 \!> \!t_4$. On one hand, in the (12)(34) channel, \eqref{eq:4ptfunc} becomes
\begin{equation}
    A_{11^\dagger22^\dagger}(x_i) = \sum_{\ell,r} C_{11^\dagger \ell}(z_{12}, x_{12},\partial_{2})C_{22^\dagger r}(z_{34}, x_{34},\partial_{4}) \mel*{\Omega}{\calO_\ell(x_2)\!\calO_r(x_4)}{\Omega}\,,
\end{equation}
where $\calO_\ell$ and $\calO_r$ are left and right genuine operators respectively and $z_{ij} = \vec{x}_{ij}^2/t_{ij}$ is the Schr\"odinger cross-ratio. However, since we used the $\calO_i \calO_i^\dagger$ OPEs, the genuine operators in the sum are necessarily massless genuine primaries, and (by hypothesis) the only such operator is the identity. Consequently, only the trivial term in each expansion survives and the (12)(34) OPE implies:
\begin{equation}\label{eq:NROPE1}
    A_{11^\dagger22^\dagger}(x_i) = \mel*{\Omega}{\calO_1(x_1)\!\calO_1^\dagger(x_2)}{\Omega}\!\!\mel*{\Omega}{\calO_2(x_3)\!\calO_2^\dagger(x_4)}{\Omega}\,.
\end{equation}

On the other hand, we can consider the OPE in the (23) channel, then the same matrix element \eqref{eq:4ptfunc} becomes
\begin{equation}\label{eq:NROPE2}
    A_{11^\dagger22^\dagger}(x_i) = \sum_n\frac{C_{12^\dagger n}(z_{23})}{|t_{23}|^{\Delta_1+\Delta_2 - \Delta_n}} (1 + \calD_{3}(x_{23}) + \dots) \mel*{\Omega}{\calO_1(x_1)\!\calO_n(x_3)\!\calO_2^\dagger(x_4)}{\Omega}\,,
\end{equation}
where we have expanded the OPE in slightly more detail here. Here $\calO_n$ is necessarily a non-genuine primary, since it appears in the OPE of the form $\calO_1^\dagger \calO_2$, and $\calD_{3}(x_{23})$ is the differential operator generating all of the descendants of $\calO_n$ in the OPE.

Now we can compare \eqref{eq:NROPE1} and \eqref{eq:NROPE2} as $x_{23}\to 0$. Clearly \eqref{eq:NROPE1} is regular as $x_2 \to x_3$: the correlation functions are completely factorized. For fixed $x_1$ and $x_4$, \eqref{eq:NROPE1} is non-vanishing, limiting to:
\begin{align}
    \lim_{x_2 \to x_3} A_{11^\dagger22^\dagger}(x_i)
        &= \lim_{x_2 \to x_3}\mel*{\Omega}{\calO_1(x_1)\!\calO_1^\dagger(x_2)}{\Omega}\!\!\mel*{\Omega}{\calO_2(x_3)\!\calO_2^\dagger(x_4)}{\Omega}\\
        &= \frac{c_1 c_2}{(t_{13})^{\Delta_1}(t_{34})^{\Delta_2}}\exp(\frac{i}{2}(M_1 z_{13} + M_2 z_{34}))\,.
\end{align}
Now we can compare this to the behaviour of \eqref{eq:NROPE2} as $x_{23}\to 0$. In this limit, the OPE in \eqref{eq:NROPE2} must also be regular and non-vanishing. Since our spectrum is well-behaved, we can determine two things: First, if there are any operators $\calO_n$ in the $\calO_1^\dagger\!\calO_2$ OPE such that $\Delta_n < \Delta_1 + \Delta_2$, then the matrix element $\mel*{\Omega}{\calO_1(x_1) \! \calO_n(x_3) \! \calO_2^\dagger(x_4)}{\Omega}$ is necessarily zero. Second, because of the constant term $1$ in $(1 + \mathcal{D}_3(x_{23}) + \dots)$, there is necessarily a primary operator with scaling dimension $\Delta_n = \Delta_1 + \Delta_2$ whose three-point function matches the correlation function
\begin{equation}\label{eq:MatrixElement}
    C_{12^\dagger n}(0) \mel*{\Omega}{\calO_1(x_1)\!\calO_n(x_3)\!\calO_2^\dagger(x_4)}{\Omega} = \mel*{\Omega}{\calO_1(x_1)\!\calO_1^\dagger(x_2)}{\Omega}\!\!\mel*{\Omega}{\calO_2(x_3)\!\calO_2^\dagger(x_4)}{\Omega}\,.
\end{equation}
This concludes our result.

In addition to finding an operator of scaling dimension $\Delta_1+\Delta_2$, we also were able to argue that three-point matrix elements $\mel*{\Omega}{\calO_1\!\calO_n\!\calO_2}{\Omega}$ must vanish if $\Delta_n < \Delta_1 + \Delta_2$. Other interesting constraints can be determined by using different states. For example, one could insert non-genuine operators $\calO_{n_1}$ and $\calO_{n_2}$ in \eqref{eq:4ptfunc}; repeating the same argument for
\begin{equation}
    \mel*{\Omega}{\calO_1(x_1) \! \calO_{n_1}(w_1) \!\calO_1^\dagger(x_2)\! \calO_2(x_3) \calO_{n_2}(w_2) \!\calO_2^\dagger(x_3)}{\Omega}
\end{equation}
would cause it to factorize into a product of two three-point functions, thereby constraining the $5$-point functions of non-genuine operators $\calO_n$ in the $\calO_1^\dagger \calO_2$ OPE, and so on. The four-point function $\mel*{\Omega}{\calO_1 \! \calO_1^\dagger\! \calO_2 \!\calO_2^\dagger}{\Omega}$ is just particularly distinguished because it is the ``first'' non-trivial state where the normal ordered product $(\calO_1^\dagger\! \calO_2)$ does not vanish. 

In \cite{Pyramids}, we consider these properties in more detail. In particular, we argue that normal ordered primaries generally have a pyramidal module structure under the action of the Schr\"odinger algebra. This allows us to link the vanishing of matrix elements to the exotic conservation laws of normal-ordered operators described in \cite{Bekaert:2011qd, Golkar:2014mwa}. Generally, we argue that the OPE is split into the form
\begin{equation}
    \calO_1^\dagger(x) \calO_2(0) \sim (\calO_1^\dagger\!\calO_2)(0)+\{\text{Aliens and Descendants}\}+ \{\text{Other Primaries}\}\,,
\end{equation}
where ``aliens and descendants'' are the alien operators of \cite{Bekaert:2011qd, Golkar:2014mwa}, and ``other primaries'' are terms that have vanishing matrix elements $\mel*{\mathcal{O}_1}{\cdots}{\mathcal{O}_2}$ like above. These structures also allow us to understand the emergence of various logarithms in theories with genuine massless particles and capture some properties and recursion relations of the structure functions $C_{ijk}(z)$ in the OPE. We leave details to \cite{Pyramids}.

Technically we have not shown that the $\calO_1^\dagger \times \calO_2$ OPE is regular. What we have shown is that there is a canonically defined primary $(\calO_1^\dagger\!\calO_2)$ with scaling dimension $\Delta_1 + \Delta_2$, whose ``first'' non-vanishing matrix element is as in \eqref{eq:MatrixElement}, and that all more singular terms must vanish in that matrix element and ever-more-complicated sequences of correlation functions. It would be nice to prove these points without arguing them in successively more complicated matrix elements.

Of course, $\calO_1^\dagger\!\calO_2$ should be the state-operator dual to $\ket{\calO_1^*}\otimes\ket{\calO_2}$: it has the right quantum numbers, algebraic properties, and physical interpretation. It might be useful to recast our discussion of the OPE and the non-renormalization theorem in the language of path-integrals to see this explicitly. This identification of the non-genuine operators with these non-trivial tensor product states in $\calH^* \otimes \calH$ also explains the non-renormalization theorem: operators $\calO^\dagger$ dual to states in $\calH$ can be renormalized, giving the corrections to energy levels in the harmonic trap. But once the scaling dimension of states in $\calH$ are determined, the scaling weight of states in the dual and $\calH^*\otimes \calH$ are completely determined from $\calH$ by the tensor product factorization. Moreover, the state-operator map essentially proves that all non-genuines are obtained in this way, as regular terms in the OPEs of genuines.

Finally, we note -- when there are no massless particles -- that we have a form of unitarity bound on non-genuine composite operators (even though they are not dual to states in $\calH$), obtained by simply adding the weakest unitarity bounds: \textit{for any non-genuine operator $\calO_n$ in a ($d+1$)-dimensional Schr\"odinger CFT without massless particles:}
\begin{equation}
    \Delta_n \geq d\,.
\end{equation}
The mass density operator $m(x)$ saturates the above bound. This is violated when there are massless particles, as shown in Section \ref{sec:CouplingTo1d}.

\subsubsection{Non-Renormalization for Generic Deformations}\label{sec:GenericFactor}
Having argued the regularity of daggered and undaggered OPEs at Schr\"odinger fixed points with no non-trivial genuine massless operators, we move to establish non-renormalization along deformations by non-genuine massless operators. Schematically, we deform the action by an interaction of the form:
\begin{equation}
    S_{\rm{int}}\left[\mathcal{O}_n\right]=ig\int d^d x\, dt\, \calO_n(t,x)\,,
\end{equation}
where $\calO_n$ is an $M=0$ non-genuine primary operator which is relevant or marginally relevant, i.e. $\Delta_n \leq d+2$. This of course includes normal ordered composites like the ``$\calO$-number density'' $(\calO^\dagger\!\calO)$.

As with any interaction, adding such a deformation necessarily leads to UV divergences. In this case, we can introduce a regularization scheme (say a hard cutoff in spacetime $|t-t'|, |x-y|^2>\epsilon$), and renormalize our relevant couplings and operators 
\begin{equation}
    g_i=g_i(\epsilon,g^R_i)\,,\quad 
    \mathcal{O}_R = Z_\mathcal{O}^{1/2}(\epsilon,g^{R}_i) \mathcal{O}\,,
\end{equation}
to obtain finite correlation functions, while breaking Schr\"odinger symmetry and triggering an RG flow between $z=2$ Schr\"odinger fixed points. 

Crucially, intermediate field theories along the RG flow between the Schr\"odinger fixed points will have Galilean symmetry. Moreover, the Galilean mass $M$ is central and is the same mass operator as in the Schr\"odinger field theories.\footnote{By comparison, the ``mass'' $M_{\mathrm{GCA}}$ in $z=1$ Galilean conformal theories is not the same as in the Galilean algebra, as explained also in Footnote \ref{footnote:MGCA}.} Thus $M$ is preserved and the mass quantum numbers of local operators are protected along the entire RG flow when deforming by massless operators (genuine or non-genuine). This means RG flows happen ``within'' the mass superselection sectors in \eqref{eq:SSSDecomp}, arranging the operator content of $M=m$ operators in the UV into the operator content of $M=m$ operators in the IR. In particular, this means that theories with no massless particles in the UV can only flow to theories with no massless particles in the IR.\footnote{On the other hand, we might expect theories with genuine massless particles in the UV to flow to theories without massless particles in the IR. Roughly speaking, our expectation is that a theory of pure genuine massless states behaves like a 1d CFT spread over all of space. Then, since deformations of non-trivial UV 1d CFTs can land on the trivial 1d CFT in the IR, we expect that the same is true in non-relativistic CFTs. The only exception we can forsee is if there exists some intrinsically non-relativistic anomaly that prevents such flows in the $M=0$ sector or that monotonicity-like theorems are false. We discuss some theories of genuine massless operators in Section \ref{sec:GenuineMassless}, but it would be interesting to study RG flows and monotonicity theorems in more detail.}

Consequently, if mass is conserved, the vacuum is the unique massless state, and the deforming operator $\calO_n$ is non-genuine, then we still retain the factorization property of correlation functions. 

For example, let us consider the ``minimal state'' measuring a number density in (time-ordered) perturbation theory i.e. we consider
\begin{equation}
    G_{11^\dagger22^\dagger}(x_i) := 
    \mel*{\Omega}{\calT\{\mathcal{O}_{R,1}(x_1)
        \mathcal{O}_{R,1}^\dagger(x_2)
        \mathcal{O}_{R,2}(x_3)
        \mathcal{O}_{R,2}^\dagger(x_4)\}}{\Omega}_g\,.
\end{equation}
We claim that, for certain configurations in time, this time-ordered correlation function factorizes.\footnote{We note that this is no different than the previous statement of factorization. We previously showed that the amplitude $A(x_i) = \mel*{\Omega}{\calO(x_1)\!\calO^\dagger(x_2)\!\calO(x_3)\!\calO^\dagger(x_4)}{\Omega}$ always factorizes into products of two point matrix elements, but we would never claim that the amplitude $\tilde{A}(x_i) = \mel*{\Omega}{\calO(x_1)\!\calO(x_2)\!\calO^\dagger(x_3)\!\calO^\dagger(x_4)}{\Omega}$ factorizes. Both matrix elements appear in a time-ordered correlation function, so the factorization of time-ordered correlators only occurs at distinguished times.} As a particular example, when $t_1 > t_2 > t_3 > t_4$, we claim that
\begin{equation}
    \label{eq:GeneralFactorization}
    G_{11^\dagger22^\dagger}(x_i)=\mel*{\Omega}{\calT\{\mathcal{O}_{R,1}(x_1)
        \mathcal{O}_{R,1}^\dagger(x_2)\}}{\Omega}_g\!\!\mel*{\Omega}{\calT\{\mathcal{O}_{R,2}(x_3)
        \mathcal{O}_{R,2}^\dagger(x_4)\}}{\Omega}_g\,.
\end{equation}
In perturbation theory, this can be seen order-by-order by repeating the factorization arguments of the previous section.

Let us start at first-order in perturbation theory, then we are interested in a term like:
\begin{equation}
    O(g^1): \quad \mel*{\Omega}{\calT\{\mathcal{O}_{1}(x_1)
        \mathcal{O}_{1}^\dagger(x_2)
        \mathcal{O}_{2}(x_3)
        \mathcal{O}_{2}^\dagger(x_4)\mathcal{O}_n(y_1)\}}{\Omega}\,.
\end{equation}
The only three non-vanishing terms in this case are when $\calO_n$ is between $\calO_1$ and $\calO_1^\dagger$; $\calO_1^\dagger$ and $\calO_2$; or $\calO_2^\dagger$ and $\calO_2$. However, in the matrix element where $\calO_n$ is between $\calO_1^\dagger$ and $\calO_2$, we can insert a complete set of states:
\begin{align}
    \bra{\Omega}\!\mathcal{O}_{1}(x_1)
        \mathcal{O}_{1}^\dagger(x_2)
        &\mathcal{O}_n(y_1)
        \mathcal{O}_{2}(x_3)
        \mathcal{O}_{2}^\dagger(x_4)\!\ket{\Omega}\\
        &= \sum_{\psi} \mel*{\Omega}{\mathcal{O}_{1}(x_1)
        \mathcal{O}_{1}^\dagger(x_2)\calO_n(y_1)}{\psi}
        \!\!\mel*{\psi}{
        \mathcal{O}_{2}(x_3)
        \mathcal{O}_{2}^\dagger(x_4)}{\Omega}\\
        &= \mel*{\Omega}{\mathcal{O}_{1}(x_1)
        \mathcal{O}_{1}^\dagger(x_2)\calO_n(y_1)}{\Omega}
        \!\!\mel*{\Omega}{
        \mathcal{O}_{2}(x_3)
        \mathcal{O}_{2}^\dagger(x_4)}{\Omega}\\
        &=0\,.
\end{align}
In going from the second line to the third line, we used the fact that $\psi$ must have $m=0$ to ensure that $\mel*{\psi}{\calO_2 \calO_2^\dagger}{\Omega}$ does not vanish. But then, since $\ket{\Omega}$ is the unique massless state, we have $\calO_n(y_1)\ket{\Omega} = 0$. Altogether, the only two surviving terms at first-order in perturbation theory are those where $\calO_n$ sits between $\calO_i$ and $\calO_i^\dagger$, and then the expression factorizes as before.

Now the only non-trivial work left is to argue that the same factorization persists at higher orders in perturbation theory, and that terms regroup to give \eqref{eq:GeneralFactorization}. In this case, we are interested in terms like
\begin{equation}
    O(g^m): \quad \mel*{\Omega}{\calT\{\mathcal{O}_{1}(x_1)
        \mathcal{O}_{1}^\dagger(x_2)
        \mathcal{O}_{2}(x_3)
        \mathcal{O}_{2}^\dagger(x_4)\mathcal{O}_n(y_1)\cdots \mathcal{O}_n(y_m)\}}{\Omega}\,.
\end{equation}
As before, non-vanishing matrix elements have $\calO_n$ sitting between $\calO_i$ and $\calO_i^\dagger$. Working out the combinatorial factors, we can choose $k$ of $m$ of them to go inbetween $\calO_1$ and $\calO_1^\dagger$, then the remaining $m-k$ must go inbetween $\calO_2$ and $\calO_2^\dagger$. Putting all the terms together, and dropping integration over internal vertices for brevity, we find that
\begin{align}
    G_{11^\dagger22^\dagger}(t_i > t_{i+1})
        &= 
    \sum_{m=0}^\infty \frac{1}{m!} \mel*{\Omega}{\calT\{\mathcal{O}_{1}(x_1)
        \mathcal{O}_{1}^\dagger(x_2)
        \mathcal{O}_{2}(x_3)
        \mathcal{O}_{2}^\dagger(x_4)\mathcal{O}_n(y_1)\cdots \mathcal{O}_n(y_m)\}}{\Omega}\\
        &= \sum_{m=0}^{\infty} \sum_{k=0}^m \frac{1}{m!} \binom{m}{k} \mel*{\Omega}{\mathcal{O}_{1}\calO_n^k
        \mathcal{O}_{1}^\dagger
        \mathcal{O}_{2}\calO_n^{m-k}
        \mathcal{O}_{2}^\dagger}{\Omega}\\
        &= \sum_{k,\ell=0}^{\infty} \frac{1}{k!\ell!} \mel*{\Omega}{\mathcal{O}_{1}\calO_n^k
        \mathcal{O}_{1}^\dagger
        \mathcal{O}_{2}\calO_n^{\ell}
        \mathcal{O}_{2}^\dagger}{\Omega}\\
        &=\mel*{\Omega}{\calT\{\mathcal{O}_{1}(x_1)
        \mathcal{O}_{1}^\dagger(x_2)\}}{\Omega}_g\!\!\mel*{\Omega}{\calT\{\mathcal{O}_{2}(x_3)
        \mathcal{O}_{2}^\dagger(x_4)\}}{\Omega}_g\,.
\end{align}

As always in perturbation theory, when we pull down vertices from the interaction in the exponential, we will have divergences from points when $\calO_n$ collides with one of the other operators, and so must regulate all of the integrals. Our argument that
\begin{equation}
    G_{11^\dagger22^\dagger}(t_i>t_{i+1})=\mel*{\Omega}{\calT\{\mathcal{O}_{R,1}(x_1)
        \mathcal{O}_{R,1}^\dagger(x_2)\}}{\Omega}_g\!\!\mel*{\Omega}{\calT\{\mathcal{O}_{R,2}(x_3)
        \mathcal{O}_{R,2}^\dagger(x_4)\}}{\Omega}_g
\end{equation}
implies that the $x_2 \to x_3$ limit is regular, even in perturbation theory, and that there exists a scheme so that the wavefunction renormalization of the composite operator is
\begin{equation}
    Z^{1/2}_{(\mathcal{O}_1^\dagger\mathcal{O}_2)}(\epsilon,g^R)=Z^{1/2}_{\mathcal{O}_1}(\epsilon,g^R)Z^{1/2}_{\mathcal{O}_2}(\epsilon,g^R)\,.
\end{equation}
This implies the non-renormalization of the canonical composites of the lowest weight operators, but one would have to work slightly harder and deal with operator mixing for higher-weight operators.

\subsubsection{Failure of Factorization in Theories with Massless States}\label{sec:failureFactorization}
All of the preceding arguments fail when the theory has non-trivial genuine massless states. Qualitatively, this is because all of the arguments hinge on conservation of particle number, not conservation of mass.

Quantitatively, in OPEs, the contribution of genuine massless operators and their conformal blocks can be resummed into usual $\mathfrak{sl}(2,\bbR)$ conformal blocks with modified $z=2$ scaling. In particular, the four-point function
\begin{equation}
    \expval*{\calO(x_1)\! \calO^\dagger(x_2)\! \calO(x_3)\! \calO^\dagger(x_4)}
\end{equation}
of external operators of dimension $\Delta_{\phi}$ and mass $m$, decomposes over terms of the form:
\begin{equation}
    \label{eq:masslesschannel}
    \frac{1}{(t_{12}t_{34})^{\Delta_\phi}}
    e^{im\left(z_{12}+z_{34}\right)/2} \times 
    \left(\frac{t_{12} t_{34}}{t_{23}t_{14}}\right)^{\frac{\Delta}{2}} \,_2F_1\left(\frac{\Delta}{2},\frac{\Delta}{2},\Delta,-\frac{t_{12}t_{34}}{t_{23}t_{14}}\right)\,.
\end{equation}
Here, the first term and exponential are the usual universal two-point contributions that are typically stripped/factored out, and the trailing terms are 1d conformal blocks, which are a function of the 1d conformal cross-ratio $t_{12} t_{34}/ t_{23} t_{14}$. If the coefficient on any (non-identity) block is non-zero, then factorization and cluster decomposition (in space) is obviously broken for the correlation function.

The ``s-channel conformal block'' in \eqref{eq:masslesschannel} is obtained by using the (12)(34) OPE. In the (23) ``cross-channel,'' the OPE $\calO^\dagger(x_2) \times \calO(x_3)$ includes a sum over non-genuine massless operators, and as $t_{23} \to 0$ we obtain an analog of a lightcone limit with characteristic logarithmic divergences in the $\mathfrak{sl}(2,\bbR)$ conformal blocks \cite{Simmons-Duffin:2016wlq}. As always, the whole infinite sum of logarithms in the cross-channel should be resummed to renormalize the non-genuine operator in the OPE. Pushing this analogy further, we expect non-genuine operators to behave like double-twist operators (indeed, we see that their form is algebraically the same) so that the identity block in the (12)(34) OPE demands the non-genuine operators in (23). Then, the same way that $1/\ell$ corrections to double-twist scaling dimensions appear from the first non-trivial operator, we see that the first non-trivial genuine massless operator corrects the scaling dimensions of the non-genuine operators here. We discuss these points in more detail in Section \ref{sec:CouplingTo1d} and \cite{Pyramids}.

\subsection{Writing Genuine Massless Theories}\label{sec:GenuineMassless}
In the previous section, we discussed non-renormalization theorems that follow from the factorization of the Schr\"odinger CFT Hilbert space. We also saw how this was violated if the theory contained non-trivial massless states in $\calH$. However, there are now two concerns:
\begin{enumerate}\setlength\itemsep{0em}
    \item Do theories with non-trivial genuine massless states exist in any suitable sense?
    \item If such theories exist, can they actually be coupled to/interact with the more familiar Schr\"odinger CFTs which do not contain any $M=0$ states?
\end{enumerate}
In this section, we comment on the first of these two questions by arguing the existence of theories with $M=0$ states and/or non-trivial genuine massless operators. Then, in Section \ref{sec:CouplingTo1d}, we will argue that an abstract massless sector can be coupled to a theory without a massless sector in a consistent way in perturbation theory.

In Section \ref{sec:MasslessLimit} we provide some arguments that non-relativistic limits of relativistic theories with massless particles should contain massless particles, but we expect difficulties for Lagrangian methods. In example \ref{sec:NullReduction} we show how massless states can emerge in null reductions, commensurate with our picture from Section \ref{sec:HarmonicTrap}. In Example \ref{sec:FreeBoson}, we show that the non-relativistic free boson does \textit{not} have genuine massless operators, as diagnosed by its four-point functions. Finally, in Example \ref{sec:BootstrapApproach}, we try to design an abstract GFF-like theory of genuine massless states by appealing to bootstrap-like axioms.

\subsubsection{Comments for Non-Relativistic Limits and Lagrangians}\label{sec:MasslessLimit}
As mentioned in the introduction, one way to obtain Schr\"odinger field theories is as the non-relativistic limit of a relativistic field theory. For example, starting with a free field $\Phi$ of mass $m$, a typical strategy is to separate the field into massive plane waves with creation and annihilation modes on top (we reinstate $c$ but leave $\hbar=1$):
\begin{equation}
    \Phi(t,\vec{x}) = \frac{1}{\sqrt{2m}}\left(e^{-imc^2t}\phi(t,\vec{x}) + e^{imc^2t}\phi^\dagger(t,\vec{x})\right)\,,
\end{equation}
and then assume that a majority of the energy is in the rest energy, i.e. $\partial^2_t \phi \ll -2imc^2 \partial_t \phi$ \cite{Baiguera:2023fus}. In this limit, we recover the free Schr\"odinger Lagrangian
\begin{equation}\label{eq:NRLag2}
    \calL_{0} = \phi^\dagger\left(i\partial_t + \tfrac{1}{2m}\nabla^2\right) \phi\,.
\end{equation}
Interactions can also be included. Such things were considered very systematically in \cite{Jensen:2014wha}.

In the preceding construction, the $m \to 0$ limit is not obviously well-defined or even unique. Indeed, in the ultrarelativistic/Carrollian limit of QFTs, one typically finds ``electric'' and ``magnetic'' limits \cite{Henneaux:2021yzg}, neither of which is preferred from an intrinsic viewpoint \cite{Ciambelli:2023xqk}, and both of which pose questions for quantization \cite{Cotler:2024xhb}. Likewise, in the Lagrangian \eqref{eq:NRLag2}, we might anticipate that the massless Schr\"odinger kinetic term possesses temporal and spatial limits\footnote{Further connections between the Schr\"odinger and Carroll algebras are discussed in \cite{Afshar:2024llh, Najafizadeh:2024imn}.}
\begin{equation}
    \calL_1 = \phi^\dagger \partial_t \phi
    \quad\text{or}\quad
    \calL_2 = \phi^\dagger \nabla^2 \phi\,.
\end{equation}
On one hand, our abstract non-Lagrangian results in Section \eqref{sec:PositiveUIRs} suggest that the first option, describing a time-dependent and spatially isotropic massless field, is preferred, as all massless particles are spatially constant. On the other hand, $\calL_1$ corresponds to $\calH_1 = 0$, so the naive $\calL_1$ is too trivial. 

We actually expect it to be non-trivial to write local covariant Lagrangians for genuine massless particles. If we want to produce states/particles which are independent of space, then a natural choice is to consider spatially constant fields $\phi(t,\vec{x}) = \phi(t)$, as mentioned above. In this sense, genuine massless fields are described by some ``mini-superspace'' if they exist. In usual relativistic CFT, the mini-superspace limit is actually rather useful for isolating technical issues of zero-modes in non-compact WZW models \cite{Teschner:1997fv}. However, one moderate side effect of assuming spatially-independent fields is that the action grows with the volume of space
\begin{equation}
    S_0 = \int_{M\times{\bbR}} \!\!\!dt\, d^d\vec{x}\, \calL(\phi(t)) = \mathrm{vol}(M) \int dt\, \calL(\phi(t))\,,
\end{equation}
and, in non-relativistic CFTs, the plane and the harmonic trap are both spatially non-compact. 

Disregarding this divergence, a more important hurdle is that we are now left with writing a Lagrangian/Hamiltonian for a 1d CFT in time $\calL(\phi(t))$.\footnote{Here there is some potential for terminological confusion. By 1d CFTs we mean 1d conformal field theories, i.e. they have a unique $\mathfrak{sl}(2,\bbR)$-invariant vacuum state, the vacuum state is the ground state of the Luscher-Mack Hamiltonian $H_{\mathrm{LM}} = \frac{1}{2}(P_0 + K_0)$, and the Hilbert space is an infinite collection of $\mathfrak{sl}(2,\bbR)$ lowest-weight modules etc. This should be contrasted to conformal quantum mechanics or AFF models \cite{deAlfaro:1976vlx, Chamon:2011xk}. These are theories of $\mathfrak{sl}(2,\bbR)$-covariant quantum mechanics, i.e. the Hilbert space is a direct sum of some (possibly one!) $\mathfrak{sl}(2,\bbR)$ reps, and $D$ symmetry is generally broken. In these theories, the generator $P_0 + K_0$ is still distinguished because it generates the compact subgroup of $SL(2,\bbR)$. A (ground) state annihilated by $P_0 + K_0$ is a ``spherical vector,'' which may or may not be normalizable depending on which reps appear in the spectrum (see \cite{Teschner:1997fv, Gaiotto:2023hda} for discussion).
} In 1d CFT, the conformal Ward identities imply that the stress-tensor vanishes $T = T_{00} = 0$, leading to theories that are either topological or non-local -- like GFF or a defect CFT. 

Here we have an out to this problem, which is also consistent with our physical understanding. In particular, in higher dimensions, the non-relativistic scale Ward identities only demand that
\begin{equation}
    z T_{00} = T^i_i\,.
\end{equation}
Hence, we can imagine that there is a spatially constant pressure in all of space, which allows the stress-tensor to be non-zero, but otherwise ``behaving like'' a 1d CFT in time kinematically. This is commensurate with our picture of genuine massless states. We take steps towards such a construction in Section \ref{sec:BootstrapApproach}.

We can also consider non-relativistic limits of propagators. Consider the massive relativistic propagator for a free scalar in ($d+1$)-dimensions
\begin{equation}
    G_m(t,\vec{x}) = \int \frac{d^d\vec{p}}{(2\pi)^d}\frac{e^{-i (E_{\vec{p}} t - \vec{p}\cdot\vec{x})}}{2E_{\vec{p}}}\,,
\end{equation}
where $E_{\vec{p}} = \sqrt{m^2c^4 + \vec{p}^2c^2}$ is the on-shell energy. Expanding in $c^{-2}$, the energy famously goes as $E_{\vec{p}} = mc^2 + \frac{\vec{p}^2}{2m} + O(c^{-2})$ and the propagator becomes
\begin{equation}
    G_m(t,\vec{x}) = \frac{e^{-i mc^2t}}{2mc^2}G_{\mathrm{Sch.},m}(t,\vec{x})\,,
\end{equation}
where $G_{\mathrm{Sch.},m}$ is the usual real-time Schr\"odinger kernel with mass $m$, i.e. the non-relativistic two-point function. Unlike the Lagrangian examples, we can actually consider the same expansion for the massless propagator:
\begin{equation}
    G_{0}(t,\vec{x}) 
        = \frac{1}{4\pi^2}\frac{1}{-c^2t^2+|\vec{x}|^2}
        =\frac{-1}{4\pi^2c^2}\left(\frac{1}{t^2} +O(c^{-2}) \right)\,,
\end{equation}
where we clearly recover the massless Schr\"odinger propagator $\sim t^{-\Delta}$ in the $c \to \infty$ expansion. Thus we believe a consistent non-relativistic limit must retain both massive and massless modes in the NR CFT description. Higher order corrections in $c^{-1}$ can be considered as perturbations away from the NR fixed point. Based on this, we expect that a consistent treatment of the $c\to\infty$ limit of a relativistic theory with massive and massless states, e.g. scalar QED, will look like a Schr\"odinger CFT with genuine massless particles, see also \cite{Chapman:2020vtn}.

\subsubsection{Example: Null Reduction}\label{sec:NullReduction}
Let us start with a general scalar primary operator $\calO$ of dimension $\Delta$ in a relativistic CFT. The Wightman two-point function is obtained as the boundary value of a Euclidean two-point function with a specific $i\epsilon$-prescription:
\begin{equation}
    \mel*{\Omega}{\calO(x_1)\!\calO(x_2)}{\Omega} 
        = \frac{1}{(x_{12}^2+i \epsilon x^0_{12})^\Delta} \,.
\end{equation}

We wish to null-reduce the correlator to obtain a Schr\"odinger CFT correlator. Kinematically, this is a conformal analogue of the famous relationship between null-reduction of Poincar\'e invariant theories and Galilean systems (see \cite{NullDefects} for more discussion). To this end, we introduce lightcone coordinates $(x^\pm,x^\perp)$ where
\begin{equation}
    x^{\pm} := x^0 \pm x^1\,.
\end{equation}
For example, in these coordinates, the Wightman two-point function is
\begin{equation}
    G_>(x_1,x_2):=\mel*{\Omega}{\calO(x_1)\!\calO(x_2)}{\Omega} 
        = \frac{1}{(-x_{12}^- x_{12}^+ + (x_{12}^\perp)^2 + i \epsilon x^0_{12})^\Delta}\,.
\end{equation}
To get a correlator in the null reduction, we compactify the null-direction $x^+ \sim x^+ + L$.\footnote{The cautious reader may worry about how we make sense of the null identification $L$. Null reductions were studied as ultraboosted limits of spacelike compactifications in \cite{Seiberg:1997ad}, but a completely satisfactory treatment is still unknown to us.} In this case, the two-point function $G_{>}(x)$ becomes
\begin{equation}
    \tilde{G}_{>}(x_{12}^+, x_{12}^-, x_{12}^\perp) = \sum_{m\in\mathbb{Z}}\frac{1}{(-x_{12}^-(x^+_{12}+mL)-z_{12}-i\epsilon (x_{12}^0 + m L))^{\Delta}}\,.
\end{equation}
Formally, the above should be resummed into some Hurwitz zeta functions for generic $\Delta$, depending on the $i\epsilon$ prescription.

However, we really want to know if there can ever be a two-point function of $m=0$ modes. Thus, in principle, we want to consider the integral
\begin{equation}\label{eq:nullReduction}
    \int_{-\infty}^{\infty} dx^+ G_{>}(x^+, x^-,z)\,.
\end{equation}
It is at this point that we make the following observation: while it is kinematically true that the correlation functions of the null reduction behave like Schr\"odinger correlation functions, it is not presently clear what the right causal structure is (but see \cite{Herzog:2008wg}). For example, it would be surprising if the null reduction of the time ordered correlation function in ($d+2$)-dimensions is related to a meaningful correlation function in the Schr\"odinger CFT, since we wouldn't expect it to have the right causal structure. Moreover, it is also unclear if the Schr\"odinger field theory correlation functions should use the Minkowski CFT vacuum $\ket{\Omega}$ or the lightcone vacuum $\ket{\Omega_{\mathrm{LC}}}$ -- although the two are believed to agree in free examples.

With this in mind, if we actually null reduce the Wightman correlator, then \eqref{eq:nullReduction} can be viewed as the lightray transform of one of the operators in the Wightman function and vanishes when $\Delta > 1$. For $\Delta=1$ we do recover 
\begin{equation}
    \expval*{\calO_{0}(x^-)\!\calO_{0}(0,\vec{x})} = \frac{a}{x^-}\,,
\end{equation}
for some constant $a$; matching our expectation for a genuine massless operator. Finally, for $\Delta < 1$ we expect the integrals to be badly divergent. Up to some lightplane supported divergences, this result is essentially independent of which $i\epsilon$ prescription is chosen for null-reduction, and different choices effectively just change the causal structure of the Schr\"odinger theory while keeping the $\sim (x^-)^{-1}$ scaling for $\Delta=1$. For higher-point functions, more care should be exercised.

Taking all of the above into consideration, the massless free scalar in $(3+1)d$ is an example of a CFT which is free, so ostensibly $\ket{\Omega} = \ket{\Omega_{\mathrm{LC}}}$, and has a primary in the spectrum with $\Delta=1$ and none of scaling dimension $\Delta < 1$. Consequently, we expect that the null reduction of the ($3+1$)d free scalar has a KK tower of operators of mass spacing
\begin{equation}
    \Delta m \sim L^{-1}
\end{equation}
and a genuine massless operator of scaling dimension $\Delta = 1$ in the spectrum.\footnote{We also speculate that the null reduction of ($3+1$)d conformal gauge theories will have similar properties, and further speculate that the null reduction of non-abelian conformal gauge theories would lead to sectors of interacting genuine massless operators. We leave studies of such examples to future works.}

\subsubsection{Non-Example: Free Boson and \texorpdfstring{$u(x) := (\phi\phi^\dagger)(x)$}{u(x) = phiphi**dag}}\label{sec:FreeBoson}
Consider the Schr\"odinger field theory describing a free boson in ($d+1$)-dimensions with mass $m$ and scaling dimension $\Delta_{\phi} = d/2$. A Lagrangian is given in \eqref{eq:NRLag2}. 

From the point of view of the non-relativistic limit or textbook QFT, we know that the field $\phi$ (and $\phi^\dagger$) can be used to generate all operators in the theory: by considering words built from $\phi$ and its derivatives. However, naive products are obviously divergent and require a weak renormalization in the form of normal-ordering -- putting all daggers ``on the left.'' In particular, we know that $\phi$ and $\phi^\dagger$ exist and act non-trivially on the vacuum on the left and right respectively, and we also know that the number operator $n(x) = (\phi^\dagger\phi)(x)$ is sensible and annihilates the vacuum on the left and on the right. And finally, we know that strings such as $u(x) = (\phi\phi^\dagger)(x)$ are meaningless and UV divergent.

In our abstract point of view we did not define daggered and undaggered operators as particular fields in a Lagrangian. Instead, we defined them by a state-operator correspondence. Moreover, we did not define $n(x)$ from some procedure where ``all the daggers are moved to the left,'' but as the regular term in the $\phi^\dagger \phi$ OPE. In this vein, one might worry that the $\phi \phi^\dagger$ OPE,
\begin{equation}
    \phi(x) \phi^\dagger(0) 
        \sim \sum_n C_{\phi\phi^\dagger n}(x, \partial)\calO_n(0) + \sum_g C_{\phi\phi^\dagger g}(x, \partial)\calO_g(0)\,,
\end{equation}
contains a genuine massless operator $u(x)$ of scaling dimension $\Delta_u = \Delta_{1} + \Delta_{2}$. 

However, the theory obviously does not contain a $u(x)$ that couples to the rest of the theory, because the four-point function of the free theory factorizes. To see this, recall that $\phi$ and $\phi^\dagger$ are like half of the modes in a relativistic free-scalar, and so Wick's theorem means that
\begin{equation}
    \mel*{\Omega}{\phi(x_1)\phi^\dagger(x_2)\phi(x_3)\phi^\dagger(x_4)}{\Omega}
        = \mel*{\Omega}{\phi(x_1)\phi^\dagger(x_2)}{\Omega}
        \!\!\mel*{\Omega}{\phi(x_3)\phi^\dagger(x_4)}{\Omega}\,.
\end{equation}
If $u(x)$ existed and coupled to the rest of the theory, this would not be possible. Another way to say this is that genuine massless operators deform the commutator $[\phi^\dagger, \phi]$.

\subsubsection{Example: A Boostrap Approach to a Genuine Massless GFF}\label{sec:BootstrapApproach}
In addition to the null example, we can also try to build a genuine massless theory by simply trying to satisfy the conformal bootstrap axioms in non-relativistic space -- specifying a CFT by literally writing all of its conformal data. In this vein, the simplest thing to do is to try to define an analogue of a generalized free theory. 

In this case, our local operators will consist of words made from the field $\phi$ and its temporal derivatives, i.e.
\begin{equation}
    \calA := \{\phi\,, \phi\phi\,, \partial_t\phi\,, \phi\partial_t \phi\,, \dots \}\,.
\end{equation}
In order to have an analogue of GFF, we can assert that $\phi$ behaves like a primary of scaling dimension $\Delta$
\begin{equation}
    \expval*{\phi(t_1,x_1)\phi(t_2,x_2)} := \frac{1}{(t_{12})^{\Delta}}\,,
\end{equation}
and all other correlation functions follow from Wicks theorem and taking derivatives. This defines the correlation functions of every local operator and we are done.

However, such an example is not necessarily ``local.'' In a Schr\"odinger CFT, to be local we expect to have a conserved primary stress tensor operator $T_{\mu\nu}$ of scaling dimension $\Delta_T = d+2$ and transforming like a symmetric operator under $SO(d)$. Moreover, the conformal Ward identity for $z=2$ Lifshitz scaling demands that
\begin{equation}
    2 T_{00} = {T^i}_i\,.
\end{equation}
In our genuine $M=0$ theory, we have topological invariance in space, so we expect that the spatial components $T_{0i}$ are zero. In a general local Schr\"odinger CFT we should also have a mass current operator $(n(x), j_i(x))$. However, since we are in the $M=0$ sector, such operators should just be identically zero. Now the question is, can we actually find a conserved primary stress tensor $T_{\mu\nu}$ satisfying the conditions above? 

Let us work in (1+1)d for simplicity and see if we can find an example where Wick's theorem still holds (a free example). Then we assert that $T_{11} = 2 T_{00}$ and the only condition is to find a $T_{00}$ which has scaling weight $3$ and satisfies $\partial_t T_{00} = 0$. First we tune the scaling weight. A straightforward computation shows that we only have a primary operator when $\Delta = \frac{1}{2}$ and
\begin{equation}
    T_{00} = \phi \partial_t \phi - \partial_t \phi \phi\,.
\end{equation}
Such a quantity is classically zero and quantum mechanically a $c$-number, and so is a-priori far more trivial than we would like. Moreover, $\partial_t T_{00} = 0$ must still be enforced by hand.

Instead we could consider upgrading our previous example, by introducing a complex bosonic field $\phi$, so that
\begin{equation}
    \expval*{\phi^*(t_1,x_1)\phi(t_2,x_2)} := (t_{12})^{-1/2}\,,
\end{equation}
and all previous assertions hold. Now a non-zero stress tensor in (1+1)d can be specified by identifying
\begin{equation}
    T_{00} = \phi^* \partial_t \phi - \partial_t \phi^* \phi\,.
\end{equation}
We still must verify conservation of the stress tensor.  We could then try to assert $\partial_t T_{00} = 0$, and interpret the vanishing as defining an ``equation of motion'' on our complex field or a shortening condition for the $\phi$ multiplet.

It would be interesting to push such analyses further to see if one could bootstrap -- even by hand -- some kinematically acceptable non-relativistic CFTs of massless operators. In principle, such solutions would be no different than finding solutions to the 1d conformal bootstrap, perhaps with some change of scaling dimensions.

\subsection{Coupling to a Massless Sector}\label{sec:CouplingTo1d}
In the previous sections we considered the implications of having genuine massless operators, and also argued that they may exist in some -- presumably non-Lagrangian -- situations. However, in our examples, the genuine massless operators were usually decoupled from the rest of the Schr\"odinger CFT. In this section, we constrain -- by way of example -- the possibility of coupling genuine massless operators to a theory while preserving conformality, at least in conformal perturbation theory.

Abstractly, we will start by considering a simple product/stack theory
\begin{equation}
    \calT = \calT_A \times \calT_0\,.
\end{equation}
We will assume that $\mathcal{T}_A$ is a ``vanilla'' Schr\"odinger CFT in ($d+1$) dimensions with no genuine massless operators, e.g. bosons or fermions at unitarity. $\calT_0$ will be one of our previous putative theories of entirely massless states. Then we will turn on a classically marginal and Schr\"odinger symmetric coupling between the two theories and try to tune it to be exactly marginal in conformal perturbation theory.

In particular, we start with a coupling of the form
\begin{equation}
    S_{\mathrm{int.}}[\calO_A,a]=g\int dt \,d^dx\, \mathcal{O}_A (t,\vec{x}) \,a(t)\,,
\end{equation}
where $\calO_A$ is a non-genuine local operator in $\calT_A$ and $a$ is a genuine massless operator in $\calT_0$. Such a deformation is rather physical, and can be viewed as turning on a generalized time-dependent ``chemical potential'' for $\calO_A$, generalizing the $\mu_\sigma |\psi_\sigma|^2$ term in \eqref{eq:fourfermi}.

Since we are specifically interested in finding Schr\"odinger CFTs with an interacting zero mass sector, we should demand the deformation to be classically marginal
\begin{equation}
    \Delta_{\mathcal{O}_A}+\Delta_a=d+2\,.
\end{equation}
A particularly nice choice is to take $\calO_A$ to be the mass density $m(x)$ (equivalently, number density, since $\calO_A$ has no non-trivial zero mass states), so that
\begin{equation}
\label{eq:marginalint}
    S_{\mathrm{int.}}[a]=g\int_{\mathbb{R}_t} dt\, M(t) a(t)\,,
\end{equation}
where $M$ is the mass operator in the Schr\"odinger algebra \eqref{eq:SchrodingerAlgebra} of $\mathcal{T}_A$, which always exists and enjoys topological properties by virtue of being conserved. Classical Schr\"odinger invariance then dictates that $\Delta_a=2$.

\subsubsection{2-Point Functions: Conformal Dimensions and Marginality}
The first step is to check whether the aforementioned deformation \eqref{eq:marginalint} is marginal quantum mechanically, seeing if it preserves the form of conformal two-point functions.\footnote{Again, we note that deformations like \eqref{eq:marginalint} are still interesting even if they are marginally relevant, we are just interested in trying to find a Schr\"odinger CFT that has interacting massless modes.} We will regularize with a hard cutoff $\epsilon$ in time, and assume that spatial cutoffs are not needed since everything is topological in space in \eqref{eq:marginalint}. Our renormalization scheme will be the ``conformal scheme'': in this scheme, correlation functions are defined to look like a conformal correlation function times a series in $g$ and logs. If the perturbed system is \textit{actually} conformal, this sum over $g$ and logs can be viewed as an anomalous shift in the scaling dimensions. This makes the scheme particularly useful for trying to find marginal deformations, and also constrains CFT coefficients, as we will see.

We are interested in calculating 2-point functions of operators $\calO$ of mass $m$ and dimension $\Delta$ in the original theory $\calT_A$. To that end, we define the renormalized operators $\calO$ and use $\calO^{(0)}$ for the original operators; we write
\begin{equation}
\label{eq:renormalizedfields}
    \calO(x)
        = Z^{\frac{1}{2}}_{\calO}(\epsilon,g) \calO^{(0)}(x)\,.
\end{equation}
The 2-point function of interest here will therefore be,
\begin{equation}
    \expval*{\calO(x)\!\calO^\dagger(0)}_g
        = Z_{\calO}(\epsilon,g)\expval*{\calO^{(0)}(x)\!\calO^{\dagger(0)}(0) \,e^{ig\int dt\, M(t)a(t)}}\,.
\end{equation}

If we expand the expectation value on the RHS to order $O(g^2)$, we get
\begin{align}
    \langle\calO^{(0)}(x)&\!\calO^{\dagger(0)}(0) \, e^{ig\int dt\, M(t)a(t)}\rangle_g\nonumber\\
        &= \expval*{\calO^{(0)}(x)\!\calO^{\dagger(0)}(0)}
        + ig \int dt_1 \, \expval*{\calO^{(0)}(x)\!\calO^{\dagger(0)}(0) M(t_1)a(t_1)}\\
        &- \frac{g^2}{2} \int dt_1\, dt_2\, \expval*{\calO^{(0)}(x)\!\calO^{\dagger(0)}(0) M(t_1)a(t_1)M(t_2)a(t_2)} + O(g^3)\nonumber\,.
\end{align}
Since we start in a product theory, the correlation function at $O(g)$ vanishes by $\expval*{a(t_1)} = 0$ and we only have the $O(g^2)$ term. At $O(g^2)$ we use our sharp cutoff scheme to find
\begin{align}
    \expval*{\calO^{(0)}(x)\!\calO^{\dagger(0)}(0) \, e^{ig\int dt\, M(t)a(t)}}\bigg\vert_{O(g^2)}
        = m^2 g^2 \left(\log(\tfrac{t}{\epsilon})-2 + \tfrac{t}{\epsilon}\right)\,.
\end{align}
Now we can absorb the $t$-independent parts into the wavefunction renormalization
\begin{equation}
    Z_{\calO}^{\frac{1}{2}}(\epsilon,g)=1-\frac{m^2g^2}{2}\left(\log(\epsilon)-2\right)+O(g^3)\,,
\end{equation}
while the $t$-dependent parts require the addition of the counterterm
\begin{equation}
   S_{\mathrm{ct}}= -i\frac{g^2}{\epsilon}\int dt \,M^2(t)+O(g^3)\,.
\end{equation}

Altogether, at $O(g^2)$ equation \eqref{eq:renormalizedfields} becomes
\begin{equation}
    \label{eq:2-ptexample}
    \expval*{\calO(x)\!\calO^{\dagger}(0)}_g = \frac{e^{i m \frac{\vec{x}^2}{2t}}}{t^{\Delta}}\left(1+ m^2 g^2 \log(t) + O(g^4)\right)\,.
\end{equation}
As previously mentioned, we interpret the trailing terms as a shift of the conformal dimension via ($\gamma(0)=0$)
\begin{equation}\label{eq:deltaExpansion}
    t^{-\Delta + \gamma(g)} = t^{-\Delta + \gamma(0)} \left(1+ g\gamma'(0)\log(t) + \tfrac{g^2}{2}\left(\gamma''(0)\log(t)+(\gamma'(0)\log(t))^2\right) + O(g^3)\right)\,.
\end{equation}
Thus we have $\gamma'(0) = 0$ and
\begin{equation}
    \gamma''(0) = 2m^2
\end{equation} 
at order $O(g^2)$. Note that if we did not have the appropriate log divergences in \eqref{eq:2-ptexample} to match \eqref{eq:deltaExpansion}, then that would signal that conformality is lost.

Physically we see that the shift in conformal dimensions of an operator $\calO$ is proportional to its mass, which makes sense since we are coupling a dynamical gauge field $a(t)$ to the number density/mass current $m(x)$, giving a dynamical chemical potential. Likewise, these same arguments will prevent $a$ from getting anomalous dimensions, since $[M,a] = 0$. This also gives us another experimental signature of genuine massless operators: from deviations of expected energy levels proportional to $m^2$.

We can continue these arguments to higher orders. At $O(g^3)$ we find a three-point function of the genuine massless operators $\expval*{a(t_1)a(t_2)a(t_3)}$. Since the genuine massless Ward identities in Section \ref{sec:M0WardIdentities} imply that they behave just as a standard three-point function, the overall result is proportional to the structure \textit{constant} $C_{aaa}$. When we complete the integrals, we find that the result includes $\log^2(t)$ terms which cannot be absorbed by $Z_{\calO}$, additional counterterms, or \eqref{eq:deltaExpansion}. To preserve conformality, this means that
\begin{equation}
    C_{aaa}=0\,,
\end{equation}
see e.g. \cite{Ginsparg:1988ui} for $c=1$ compact boson CFT. Going further to $O(g^4)$ does not add any dramatically new constraint, we simply find that the 4-point function of $\expval*{aaaa}$ must decompose over conformal blocks that are not just the identity, as we would expect for a general global-conformally invariant CFT.

\subsubsection{3-Point Functions: Genuine Massless OPE Channels}
The second step is rather easy: establish that the interaction correctly reproduces a 3-point function of the form constrained by Schr\"odinger symmetries. As before, we write
\begin{equation}
    \expval*{\calO(x)\!\calO^\dagger(0)a(t')}_g
        = Z_{\calO}(\epsilon,g)\expval*{\calO^{(0)}(x)\!\calO^{\dagger(0)}(0) a(t')\,e^{ig\int dt\, M(t)a(t)}}\,.
\end{equation}
Identically to before, we expand to order $O(g^2)$ and find
\begin{equation}
    \expval*{\calO(x)\!\calO^\dagger(0)a(t')}_g
        = i g m \frac{e^{i m \frac{\vec{x}^2}{2t}}}{t^{\Delta-1}t'(t-t')} + O(g^3)\,,
\end{equation}
which matches our results in Section \ref{sec:M0WardIdentities}. This also identifies\footnote{We note the $i$ is from the Lorentzian exponential, not signalling a breakdown of unitarity.}
\begin{equation}
    C_{\calO \!\calO^\dagger \! a} = i g m + O(g^3)\,,
\end{equation}
and imposes a renormalization condition on $a$:
\begin{equation}
    a = a^{(0)} + \frac{i g}{\epsilon} M + O(g^2)\,.
\end{equation}

\subsubsection{4-Point Functions: Violation of Factorization}
Finally, we can check the four-point function and the violation of factorization very directly. A direct computation of the four-point function at $O(g^2)$ in perturbation theory, with $t_1 > \dots > t_4$ gives
\begin{align}
    \label{eq:factorizationviolationexample}
    \begin{split}
        \langle\calO(x_1)\!\calO^{\dagger}(x_2)\!&\calO(x_3)\!\calO^{\dagger}(x_4)\rangle
            = \expval*{\calO(x_1)\!\calO^{\dagger}(x_2)}\!\!\expval*{\calO(x_3)\!\calO^{\dagger}(x_4)}\\
            &-m^2g^2\frac{e^{im\frac{\vec{x}^2_{12}}{2t_{12}}}e^{im\frac{\vec{x}^2_{34}}{2t_{34}}}}{t^{\Delta}_{12}t^{\Delta}_{34}}\int_{t_2}^{t_1} dt\int_{t_4}^{t_3} dt' \frac{1}{(t-t')^2}+O(g^4)\,.
    \end{split}
\end{align}
Thus we see that the four-point function factorizes at $O(g^2)$ up to this final integral. Directly evaluating the integral gives
\begin{equation}
    \int_{t_2}^{t_1} dt \int_{t_4}^{t_3} dt' \frac{1}{(t-t')^2} = \log(\frac{t_{13}t_{24}}{t_{23}t_{14}})\,.
\end{equation}
Happily, we have already seen this result in Section \ref{sec:failureFactorization}: this logarithm and violation of factorization is just the conformal block:
\begin{equation}
 \begin{matrix}
     \text{Violation of}\\
     \text{Factorization}
 \end{matrix}\,\,
    =   -m^2g^2\frac{e^{im\frac{\vec{x}^2_{12}}{2t_{12}}}e^{im\frac{\vec{x}^2_{34}}{2t_{34}}}}{t^{\Delta}_{12}t^{\Delta}_{34}}\frac{t_{12}t_{34}}{t_{14}t_{23}}\,_2F_1\left(1,1;2;- \frac{t_{12}t_{34}}{t_{14}t_{23}}\right)\,.
\end{equation}

There is actually a neat diagrammatic picture of this violation of factorization. Without the genuine massless operators, factorization occurs as explained in \eqref{sec:GenericFactor} because of the splitting into daggered and undaggered operators. Since genuine massless operators are neither creation nor annihilation operators, in time-ordering configurations where they fall between $(t_2,t_1)$ and $(t_4,t_3)$ they break factorization, allowing for communication between the $\calO_1\calO_1^\dagger$ pair and $\calO_2\calO_2^\dagger$ pair.

\acknowledgments We would like to thank Jacob Abajian, Omar Abdelghani, Philip Argyres, Diego Delmastro, Lorenzo Di Pietro, Rajeev Erramilli, Davide Gaiotto, Jaume Gomis, Simeon Hellerman, Kristan Jensen, Zohar Komargodski, Fedor Popov, Leonardo Rastelli, Avia Raviv-Moshe, Adar Sharon, and Siwei Zhong for useful conversations and feedback on the draft. The work of JK is supported by the NSERC PDF program. The work of MB is supported by the FRQNT doctoral training scholarship. 

\bibliographystyle{JHEP}
\bibliography{refs}
\end{document}